\documentclass[twoside,11pt]{article}

\usepackage{jmlr2e}
\usepackage{setspace} 
\usepackage[english]{babel}
\usepackage[margin=1in]{geometry}
\usepackage[OT1]{fontenc}
\usepackage{algorithmic}
\usepackage{rotating}
\usepackage[utf8]{inputenc}
\usepackage{amsmath, amsfonts}
\usepackage[ruled]{algorithm2e}
\usepackage{bm}
\usepackage{color}
\usepackage{bbold}
\usepackage[table]{xcolor}
\definecolor{orangec}{rgb}{0.83, 0.83,0.83}
\definecolor{bluette}{rgb}{0.6,0.6,0.6}
\usepackage{floatrow}
\newfloatcommand{capbtabbox}{table}[][\FBwidth]
\usepackage[font=small,labelfont=bf,tableposition=top]{caption}
\newcolumntype{H}{>{\setbox0=\hbox\bgroup}c<{\egroup}@{}}
\DeclareCaptionLabelFormat{andtable}{#1~#2  \&  \tablename~\thetable}
\usepackage{blindtext}
\newtheorem{mydef}{Definition}

\newtheorem{myprop}{Proposition}
\newtheorem{mycor}{Corollary}

\newcommand{\argmax}{\operatornamewithlimits{argmax}}
\usepackage{multirow}

\jmlrheading{x}{xxxx}{x-x}{x/xx}{x/xx}{Valeria Vitelli, {\O}ystein S{\o}rensen, Marta Crispino, Arnoldo Frigessi and Elja Arjas}

\ShortHeadings{Probabilistic preference learning with the Mallows rank model}{Vitelli, S{\o}rensen, Crispino, Frigessi and Arjas}
\firstpageno{1}

\usepackage{xr}
\begin{document}

\title{Probabilistic preference learning with the Mallows rank model}

\author{\name Valeria Vitelli \email valeria.vitelli@medisin.uio.no \\
       \addr Oslo Centre for Biostatistics and Epidemiology, \\
       Department of Biostatistics, University of Oslo,\\
       P.O.Box 1122 Blindern, NO-0317, Oslo, Norway
       \AND
       \name {\O}ystein S{\o}rensen \email oystein.sorensen.1985@gmail.com \\
       \addr Oslo Centre for Biostatistics and Epidemiology, \\
       Department of Biostatistics, University of Oslo,\\
       P.O.Box 1122 Blindern, NO-0317, Oslo, Norway
       \AND
       \name Marta Crispino \email marta.crispino@phd.unibocconi.com \\
       \addr 
       Department of Decision Sciences, Bocconi University,\\
       via R\"ontgen 1, 20100, Milan, Italy
       \AND
       \name Arnoldo Frigessi \email arnoldo.frigessi@medisin.uio.no \\
       \addr Oslo Centre for Biostatistics and Epidemiology,\\
       University of Oslo and Oslo University Hospital,\\
       P.O.Box 1122 Blindern, NO-0317, Oslo, Norway
       \AND
       \name Elja Arjas \email elja.arjas@helsinki.fi \\
       \addr Oslo Centre for Biostatistics and Epidemiology, \\
       Department of Biostatistics, University of Oslo,\\
       P.O.Box 1122 Blindern, NO-0317, Oslo, Norway}

\editor{}

\maketitle

\begin{abstract}
Ranking and comparing items is crucial for collecting information 
about preferences in many areas, from marketing to politics. 
The Mallows rank model is among the most successful approaches to 
analyse rank data, but its computational complexity has limited its 
use to a particular form based on Kendall distance.
We develop new computationally tractable methods for Bayesian inference 
in Mallows models that work with any right-invariant distance. Our method 
performs inference on the consensus ranking of the items, also when 
based on partial rankings, such as top-$k$ items or pairwise 
comparisons. We prove that items that none of the assessors has 
ranked do not influence the maximum a posteriori consensus ranking, 
and can therefore be ignored. When assessors are many or 
heterogeneous, we propose a mixture model for clustering them in 
homogeneous subgroups, with cluster-specific consensus rankings. We develop approximate stochastic 
algorithms that allow a fully probabilistic analysis, leading to 
coherent quantifications of uncertainties. We make probabilistic 
predictions on the class membership of assessors based on their 
ranking of just some items, and predict missing individual 
preferences, as needed in recommendation systems. We test our 
approach using several experimental and benchmark datasets.
\end{abstract}


\section{Introduction}\label{sec:intro}
Various types of data have ranks as their natural scale. Companies recruit panels to rank novel products, market studies are often based on interviews where competing services or items are compared or ranked. In recent years, analyzing preference data collected over the internet (for example, movies, books, restaurants, political candidates) has been receiving much attention, and often these data are in the form of partial rankings.

Some typical tasks for rank or preference data are: (i) aggregate, merge, summarize multiple individual rankings to estimate the consensus ranking; (ii) predict the ranks of unranked items at individual level; (iii) partition the assessors into classes, each sharing a consensus ranking of the items, and classify new assessors to a class. In this paper we phrase all these tasks (and their combinations) in a unified Bayesian inferential setting, which allows us to also quantify posterior uncertainty of the estimates. Uncertainty evaluations of the estimated preferences and class memberships are a fundamental aspect of information in marketing and decision making. When predictions are too unreliable, actions based on these might better be postponed until more data are available and safer predictions can be made, so as not to unnecessarily annoy users or clients. 

There exist many probabilistic models for ranking data which differ both in the data generation mechanism and in the parametric space. Two of the most commonly used are the Plackett-Luce, PL, \citep{Luce1959,Plackett1975} and the Mallows models \citep{Mallows1957}. The PL model is a stage-wise probabilistic model on permutations, while the Mallows model is based on a distance function between rankings. Inferring the parameters of the PL distribution is typically done by maximum likelihood estimation, using a minorize/maximize algorithm \citep{Hunter2004}. A Bayesian approach was first proposed by \citet{guiver2009}. \citet{Caron2012} perform Bayesian inference in a Plackett-Luce model with time-dependent preference probabilities, and further develop the framework in \citet{Caron2014}, where a Dirichlet process mixture is used to cluster assessors based on their preferences. The parameters in the PL model are continuous, which gives to this model much flexibility. \citet{Volkovs2014} develop a generalization of the PL model, called multinomial preference model, which deals with pairwise preferences, even inconsistent ones, and extends to supervised problems. One difficulty of this method is the use of gradient optimization in a non-convex problem (which can lead to local optima), and the somewhat arbitrary way of imputing missing ranks. Compared to the PL model, the Mallows model has the advantage of being flexible in the choice of the distance function between permutations. It is also  versatile in its ability to adapt to different kinds of data (pairwise comparisons, partial rankings). However, for some distances exact inference is very demanding, because the partition function normalizing the model is very expensive to compute. Therefore most work on the Mallows has been limited to a few particular distances, like the Kendall distance, for which the partition function can be computed analytically. Maximum Likelihood inference about the consensus ranking in the Mallows model is generally very difficult, and in many cases NP-hard, which  lead to the development of heuristic algorithms. The interesting proposal of \citet{Lu2015} makes use of the Generalized Repeated Insertion Model (GRIM), based on the EM algorithm, and allows also for data in the form of pairwise preferences. Their model focuses on the Kendall distance only, and it provides no uncertainty quantification. Another interesting EM-based approach is \citet{seegel2014}, which is driven by expectation propagation approximate inference, and scales to very large datasets without requiring strong factorization assumptions. Among probabilistic approaches, \citet{Meila2010} use Dirichlet process mixtures to perform Bayesian clustering of assessors in the Mallows model, but they again focus on the Kendall distance only. \citet{Jacques2014} also propose clustering based on partial rankings, but in the context of the Insertion Sorting Rank (ISR) model. Hence, the approach is probabilistic but it is far from the general form of the Mallows, even though it has connections with the Mallows with Kendall distance. See Section \ref{sec:RelatedWork} for a more detailed presentation of related work. For the general background on statistical methods for rank data, we refer to the excellent monograph by \citet{Marden1995}, and to the book by \citet{AlvoYu2014}.

The contributions of this paper are summarized as follows. We develop a Bayesian framework for inference in Mallows models that works with any right-invariant metric. In particular, the method is able to handle some of the right-invariant distances poorly considered in the existing literature, because of their well-known intractability. In this way the main advantage of the Mallows models, namely its flexibility in the choice of the distance, is fully exploited. We propose a Metropolis-Hastings iterative algorithm, which converges to the Bayesian posterior distribution, if the exact partition function is available. In case the exact partition function is not available, we propose to approximate it using an off-line importance sampling scheme, and we document the quality and efficiency of this approximation. Using data augmentation techniques, our method handles incomplete rankings, like the important cases of top-$k$ rankings, pairwise comparisons, and ranks missing at random. For the common situation when the pool of assessors is heterogeneous, and cannot be assumed to share a common consensus, we develop a Bayesian clustering scheme which embeds the Mallows model. Our approach unifies clustering, classification and preference prediction in a single inferential procedure, thus leading to coherent posterior credibility levels of learned rankings and predictions. The probabilistic Bayesian setting allows us to naturally compute complex probabilities of interest, like the probability that an item has consensus rank higher than a given level, or the probability that the consensus rank of an item is higher than that of another item of interest. For incomplete rankings this can be done also at the individual assessor level, allowing for individual recommendations.

In Section \ref{sec:MallowsModel}, we introduce the Bayesian Mallows model for rank data. In Section \ref{sec:Distances}, we discuss how the choice of the distance function influences the calculation of the partition function, and Section \ref{sec:Priors} is devoted to the choice of the prior distributions. In Sections \ref{sec:Inference} and \ref{sec:MHAlgorithm}, we show how efficient Bayesian computation can be performed for this model, using a novel leap-and-shift proposal distribution. The tuning of the hyperparameters is discussed in the Supplementary Material, Section \ref{subsec:tuning}. In Section \ref{sec:approximatingZ} we develop and test an importance sampling scheme for computing the partition function, based on a pseudo-likelihood approximation of the Mallows model. We carefully test and study this importance sampling estimation of the partition function (Section \ref{sec:ISConvergence}), and the effect of this estimation on inference, both theoretically (Section \ref{sec:EffectnormalisingConstant}) and by simulations (Section \ref{subsec:modelTesting}). Section \ref{sec:extensions} is dedicated to partial rankings and clustering of assessors. In Section \ref{sec:PartialRanks} we extend the Bayesian Mallows approach to partial rankings, and we prove some results on the effects of unranked items on the consensus ranking (Section \ref{sec:Preselection}). Section \ref{sec:Pairwise} considers data in the form of ordered subsets or pairwise comparisons of items. In Section \ref{sec:Clustering} we describe a mixture model to deal with the possible heterogeneity of assessors, finding cluster-specific consensus rankings. Section \ref{sec:Prediction} is dedicated to prediction in a realistic setup, which requires both the cluster assignment and personalized preference learning. We show that our approach works well in a simulation context. In Section \ref{sec:RelatedWork} we review related methods which have been proposed in the literature, and compare by simulation some algorithms with our procedure (Section \ref{subsec:comparisons}). In Section \ref{sec:Experiments}, we then move to the illustration of the performance of our method on real data: the selected case studies illustrate the different incomplete data situations considered. This includes the Sushi (Section \ref{sec:SushiData}) and Movielens  (Section \ref{subsec:Movielens}) benchmark data. Section \ref{sec:Discussion} presents some conclusions and extensions. 

\section{A Bayesian Mallows Model for Complete Rankings}\label{sec:MallowsModel}

Assume we have a set of $n$ items, labelled $\mathcal{A} = \{A_{1}, A_{2}, \dots,A_{n}\}$. We first assume that each of $N$ assessors ranks all items individually with respect to a considered feature. The ordering provided by assessor $j$ is represented by $\mathbf{X}_{j}$, whose $n$ components are items in $\mathcal{A}$. The item with rank $1$ appears as the first element, up to the item with rank $n$ appearing as the $n$-th element. The observations $\mathbf{X}_1,\ldots,\mathbf{X}_N$ are hence $N$ permutations of the labels in $\mathcal{A}$. Let $R_{ij} = \mathbf{X}_{j}^{-1}(A_{i}), ~ i=1,\dots,n,~ j=1,\dots,N$, denote the rank given to item $A_{i}$ by assessor $j$, and let $\mathbf{R}_{j} = \left(R_{1j}, R_{2j},\dots, R_{nj}\right)$, $j=1,\dots,N$, denote the ranking (that is the full set of ranks given to the items), of assessor $j$. Letting $\mathcal{P}_{n}$ be the set of all permutations of $\{1,\dots,n\}$, we have $\mathbf{R}_{j} \in \mathcal{P}_{n}, ~ j=1,\dots,N$. Finally, let $d(\cdot,\cdot):\mathcal{P}_n\times\mathcal{P}_n\rightarrow[0,\infty)$ be a distance function between two rankings. 

The Mallows model \citep{Mallows1957} is a class of non-uniform joint distributions for a ranking $\mathbf{r}$ on $\mathcal{P}_{n}$, of the form $P(\mathbf{r} | \alpha, \bm{\rho} ) = Z_{n}(\alpha,\bm{\rho})^{-1}\exp\{-(\alpha/n)d( \mathbf{r},\bm{\rho})\} 1_{\mathcal{P}_n}(\mathbf{r})$, where $\bm{\rho}\in\mathcal{P}_n$ is the latent consensus ranking, $\alpha$ is a scale parameter, assumed positive for identification purposes, $Z_{n}(\alpha,\bm{\rho})=\sum_{\mathbf{r} \in \mathcal{P}_{n}} e^{-\frac{\alpha}{n}d(\mathbf{r}, \bm{\rho} )}$ is the partition function, and $1_S(\cdot)$ is the indicator function of the set $S$. We assume that the $N$ observed rankings $\mathbf{R}_{1},\dots,\mathbf{R}_{N}$ are conditionally independent given $\alpha$ and $\bm{\rho}$, and that each of them is distributed according to the Mallows model with these parameters. The likelihood takes then the form
\begin{equation}\label{eq:Joint}
P\left(\mathbf{R}_{1},\dots,\mathbf{R}_{N} | \alpha,\bm{\rho}\right)  = \frac{1}{Z_{n}(\alpha,\bm{\rho})^{N}}\exp\left\{-\frac{\alpha}{n} \sum_{j=1}^N d(\mathbf{R}_{j},\bm{\rho}) \right\}\prod_{j=1}^N\left\{1_{\mathcal{P}_n}(\mathbf{R}_{j})\right\}.
\end{equation}
For a given $\alpha$, the maximum likelihood estimate of $\bm{\rho}$ is obtained by computing 
\begin{equation}\label{eq:MLE}
\argmax_{\bm{\rho}\in \mathcal{P}_n} \frac{\exp\left\{-\frac{\alpha}{n} \sum_{j=1}^N d(\mathbf{R}_{j},\bm{\rho}) \right\}}{Z_{n}(\alpha,\bm{\rho})^{N}}.
\end{equation}
For large $n$ this optimization problem is not feasible, because the space of permutations has $n!$ elements. This has impact both on the computation of $Z_{n}(\alpha,\bm{\rho}),$ and on the minimization of the sum in the exponential of (\ref{eq:MLE}), which is NP-hard \citep{bartholdi1989voting}.

\subsection{Distance Measures and Partition Function}\label{sec:Distances}

Right-invariant distances \citep{Diaconis1988} play an important role in the Mallows models. A right-invariant distance is unaffected by a relabelling of the items, which is a reasonable assumption in many situations. For any right-invariant distance it holds  $d(\bm{\rho_1},\bm{\rho_2})= d(\bm{\rho_1}\bm{\rho_2}^{-1},\mathbf{1}_n)$, where $\mathbf{1}_n=\{1,2,...,n\}$, and therefore the partition function $Z_{n}(\alpha, \bm{\rho})$ of (\ref{eq:Joint}) is independent on the latent consensus ranking $\bm{\rho}$. We write $Z_{n}(\alpha, \bm{\rho}) = Z_{n}(\alpha) = \sum_{\mathbf{r} \in \mathcal{P}_{n}} \exp\{-\frac{\alpha}{n}d(\mathbf{r},\mathbf{1}_n)\}$. All distances considered in this paper are right-invariant. Importantly, since the partition function $Z_{n}(\alpha)$ does not depend on the latent consensus $\bm{\rho}$, it can be computed off-line over a grid for $\alpha$, given $n$ (details in Section \ref{sec:normalisingConstant}).
For some choices of right-invariant distances, the partition function can be analytically computed. For this reason, most of the literature considers the Mallows model with Kendall distance \citep{Lu2015,Meila2010}, for which a closed form of $Z_n(\alpha)$ is given in \citet{Fligner1986},
or with the Hamming \citep{irurozki2014Ham}
and Cayley \citep{irurozki2016sampling} distances.
There are important and natural right-invariant distances for which the computation of the partition function is NP-hard, in particular
%
the footrule (${l}_1$)
and the Spearman's (${l}_2$) distances. For precise definitions of all distances involved in the Mallows model we refer to \citet{Marden1995}.
Following \citet{irurozki2016permallows}, $Z_n(\alpha)$ can be written in a more  convenient way. Since $d(\mathbf{r},\mathbf{1}_n)$ takes only the finite number of discrete values $\mathcal{D}=\{d_1,...,d_a\}$, where $a$ depends on $n$ and on the distance $d(\cdot,\cdot)$, we define $L_i=\{\mathbf{r}\in\mathcal{P}_n: d(\mathbf{r},\mathbf{1}_n)=d_i\}\subset\mathcal{P}_n$, $i=1,...,a$, to be the set of permutations at the same given distance from $\mathbf{1}_n$, and  $|L_i|$ corresponds to its cardinality. Then
\begin{equation}
Z_n(\alpha)=\sum_{{d_i}\in\mathcal{D}}|L_i|\exp\{-(\alpha/n)d_i\}.
\end{equation}
In order to compute $Z_n(\alpha)$ one thus needs $|L_i|$, for all values $d_i\in\mathcal{D}$. In the case of the footrule distance, the set $\mathcal{D}$ includes all even numbers, from 0 to $\lfloor n^2/2 \rfloor$, and $|L_i|$ corresponds to the sequence A062869 available for $n\leq 50$ on the On-Line Encyclopedia of Integer Sequences (OEIS) \citep{sloane}. 
In the case of {Spearman's distance}, the set $\mathcal{D}$ includes all even numbers, from 0 to  $2\binom{n}{3}$, and $|L_i|$ corresponds  to  the sequence  A175929 available for $n\leq 14$ in the OEIS. 
When the partition function is needed for larger values of $n$, we suggest an importance sampling scheme which efficiently approximates $Z_{n}(\alpha)$ to an arbitrary precision (see Section \ref{sec:normalisingConstant}). An interesting asymptotic approximation for $Z_n(\alpha)$, when $n \rightarrow \infty$, has been studied in \citet{mukherjee2016}, and we apply it in an example where $n=200$ (see  Section \ref{subsec:Movielens}, and Section \ref{subsec:Zlim} in the Supplementary Material).

\subsection{Prior Distributions}\label{sec:Priors}
To complete the specification of the Bayesian model for the rankings $\mathbf{R}_{1},\dots,\mathbf{R}_{N}$, a prior for its parameters is needed. We assume a priori that $\alpha$ and $\bm{\rho}$ are independent.

An obvious choice for the prior for $\bm{\rho}$ in the context of the Mallows likelihood is to utilize the Mallows model family also in setting up a prior for $\bm{\rho}$, and let $\pi(\bm{\rho})=\pi(\bm{\rho}|\alpha_0,\bm{\rho}_0)\propto\exp\left\{-\frac{\alpha_0}{n} d(\bm{\rho},\bm{\rho}_0) \right\}$. Here $\alpha_0$  and $\bm{\rho}_0$ are fixed hyperparameters, with $\bm{\rho}_0$ specifying the ranking that is a priori thought most likely, and $\alpha_0$ controlling the tightness of the prior around $\bm{\rho}_0$. Since $\alpha_0$ is fixed, $Z_n(\alpha_0)$ is a constant. Note that combining the likelihood with the prior $\pi(\bm{\rho}|\alpha_0,\bm{\rho}_0)$ above has the same effect on inference as involving an additional hypothetical assessor $j = 0$, say, who then provides the ranking $\mathbf{R}_{0}=\bm{\rho}_0$ as data, with $\alpha_0$ fixed. 

If we were to elicit a value for $\alpha_0$, we could reason as follows. Consider, for $\bm{\rho}_0 $ fixed, the prior expectation $g_n(\alpha_0) := E_{\pi(\bm{\rho})} (d(\bm{\rho},\bm{\rho}_0) | \alpha_0, \bm{\rho}_0)$. Because of the assumed right invariance of the distance $d(\cdot,\cdot)$, this expectation is independent of $\bm{\rho}_0$, which is why $g_n(\cdot)$ depends only on $\alpha_0$. Moreover, $g_n(\alpha_0)$ is obviously decreasing in $\alpha_0$. For the footrule and Spearman distances, which are defined as sums of item specific deviations $|\rho_{0i} -\rho_{i}|$ or $|\rho_{0i} -\rho_{i}|^2$, $g_n(\alpha_0)$ can be interpreted as the expected (average, per item) error in the prior ranking $\pi(\bm{\rho}|\alpha_0,\bm{\rho}_0)$ of the consensus. A value for $\alpha_0$ is now elicited by first choosing a target level $\tau_0$, say, which would realistically correspond to such an a priori expected error size, and then finding the value $\alpha_0$ such that $g_n(\alpha_0) = \tau_0$. This procedure requires numerical evaluation of the function $g_n(\alpha_0)$ over a range of suitable $\alpha_0$ values.
In this paper, we employ only the uniform prior $\pi( \bm{\rho}) = (n!)^{-1} 1_{\mathcal{P}_{n}}(\bm{\rho})$ in the space $\mathcal{P}_{n}$ of $n-$dimensional permutations, corresponding to $\alpha_0=0$. 

For the scale parameter $\alpha$ we have in this paper used the exponential prior, with density $\pi(\alpha|\lambda)=\lambda e^{-\lambda\alpha}1_{[0,\infty)}(\alpha)$. We show in Figure \ref{fig:sim1_1_rho} of Section \ref{subsec:modelTesting} on simulated data, that the inferences on $\bm{\rho}$ are almost completely independent of  the choice of the value of $\lambda$. Also a theoretical argument for this is provided in that same section, although it is tailored more specifically to the numerical approximations of $Z_n(\alpha)$. For these reasons, in all our data analyses, we assigned $\lambda$ a fixed value. We chose $\lambda = 0.1$ or $\lambda=0.05$, depending on the complexity of the data, thus implying a prior density for $\alpha$ which is quite flat in the region supported in practice by the likelihood. If a more elaborate elicitation of the prior for $\alpha$ for some reason were preferred, this could be achieved by computing, by numerical integration, values of the function $E_{\pi(\alpha)} (g_n(\alpha) | \lambda)$,  selecting a realistic target $\tau$, and solving $E_{\pi(\alpha)} (g_n(\alpha) | \lambda)=\tau$ for $\lambda$. In a similar fashion as earlier, also $E_{\pi(\alpha)} (g_n(\alpha) | \lambda)$ can be interpreted as an expected (average, per item) error in the ranking, but now by {\em errors} is meant those made by the assessors, relative to the consensus, and expectation is with respect to the exponential prior $\pi(\alpha|\lambda)$.

\clearpage

\subsection{Inference}\label{sec:Inference}
Given the prior distributions $\pi(\bm{\rho})$ and $\pi(\alpha)$, and assuming prior independence of these variables, the posterior distribution for $\bm{\rho}$ and $\alpha$ is given by
\begin{equation} \label{eq:posterior}
P\left(\bm{\rho},\alpha|\mathbf{R}_1,\dots,\mathbf{R}_N\right) \propto \frac{\pi\left(\bm{\rho} \right) \pi\left(\alpha\right)}{Z_{n}\left(\alpha\right)^{N}} \exp\left\{-\frac{\alpha}{n}\sum_{j=1}^{N}d\left(\mathbf{R}_{j},\bm{\rho}\right)\right\}  .
\end{equation}
Often one is interested in computing posterior summaries of this distribution. One such summary is the marginal posterior mode of $\bm{\rho}$ (the maximum a posteriori, MAP) from (\ref{eq:posterior}), which does not depend on $\alpha$, and in case of uniform prior for $\bm{\rho}$ coincides with the ML estimator of $\bm{\rho}$ in (\ref{eq:MLE}). The marginal posterior distribution of $\bm{\rho}$ is given by
\begin{equation} \label{eq:posteriorrho}
P\left(\bm{\rho}|\mathbf{R}_1,\dots,\mathbf{R}_N\right) \propto \pi\left(\bm{\rho} \right) \int_0^\infty \frac{
\pi\left(\alpha\right)}{Z_{n}\left(\alpha\right)^{N}} 
\exp\left\{-\frac{\alpha}{n}\sum_{j=1}^{N}d\left(\mathbf{R}_{j},\bm{\rho}\right)\right\} \text{d}\alpha.
\end{equation}
Given the data, $\mathbf{R}=\{\mathbf{R}_1,\dots,\mathbf{R}_N\}$ and the consensus ranking $\bm{\rho}$, the sum of distances, $T(\bm{\rho},\;\mathbf{R})=\sum_{j=1}^{N}d\left(\mathbf{R}_{j},\bm{\rho}\right)$, takes only a finite set of discrete values $\{t_1, t_2, ... t_m\}$, where $m$ depends on the distance $d(\cdot,\cdot)$, on the sample size $N$, and on $n$. Therefore, the set of all  permutations $\mathcal{P}_{n}$ can be partitioned into the sets $H_i =\{ \bm{r} \in \mathcal{P}_{n} : 
T(\bm{r},\;\mathbf{R})=t_i\}$ for each distance $t_i$. These sets are level sets of the posterior marginal distribution in (\ref{eq:posteriorrho}), as all $\bm{r} \in H_i$ have the same posterior marginal probability. The level sets do not depend on $\alpha$ but the posterior distribution shared by the permutations in each set does. 

In applications, the interest often lies in computing posterior probabilities of more complex functions of the consensus $\bm{\rho}$, for example the posterior probability that a certain item has consensus rank lower than a given level (``among the top 5", say), or that the consensus rank of a certain item is higher than the consensus rank of another one. These probabilities cannot be readily obtained within the maximum likelihood approach, while the Bayesian setting very naturally allows to approximate any posterior summary of interest by means of a Markov Chain Monte Carlo algorithm, which at convergence samples from the posterior distribution (\ref{eq:posterior}). 

\subsection{Metropolis-Hastings Algorithm for Complete Rankings}\label{sec:MHAlgorithm}

In order to obtain samples from the posterior in equation (\ref{eq:posterior}), we iterate between two steps. In one step we update the consensus ranking. Starting with $\alpha \geq 0$ and $\bm{\rho} \in \mathcal{P}_{n}$, we first update $\bm{\rho}$ by proposing $\bm{\rho}^{\prime}$ according to a distribution which is centered around the current rank $\bm{\rho}$. 
\begin{mydef}{Leap-and-Shift Proposal (L\normalfont{\&}S).}\label{def:LeapShift}
Fix an integer $L\in\{1,\ldots,\lfloor (n-1)/2 \rfloor\}$ and draw a random number $u\sim\mathcal{U}\{1,\dots,n\}$. Define, for a given $\bm{\rho}$, the set of integers $\mathcal{S} = \{\max(1, \rho_u - L), \min(n, \rho_u + L)\}\setminus \{\rho_{u}\},$ $\mathcal{S}\subseteq\{1,\ldots,n\}$, and draw a random number $r$ uniformly in $\mathcal{S}$. Let $\bm{\rho}^{*}\in \{1,2,...n\}^{n}$ have elements $\rho_{u}^{*}=r$ and $\rho_{i}^{*} = \rho_{i}$ for $i\in \{1,\dots,n\}\setminus \{u\},$ constituting the leap step.
Now, define $\Delta = \rho_{u}^{*} - \rho_{u}$ and the proposed $\bm{\rho}^\prime\in\mathcal{P}_n$ with elements
\begin{equation*}
\rho_{i}^{\prime} =
\begin{cases}
\rho_{u}^{*}&\text{ if }\rho_i = \rho_u \\
\rho_{i} - 1 &\text{ if } \rho_{u} < \rho_{i} \leq \rho_{u}^{*} \text{ and } \Delta > 0 \\
\rho_{i} + 1 &\text{ if } \rho_{u} > \rho_{i} \geq \rho_{u}^{*} \text{ and } \Delta < 0 \\
\rho_{i} &\text{ else },
\end{cases}
\end{equation*}
for $i=1,\dots,n$, constituting the shift step.
\end{mydef}
The probability mass function associated to the transition is given by
\begin{align*}
&P_L(\bm{\rho}^{\prime} | \bm{\rho}) =   \sum_{u=1}^{n} P_L(\bm{\rho}^{\prime}|U=u,\bm{\rho}) P(U=u) \\ \nonumber
& = \frac{1}{n} \sum_{u=1}^{n} \left\{ 1_{\left\{\bm{\rho}_{-u}\right\}}(\bm{\rho}_{-u}^{*}) \cdot 
1_{\left\{0< | \bm{\rho}_{u} - \bm{\rho}_{u}^{*} | \leq L\right\}}(\bm{\rho}_{u}^{*}) \cdot
\left[ \frac{1_{\left\{L+1,\ldots,n-L\right\}}(\bm{\rho}_{u})}{2L} + \sum_{l=1}^L \frac{1_{\left\{l\right\}}(\bm{\rho}_{u}) + 1_{\left\{n-l+1\right\}}(\bm{\rho}_{u})}{L+l-1} \right] \right\} \\ \nonumber
& + \frac{1}{n} \sum_{u=1}^{n} \left\{ 1_{\left\{\bm{\rho}_{-u}\right\}}(\bm{\rho}_{-u}^{*}) \cdot 1_{\left\{ | \bm{\rho}_{u} - \bm{\rho}_{u}^{*} | = 1\right\}}(\bm{\rho}_{u}^{*}) \cdot 
\left[ \frac{1_{\left\{L+1,\ldots,n-L\right\}}(\bm{\rho}_{u}^{*})}{2L} + \sum_{l=1}^L \frac{1_{\left\{l\right\}}(\bm{\rho}_{u}^{*}) + 1_{\left\{n-l+1\right\}}(\bm{\rho}_{u}^{*})}{L+l-1} \right] \right\},
\end{align*}
where $\bm{\rho}_{-u} = \{\rho_{i}; ~ i\neq u\}$. 

\begin{myprop}\label{prop:LeapShift}
The leap-and-shift proposal $\bm{\rho}^{\prime} \in \mathcal{P}_{n}$ is a local perturbation of $\bm{\rho}$, separated from $\bm{\rho}$ by a Ulam distance $1$ .
\end{myprop}
\begin{proof}
From the definition and by construction, $\bm{\rho}^{*}\notin \mathcal{P}_n$, since there exist two indices $i\neq j$ such that $\rho_{i}^{*} = \rho_{j}^{*}$. The shift of the ranks by $\Delta$ brings $\bm{\rho}^{*}$ to $\bm{\rho}^{\prime}$ back into $\mathcal{P}_{n}$. The Ulam distance $d(\bm{\rho}, \bm{\rho}^{\prime})$ is the number of edit operations needed to convert $\bm{\rho}$ to $\bm{\rho}^{\prime}$, where each edit operation involves deleting a character and inserting it in a new place. This is equal to 1, following \cite{Gopalan2006}.
\end{proof}

The acceptance probability when updating $\bm{\rho}$ in the Metropolis-Hastings algorithm is
\begin{equation}\label{eq:MHrho}
\text{min}\left\{1,\frac{P_L(\bm{\rho} | \bm{\rho}^{\prime})\pi\left(\bm{\rho}^{\prime}\right) }{P_L(\bm{\rho}^{\prime} | \bm{\rho}) \pi\left(\bm{\rho}\right) }\exp\left[-\frac{\alpha}{n} \sum_{j=1}^{N}\left\{d\left(\mathbf{R}_{j},\bm{\rho}^{\prime}\right) - d\left(\mathbf{R}_{j},\bm{\rho}\right) \right\}\right] \right\}.
\end{equation}
The leap-and-shift proposal is not symmetric, thus the ratio $P_L(\bm{\rho} | \bm{\rho}^{\prime})/P_L(\bm{\rho}^{\prime} | \bm{\rho})$ does not cancel in (\ref{eq:MHrho}). The parameter $L$ is used for tuning this acceptance probability.

The term $\sum_{j=1}^{N} \left\{d\left(\mathbf{R}_{j}, \bm{\rho}' \right) - d\left(\mathbf{R}_{j}, \bm{\rho} \right)\right\}$ in (\ref{eq:MHrho}) can be computed efficiently, since most elements of $\bm{\rho}$ and $\bm{\rho}'$ are equal. Let $\rho_{i} = \rho_{i}'$ for $i \in E \subset\left\{ 1,\dots,n\right\}$, and $\rho_{i} \neq \rho_{i}'$ for $i \in E^{c}$. For the footrule and Spearman distances, we then have
\begin{equation}\label{eq:FormulaPDistance}
\sum_{j=1}^{N} \left\{d\left(\mathbf{R}_{j}, \bm{\rho}' \right) - d\left(\mathbf{R}_{j}, \bm{\rho} \right)\right\} = \sum_{j=1}^{N} \left\{ \sum_{i \in E^{c}} \left|R_{ij} - \rho_{i}' \right|^{p} - \sum_{i \in E^{c}} \left|R_{ij} - \rho_{i} \right|^{p}\right\},
\end{equation}
for $p \in \left\{1,2\right\}$. For the Kendall distance, instead, we get
\begin{align*} 
&\sum_{j=1}^{N} \left\{d\left(\mathbf{R}_{j}, \bm{\rho}' \right) - d\left(\mathbf{R}_{j}, \bm{\rho} \right)\right\}= \\
&= \sum_{j=1}^{N} \left\{ \sum_{1 \leq k < l \leq n} 1\left[\left(R_{kj} - R_{lj} \right) \left( \rho_{k}' - \rho_{l}'\right) >0 \right] - 1\left[\left(R_{kj} - R_{lj} \right) \left( \rho_{k} - \rho_{l}\right) >0 \right]\right\}  \\
&= \sum_{j=1}^{N} \left\{ \sum_{k \in E^{c} \setminus \{n\}} \sum_{l \in \{E^{c} \cap \{l > k\}\} } 1\left[\left(R_{kj} - R_{lj} \right) \left( \rho_{k}' - \rho_{l}'\right) >0 \right] - 1\left[\left(R_{kj} - R_{lj} \right) \left( \rho_{k} - \rho_{l}\right) >0 \right]\right\}.
\end{align*}
Hence, by storing the set $E^{c}$ at each MCMC iteration, the computation of (\ref{eq:MHrho}) involves a sum over fewer terms, speeding up the algorithm consistently.

The second step of the algorithm updates the value of $\alpha$. We sample a proposal $\alpha'$ from a lognormal distribution $\log\mathcal{N}(\alpha, \sigma_{\alpha}^{2})$ and accept it with probability
\begin{equation}\label{eq:MHalpha}
\text{min}\left\{1, \frac{Z_{n}\left(\alpha\right)^{N} \pi\left(\alpha^{\prime}\right)\alpha^\prime}{Z_{n}\left(\alpha^{\prime}\right)^{N}  \pi\left(\alpha\right)\alpha} \exp\left[-\frac{\left(\alpha^{\prime}-\alpha\right)}{n} \sum_{j=1}^{N}d\left(\mathbf{R}_{j},\bm{\rho}\right)\right] \right\},
\end{equation}
where $\sigma_{\alpha}^{2}$ can be tuned to obtain a desired acceptance probability. A further parameter, named $\alpha_\text{jump},$ can be used to update $\alpha$ only every $\alpha_\text{jump}$ updates of $\bm{\rho}$: the possibility to tune this parameter ensures a better mixing of the MCMC in the different sparse data applications. The above described MCMC algorithm is summarized as Algorithm \ref{algo:basicMH} of Appendix \ref{sec:algoSupp}.

\begin{myprop}{Convergence of the MCMC algorithm for exact $Z_{n}\left(\alpha\right)$.}
The MCMC Algorithm \ref{algo:basicMH} using the exact partition function $Z_{n}\left(\alpha\right)$ samples from the Mallows posterior in equation (\ref{eq:posterior}), as the number of MCMC iterations tends to infinity.
\end{myprop}

\begin{proof}
Because of reversibility of the proposals, detailed balance holds for the Markov chain. Ergodicity follows by aperiodicity and positive recurrence. 
\end{proof}

Section \ref{sec:approximatingZ} investigates approximations of $Z_{n}\left(\alpha\right),$ and how they affect the MCMC and the estimate of the consensus $\bm{\rho}$. In Section \ref{subsec:tuning} of the Supplementary Material we instead focus on aspects related to the practical choices involved in the use of our MCMC algorithm, and in particular we aim at defining possible strategies for tuning the MCMC parameters $L$ and $\sigma_\alpha$.

\section{Approximating the partition function $Z_n(\alpha)$ via off-line importance sampling}\label{sec:approximatingZ}\label{sec:normalisingConstant}

For Kendall's, Hamming and Cayley distances, the partition function $Z_{n}\left(\alpha\right)$ is available in close form, but this is not the case for footrule and Spearman distances. To handle these cases, we propose an approximation of the partition function $Z_n(\alpha)$ based on importance sampling. Since we focus on right-invariant distances, the partition function does not depend on $\bm{\rho}.$ Hence, we can obtain an off-line approximation of the partition function on a grid of $\alpha$ values, interpolate it to yield an estimate of $Z_n(\alpha)$ over a continuous range, and then read off needed values to compute the acceptance probabilities very rapidly. 

We study the convergence of the importance sampler theoretically (Section \ref{sec:EffectnormalisingConstant}) and numerically (Sections \ref{sec:ISConvergence}, \ref{subsec:modelTesting}), with a series of experiments aimed at demonstrating the quality of the approximation, and its impact in inference. We here show the results obtained with the footrule distance, but we obtained similar results with the Spearman distance. We also summarize in the Supplementary Material (Section  \ref{subsec:Zlim}) a further possible approximation of $Z_n(\alpha)$, namely the asymptotic proposal in \cite{mukherjee2016}. 

We briefly discuss the pseudo-marginal approaches for tackling intractable Metropolis-Hastings ratios, which could  in principle be an interesting alternative. We refer to \citet{Andrieu2009}, \citet{murray2012mcmc} and  \citet{Sherlock2015} for a full description of the central methodologies. The idea is to replace $P(\bm{\rho}, \alpha |\mathbf{R})$ in (4) with a non-negative unbiased estimator $\hat P$, such that for some $C>0$ we have $\mathbb{E}[\hat P]=C P$. The approximate acceptance ratio then uses  $\hat P$, but this results in an algorithm still targeting the exact posterior. An unbiased estimate of the posterior $P$ can be obtained via importance sampling if it is possible to simulate directly from the likelihood. This is not the case in our model, as there are no algorithms available to sample from the Mallows model with, say, the footrule distance. Neither is use of exact simulation possible for our model.
The approach in \citet{murray2012mcmc} extends the model by introducing an auxiliary variable, and uses a proposal distribution in the MCMC such that the partition functions cancel. A useful  proposal for this purpose would in our case be based on the Mallows likelihood, so that again one would need to be able to sample from it, which is not feasible.


Our suggestion is instead to estimate the partition function directly, using an Importance Sampling (IS) approach. 
For $K$ rank vectors $\mathbf{R}^{1},\dots,\mathbf{R}^{K}$ sampled from an IS auxiliary distribution $q(\mathbf{R})$, the unbiased IS estimate of $Z_{n}(\alpha)$ is given by
\begin{equation}\label{eq:Zhat}
\hat{Z}_{n}(\alpha) = K^{-1}\sum_{k=1}^{K} \exp\{-(\alpha/n)d(\mathbf{R}^{k}, \mathbf{1}_n )\}q(\mathbf{R}^{k})^{-1}.
\end{equation}
The more $q(\mathbf{R})$ resembles the Mallows likelihood (\ref{eq:Joint}), the smaller is the variance of $\hat{Z}_{n}(\alpha)$. On the other hand, it must be computationally feasible to sample from $q(\mathbf{R})$. We use the following pseudo-likelihood approximation of the target (\ref{eq:Joint}). Let $\{i_1,\ldots,i_n\}$ be a uniform sample from $\mathcal{P}_n,$ which gives the order of the pseudo-likelihood factorization. Then
$$P
(\mathbf{R}| \bm{1}_n ) = P(R_{i_1} | R_{i_2},\dots,R_{i_n},\bm{1}_n)  P(R_{i_2} | R_{i_3},\dots,R_{i_n}, \bm{1}_n)\cdots P(R_{i_{n-1}} |  R_{i_n}, \bm{1}_n)P(R_{i_n}|\bm{1}_n),
$$ 
and the conditional distributions are given by
\begin{align*}
&P\left(R_{i_n}|\bm{1}_n\right) = \frac{\exp\left\{-(\alpha/n)  d\left(R_{i_n} , i_n\right) \right\} \cdot 1_{\left[1,\dots,n\right]}(R_{i_n})}{\sum_{r_n \in \{1,\dots,n\}} \exp\left\{-(\alpha/n)  d\left(r_n , i_n\right) \right\}},
\\
 &P\left(R_{i_{n-1}} | R_{i_n},\bm{1}_n \right) = \frac{\exp\left\{-(\alpha/n) d\left(R_{i_{n-1}}, i_{n-1}\right) \right\} \cdot 1_{\left[\{1,\dots,n\} \setminus \{R_{i_n}\} \right]}(R_{i_{n-1}})}{\sum_{r_{n-1} \in \{1,\dots,n\} \setminus \{R_{i_n}\}} \exp\left\{-(\alpha/n)  d\left(r_{n-1} , i_{n-1} \right) \right\}}, \\ \nonumber
& \vdots
\\
&
 P\left(R_{i_2} | R_{i_3},\dots,R_{i_n}, \bm{1}_n\right) = \frac{\exp\left\{-(\alpha/n)  d\left(R_{i_2} , i_2\right) \right\} \cdot 1_{\left[\{1,\dots,n\} \setminus \{R_{i_3},\dots,R_{i_n}\}\right]}(R_{i_2})}{\sum_{r_{2} \in \{1,\dots,n\} \setminus \{R_{i_3},\dots,R_{i_n}\} }\exp\left\{-(\alpha/n)  d\left(r_{2} , i_2 \right) \right\}},
\\
&
 P\left(R_{i_1} | R_{i_2},\dots,R_{i_n},\bm{1}_n\right) = 1_{\left[\{1,\dots,n\} \setminus \{R_{i_2},\dots,R_{i_n}\}\right]}(R_{i_1}).
\end{align*}
Each factor is a simple univariate distribution. We sample $R_{i_n}$ first, and then conditionally on that, $R_{i_{n-1}}$ and so on. The $k$-th full sample $\mathbf{R}^{k}$ has probability $q(\mathbf{R}^{k}) = \penalty 0 P(R^{k}_{i_n}|\bm{1}_n) \penalty 0 P(R^{k}_{i_{n-1}} | R^{k}_{i_n},\bm{1}_n) \penalty 0 \cdots \penalty 0 P(R^{k}_{i_2} | R^{k}_{i_3},\dots,R^{k}_{i_n},\bm{1}_n)$. We observe that this pseudo-likelihood construction is similar to the sequential representation of the Plackett-Luce model with a Mallows parametrization of probabilities.

Note that, in principle, we could sample rankings $\mathbf{R}^k$ from the Mallows model with a different distance than the one of the target model (for example Kendall), or use the pseudo-likelihood approach with a different ``proposal distance'' other than the target distance. We experimented with these alternatives, but keeping the pseudo-likelihood with the same distance as the one in the target was most accurate and efficient (results not shown). In what follows the distance in (\ref{eq:Zhat}) is the same as the distance in (\ref{eq:posterior}).

\subsection{Testing the Importance Sampler}\label{sec:ISConvergence}

\begin{figure}[h!]
\centering
\includegraphics[scale=.73]{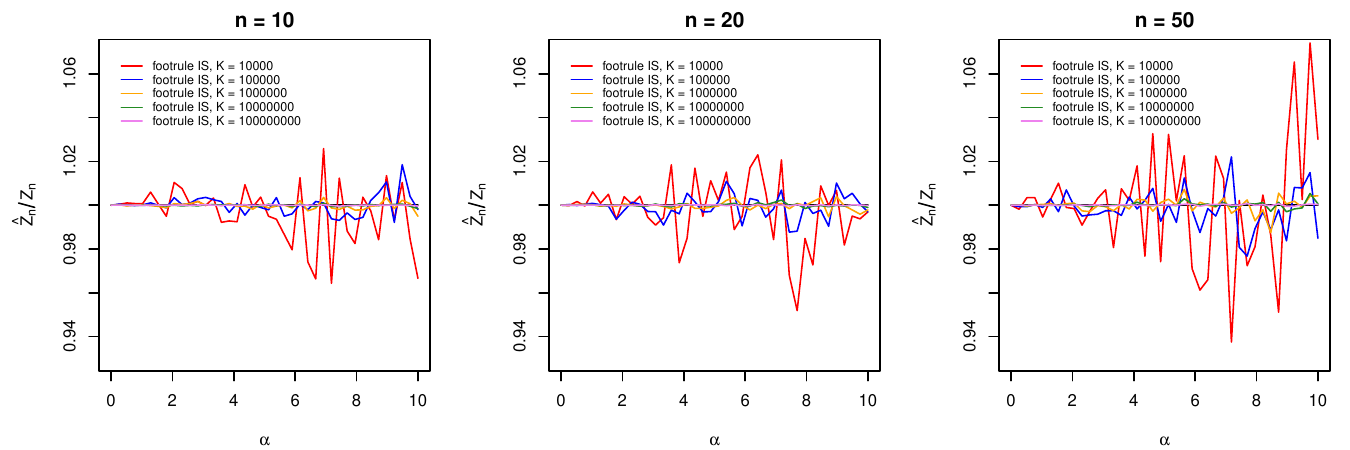}
\caption{Ratio of the approximate partition function computed via IS to the exact, $\hat{Z}_n(\alpha)/Z_n(\alpha)$, as a function of $\alpha$, when using the footrule distance. From left to right, $n=10, 20, 50;$ different colors refer to different values of $K,$ as stated in the legend.}
\label{fig:Zratio}
\end{figure}

We experimented by increasing the number $K$ of importance samples in powers of ten, over a discrete grid of $100$ equally spaced $\alpha$ values between $0.01$ and $10$ (this is the range of $\alpha$ which turned out to be relevant in all our applications, typically $\alpha < 5$). We produced a smooth partition function simply using a polynomial of degree 10.  The ratio $\hat{Z}^K_n(\alpha)/Z_n(\alpha)$ as a function of $\alpha$ is shown in Figure \ref{fig:Zratio} for $n=10,20,50$ and when using different values of $K$: the ratio quickly approaches 1 when increasing $K$; for larger $n,$ a larger $K$ is needed to ensure precision, but $K=10^6$ seems enough to give very precise estimates.

When $n$ is larger than $50,$ no exact expression for $Z_n(\alpha)$ is available. Then, we directly compare the estimated $\hat{Z}^K_n(\alpha)$ for increasing $K,$ to check whether the estimates stabilize. We thus inspect the maximum relative error 
\begin{align}\label{eq:RelativeError}
\epsilon_K = \underset{\alpha}{\text{max}} \left[\frac{\left|\hat{Z}^{K}_{n}(\alpha) - \hat{Z}^{K/10}_{n}(\alpha) \right|}{\left|\hat{Z}^{K/10}_{n}(\alpha)\right|}\right]
\end{align}
for $K=10^{2},\ldots,10^{8}$. Results are shown in Table \ref{tab:Zerr} for $n=75$ and 100. For both values of $n$ we see that the estimates quickly stabilize, and $K=10^6$ appears to give good approximations. The computations shown here were performed on a desktop computer, and the off-line computation with $K=10^{6}$ samples for $n=10$ took less than 15 minutes, with no efforts for parallelizing the algorithm, which would be easy and beneficial. $K = 10^6$ samples for $n = 100$ were obtained on a 64-cores computing cluster in $12$ minutes.

\begin{table}[h!]
\center
\begin{tabular}{|c|c|c|c|c|c|c|c|}
\hline
$K$ & $10^{2}$ &$10^{3}$ &$10^{4}$&$10^{5}$&$10^{6}$ &$10^{7}$ &$10^{8}$\\\hline
$n=75$ &  $152.036$ &$0.921$ &  $0.373$& $0.084$ & $0.056$&$0.005$ &$0.004$\\
$n=100$ & $67.487$ &$1.709$&$0.355$&$0.187$&$0.045$&$0.018$&$0.004$\\\hline

\end{tabular}
\caption{Approximation of the partition function via the IS for the footrule model: maximum relative error $\epsilon_K$ from equation \eqref{eq:RelativeError}, between the current and the previous $K$, for $n=75$ and 100.}
\label{tab:Zerr}
\end{table}

\subsection{Effect of $\hat{Z}_n(\alpha)$ on the MCMC}\label{sec:EffectnormalisingConstant}

In this Section we report theoretical results regarding the convergence of the MCMC, when using the IS approximation of the partition function. 
\begin{myprop}
Algorithm \ref{algo:basicMH} of Appendix \ref{sec:algoSupp} using $\hat{Z}_n(\alpha)$ in (\ref{eq:Zhat}) instead of $Z_n(\alpha)$ converges to the posterior distribution proportional to
\begin{equation}\label{eq:postHat}
\frac{1}{\hat{C}(\mathbf{R})}\pi(\bm{\rho})\pi(\alpha) \hat{Z}_{n}(\alpha)^{-N}\exp\left\{-\frac{\alpha}{n} \sum_{j=1}^N d(\mathbf{R}_{j},\bm{\rho}) \right\},
\end{equation}
with the normalizing factor $\hat{C}(\mathbf{R}) = \int_0^\infty \pi(\bm{\rho})\pi(\alpha) \hat{Z}_{n}(\alpha)^{-N} \sum_{\bm{\rho}\in\mathcal{P}_n}\exp\left\{-\frac{\alpha}{n} \sum_{j=1}^N d(\mathbf{R}_{j},\bm{\rho}) \right\} d\alpha$.
\end{myprop}
\proof
The acceptance probability of the MCMC in Algorithm 1 with the approximate partition function is given by (\ref{eq:MHalpha}) using $\hat{Z}_n(\alpha)$ in (\ref{eq:Zhat}) instead of $Z_n(\alpha)$, which is exactly the acceptance probability needed for (\ref{eq:postHat}).
\endproof
The IS approximation $\hat{Z}_n(\alpha)$ converges to $Z_n(\alpha)$ as the number $K$ of IS samples converges to infinity. In order to study this limit, let us change the notation to explicitly show this dependence and write $\hat{Z}^{K}_n(\alpha)$. Clearly, the approximate posterior (\ref{eq:postHat}) converges to the correct posterior (\ref{eq:posterior}) if $K$ increases with $N$, $K=K(N),$ and 
\begin{equation}\label{eq:Zlimit}
\lim_{N\rightarrow \infty} \left( \frac{\hat{Z}^{K(N)}_n(\alpha)}{Z_n(\alpha)} \right)^N = 1,\quad \text{for all } \alpha.
\end{equation}

\begin{myprop}\label{prop:Zlimit}
There exists a factor $c(\alpha,n,d(\cdot,\cdot))$ not depending on $N$, such that, if $K=K(N)$ tends to infinity as $N\rightarrow\infty$ faster than $c(\alpha,n,d(\cdot,\cdot))\cdot N^2$, then (\ref{eq:Zlimit}) holds.
\end{myprop}


\proof
We see that
$$
\left( \frac{\hat{Z}^{K(N)}_n(\alpha)}{Z_n(\alpha)} \right)^N = \exp\left\{ N \log\left( 1+\frac{\hat{Z}^{K(N)}_n(\alpha) - Z_n(\alpha)}{Z_n(\alpha)} \right) \right\} 
$$
tends to 1 in probability as $K(N)\rightarrow \infty$ when $N\rightarrow\infty$ if
\begin{equation}\label{eq:pippo}
\frac{\hat{Z}^{K(N)}_n(\alpha) - Z_n(\alpha)}{Z_n(\alpha)}
\end{equation}
tends to 0 in probability faster than $1/N.$ Since (\ref{eq:Zhat}) is a sum of i.i.d. variables, there exists a constant $c=c(\alpha,n,d(\cdot,\cdot))$ depending on $\alpha,$ $n$ and the distance $d(\cdot,\cdot)$ (but not on $N$) such that
$$
\sqrt{K(N)}(\hat{Z}^{K(N)}_n(\alpha) - Z_n(\alpha) ) \overset{\mathcal{L}}{\rightarrow} \mathcal{N}(0,c^2),
$$
in law as $K(N)\rightarrow\infty.$ Therefore, for (\ref{eq:pippo}) tending to 0 faster than $1/N$, it is sufficient that $K(N)$ grows faster than $N^2.$ The speed of convergence to 1 of (\ref{eq:Zlimit}) depends on $c=c(\alpha,n,d(\cdot,\cdot))$.
\endproof

\subsection{Testing approximations of the MCMC in inference}\label{subsec:modelTesting}

We report results from extensive simulation experiments carried out in several different parameter settings, to investigate if our algorithm provides correct posterior inferences. In addition, we study the sensitivity of the posterior distributions to differences in the prior specifications, and demonstrate their increased precision when the sample size $N$ grows. We explore the robustness of inference when using approximations of the partition function $Z_n(\alpha)$, both when obtained by applying our IS approach, and when using, for large $n$, the asymptotic approximation $Z_{\lim}(\alpha)$ proposed in \cite{mukherjee2016}. We focus here on the footrule distance since it allows us to explore all these different settings, being also the preferred distance in the experiments reported in Section \ref{sec:Experiments}. Some model parameters are kept fixed in the various cases: $\alpha_\text{jump}=10,$ $\sigma_\alpha=0.15,$ and $L=n/5$ (for the tuning of the two latter parameters, see the simulation study in the Supplementary Material, Section \ref{subsec:tuning}). Computing times for the simulations,  performed on a laptop computer, varied depending on the value of $n$ and $N$, from a minimum of $24''$ in the smallest case with $n=20$ and $N=20,$ to a maximum of $3'22''$ for $n=100$ and $N=1000$.

\begin{figure}[h!]
\centering
\includegraphics[scale=.55]{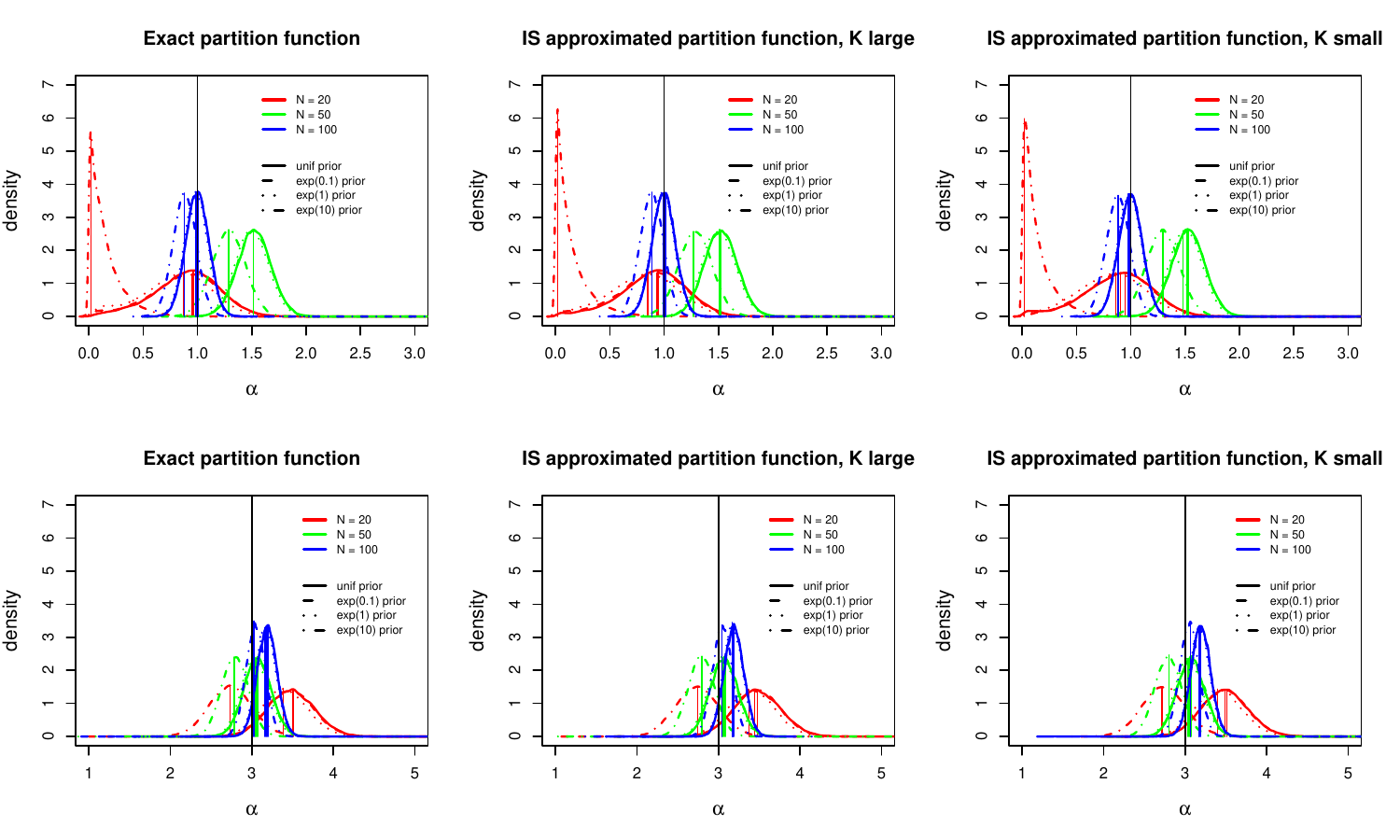}
\caption{Results of the simulations described in Section \ref{subsec:modelTesting}, when $n=20$. In each plot, posterior density of $\alpha$ (the black vertical line indicates $\alpha_{\text{true}}$) obtained for various choices of $N$ (different colors), and for different choices of the prior for $\alpha$ (different line types), as stated in the legend. From left to right, MCMC run with the exact $Z_n(\alpha)$, with the IS approximation $\hat{Z}_n^K(\alpha)$ with $K=10^8$, and with the IS approximation $\hat{Z}_n^K(\alpha)$ with $K=10^4$. First row: $\alpha_{\text{true}}=1;$ second row: $\alpha_{\text{true}}=3.$}
\label{fig:sim1_1_alpha}
\end{figure}

\begin{figure}[h!]
\centering
\includegraphics[scale=.55]{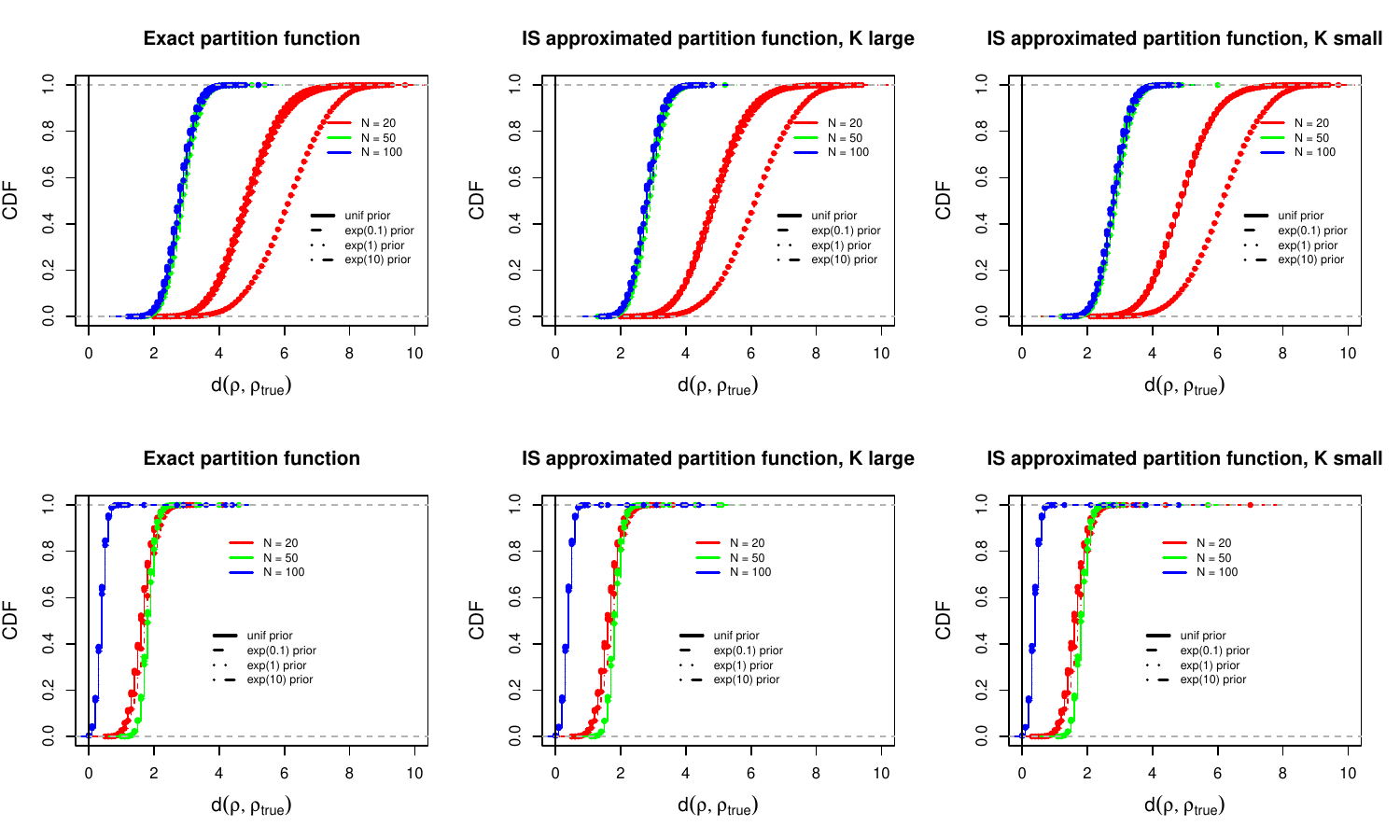}
\caption{Results of the simulations described in Section \ref{subsec:modelTesting}, when $n=20$. In each plot, posterior CDF of $d(\bm{\rho},\bm{\rho_{\text{true}}})$ obtained for various choices of $N$ (different colors), and for different choices of the prior for $\alpha$ (different line types), as stated in the legend. From left to right, MCMC run with the exact $Z_n(\alpha)$, with the IS approximation $\hat{Z}_n^K(\alpha)$ with $K=10^8$, and with the IS approximation $\hat{Z}_n^K(\alpha)$ with $K=10^4$. First row: $\alpha_{\text{true}}=1;$ second row: $\alpha_{\text{true}}=3.$}
\label{fig:sim1_1_rho}
\end{figure}

First, we generated data from a Mallows model with $n = 20$ items, using samples from $N = 20, 50$, and $100$ assessors, a setting of moderate complexity. The value of $\alpha_{\text{true}}$ was chosen to be either 1 or 3, and $\bm{\rho}_{\text{true}}$ was fixed at $(1,\ldots,n)$. To generate the data, we run the MCMC sampler (see Appendix \ref{sec:sampling}) for $10^5$ burn-in iterations, and collected one sample every 100 iterations after that (these settings were kept in all data generations). In the analysis, we considered the performance of the method when using the IS approximation $\hat{Z}^K_n(\alpha)$ with $K = 10^4$ and $10^8$, then comparing the results with those based on the exact $Z_n(\alpha)$. In each case, we run the MCMC for $10^6$ iterations, with $10^5$ iterations for burn-in. Finally, we varied the prior for $\alpha$ to be either the nonintegrable uniform or the exponential using hyperparameter values $\lambda=0.1,1$ and 10. The results are shown in Figures \ref{fig:sim1_1_alpha} for $\alpha$ and \ref{fig:sim1_1_rho} for $\bm{\rho}$. As expected, we can see the precision and the accuracy of the marginal posterior distributions increasing, both for $\alpha$ and $\bm{\rho}$, with $N$ becoming larger. For smaller values of $\alpha_{\text{true}},$ the marginal posterior for $\alpha$ is more dispersed, and $\bm{\rho}$ is stochastically farther from $\bm{\rho}_{\text{true}}.$ These results are remarkably stable against varying choices of the prior for $\alpha$, even when the quite strong exponential prior with $\lambda=10$ was used (with one exception: in the case of $N = 20$ the rather dispersed data generated by $\alpha_{\text{true}}=1$ were not sufficient to overcome the control of the exponential prior with $\lambda=10$, which favored even smaller values of $\alpha$; see Figure \ref{fig:sim1_1_alpha}, top panels). Finally and most importantly, we see that inference on both $\alpha$ and $\bm{\rho}$ is completely unaffected by the approximation of $Z_n(\alpha)$ already when $K = 10^4$. 

\begin{figure}[t]
\centering
\includegraphics[scale=.55]{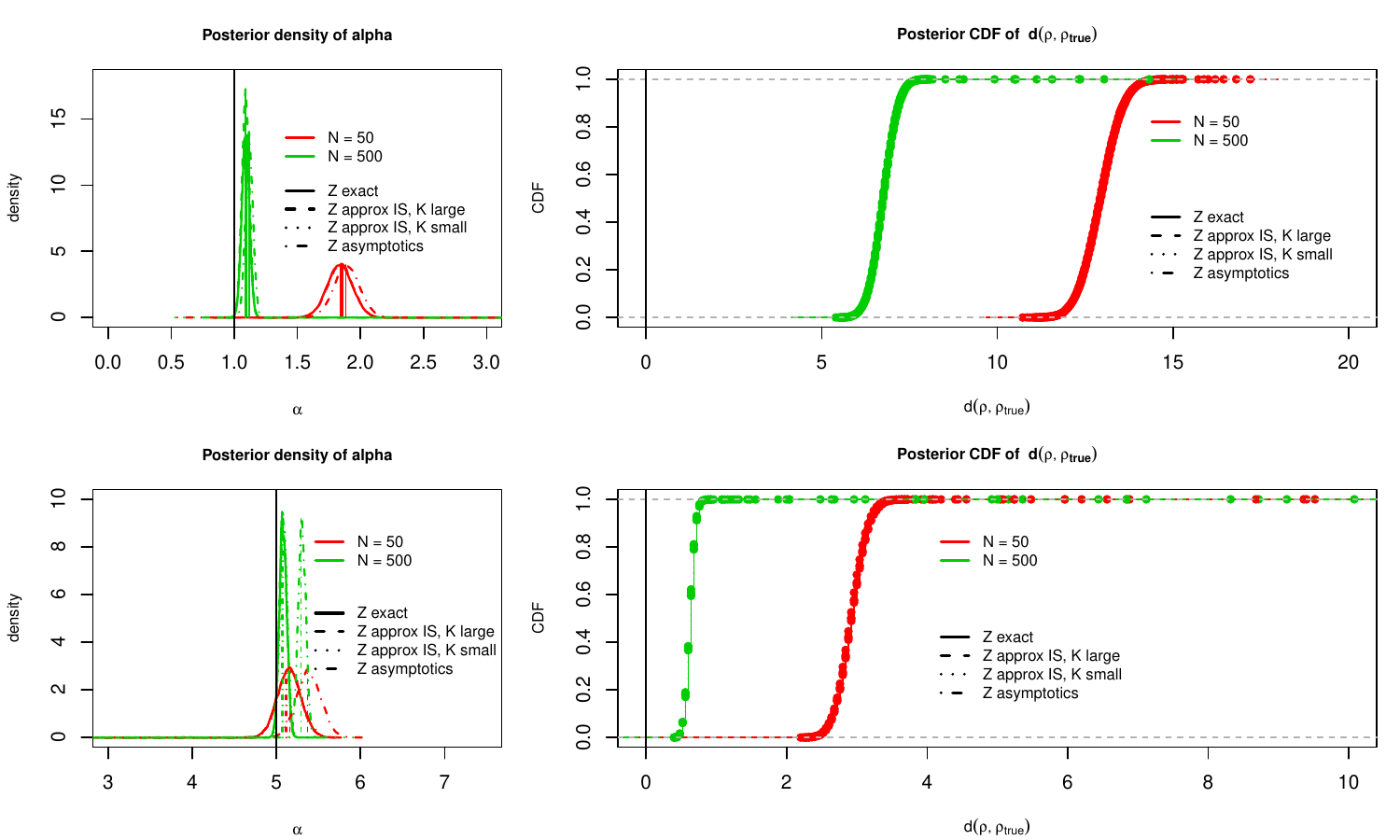}
\caption{Results of the simulations described in Section \ref{subsec:modelTesting}, when $n=50$. Left, posterior density of $\alpha$ (the black vertical line indicates $\alpha_{\text{true}}$) obtained for various choices of $N$ (different colors), and when using the exact, or different approximations to the partition function (different line types), as stated in the legend. Right, posterior CDF of $d(\bm{\rho},\bm{\rho_{\text{true}}})$ in the same settings. First row: $\alpha_{\text{true}}=1;$ second row:  $\alpha_{\text{true}}=5.$}
\label{fig:sim1_2_alphaRho}
\end{figure}

\begin{figure}[t]
\centering
\includegraphics[scale=.55]{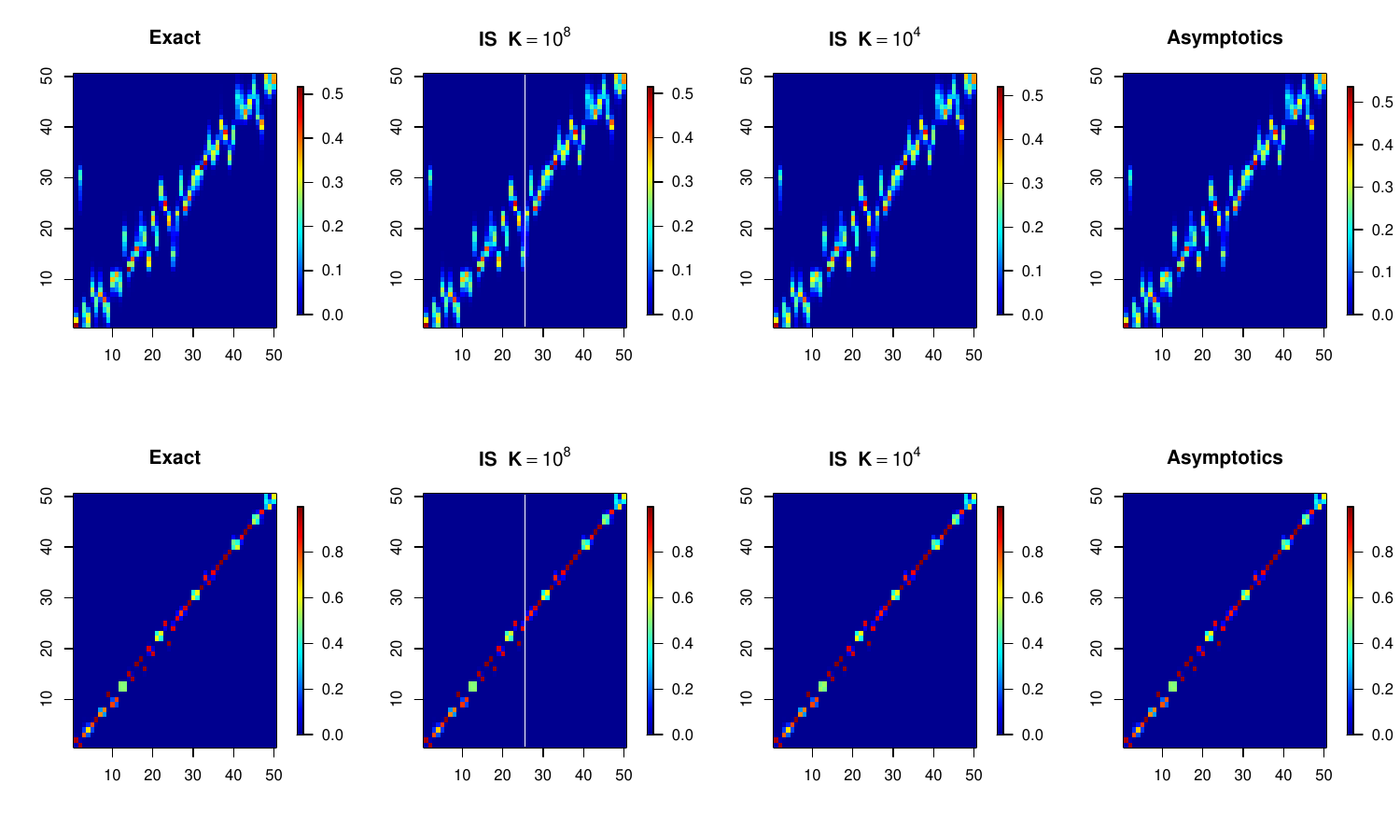}
\caption{Results of the simulations described in Section \ref{subsec:modelTesting}, when $n=50$ and $\alpha_{\text{true}}=5$. In the x-axis items are ordered according to the true consensus $\bm{\rho}_{\text{true}}.$ Each column $j$ represents the posterior marginal density of item $j$ in the consensus $\bm{\rho}.$ Concentration along the diagonal is a sign of success of inference. From left to right, results obtained with the exact $Z_n(\alpha)$, with the IS approximation $\hat{Z}_n^K(\alpha)$ with $K=10^8$, with the IS approximation $\hat{Z}_n^K(\alpha)$ with $K=10^4,$ and with $Z_{\lim}(\alpha).$ First row: $N=50;$ second row: $N=500.$}
\label{fig:sim1_2_marginalRho}
\end{figure}

In a second experiment, we generated data using $n = 50$ items, $N = 50$ or 500 assessors, and scale parameter  $\alpha_{\text{true}}=1$ or 5. This increase in the value of $n$ gave us some basis for comparing the results obtained by using the IS approximation of $Z_n(\alpha)$ with those from the asymptotic approximation $Z_{\lim}(\alpha)$ of \cite{mukherjee2016}, while still retaining also the possibility of using the exact $Z_n(\alpha)$. For the analysis, all the previous MCMC settings were kept, except for the prior for $\alpha$: since results from $n = 20$ turned out to be independent of the choice of the prior, here we used the same exponential prior with $\lambda = 0.1$ in all comparisons (see the discussion in Section \ref{sec:Priors}). The results are shown in Figures \ref{fig:sim1_2_alphaRho} and \ref{fig:sim1_2_marginalRho}. Again, we observe substantially more accurate results for larger values of $N$ and $\alpha_{\text{true}}$. Concerning the impact of approximations to $Z_n(\alpha)$, we notice that, even in this case of larger $n$, the marginal posterior of $\bm{\rho}$ appears completely unaffected by the partition function not being exact (see Figure \ref{fig:sim1_2_alphaRho}, right panels, and Figure \ref{fig:sim1_2_marginalRho}). In the marginal posterior for $\alpha$ (Figure \ref{fig:sim1_2_alphaRho}, left panels), there are no differences between using the IS approximations and the exact, but there is a difference between $Z_{\lim}$ and the other approximations: $Z_{\lim}$ appears to be systematically slightly worse.

Finally, we generated data from the Mallows model with $n = 100$ items, $N = 100$ or 1000 assessors, and using $\alpha_{\text{true}}=5$ or 10. Because of this large value of $n$ we were no longer able to compute the exact $Z_n(\alpha)$, hence we only compared results from the different approximations. We kept the same MCMC settings as for $n = 50$, both in data generation and analysis. The results are shown in Figures \ref{fig:sim1_3_alphaRho} and \ref{fig:sim1_3_marginalRho} of the Supplementary Material, Section 3. Also in this case, we observe substantially more accurate estimates with larger values of $N$ and $\alpha_{\text{true}},$ establishing an overall stable performance of the method. 
Here, using the small number $K = 10^4$ of samples in the IS approximation has virtually no effect on the accuracy of the marginal posterior for $\alpha$, while a small effect can be detected from using the asymptotic approximation (Figure \ref{fig:sim1_3_alphaRho} of the Supplementary Material, left panels). However, again, the marginal posterior for $\bm{\rho}$ appears completely unaffected by the considered approximations in the partition function (Figure \ref{fig:sim1_3_alphaRho}, right panels, and Figure \ref{fig:sim1_3_marginalRho} of the Supplementary Material). 

In conclusion, the main positive result from the perspective of practical applications was the relative lack of sensitivity of the posterior inferences to the specification of the prior for the scale parameter $\alpha$, and the apparent robustness of the marginal posterior inferences on $\bm{\rho}$ on the choice of the approximation of the partition function $Z_n(\alpha)$. The former property was not an actual surprise, as it can be understood to be a consequence of the well-known Bernstein-von Mises principle: with sufficient amounts of data, the likelihood dominates the influence of the prior.

The second observation deserves a somewhat closer inspection, however. The marginal posterior $P(\alpha|\mathbf{R}),$ considered in Figures \ref{fig:sim1_1_alpha} and \ref{fig:sim1_2_alphaRho} (left), and in Figure 3 (left) of the Supplementary Material, is obtained from the joint posterior (\ref{eq:posterior}) by simple summation over $\bm{\rho},$ then getting the expression
\begin{equation}\label{eq:A1}
P(\alpha|\mathbf{R}) \sim_{(\alpha)} \pi(\alpha) C(\alpha; \mathbf{R})/(Z_n(\alpha))^N,
\end{equation}
where $C(\alpha; \mathbf{R}) = \sum_{\bm{\rho}\in\mathcal{P}_n}\exp\left\{-\frac{\alpha}{n}\sum_{j=1}^N d(\mathbf{R}_j, \bm{\rho})\right\}$ is the required normalization. For a proper understanding of the structure of the joint posterior and its modification (\ref{eq:postHat}), it is helpful to first factorize (\ref{eq:posterior}) into the product
\begin{equation}\label{eq:A3}
P(\alpha,\bm{\rho}|\mathbf{R}) = P(\alpha|\mathbf{R}) P(\bm{\rho}|\alpha,\mathbf{R}),
\end{equation}
where then 
\begin{equation}\label{eq:A4}
P(\bm{\rho}|\alpha,\mathbf{R})=[C(\alpha; \mathbf{R})]^{-1}\exp\left\{-\frac{\alpha}{n}\sum_{j=1}^N d(\mathbf{R}_j, \bm{\rho})\right\}.
\end{equation}
The joint posterior (\ref{eq:postHat}), which arises from replacing the partition function $Z_n(\alpha)$ by its approximation $\hat{Z}_n(\alpha)$, can be similarly expressed as the product
\begin{equation}\label{eq:A5}
\hat{P}(\alpha,\bm{\rho}|\mathbf{R})=\hat{P}(\alpha|\mathbf{R}) P(\bm{\rho}|\alpha,\mathbf{R}),
\end{equation}
where
\begin{equation}\label{eq:A6}
\hat{P}(\alpha|\mathbf{R})=[\hat{C}(\mathbf{R})]^{-1}(Z_n(\alpha)/\hat{Z}_n(\alpha))^N P(\alpha|\mathbf{R}).
\end{equation}
This requires that the normalizing factor $\hat{C}(\mathbf{R})$ already introduced in (\ref{eq:postHat}), and here expressed as 
\begin{equation}\label{eq:A7}
\hat{C}(\mathbf{R}) \equiv \int_0^\infty (Z_n(\alpha)/\hat{Z}_n(\alpha))^N P(\alpha|\mathbf{R}) d\alpha,
\end{equation}
is finite. By comparing (\ref{eq:A3}) and (\ref{eq:A5}) we see that, under this condition, the posterior $\hat{P}(\alpha,\bm{\rho}|\mathbf{R})$ arises from $P(\alpha,\bm{\rho}|\mathbf{R})$ by changing the expression (\ref{eq:A1}) of the marginal posterior for $\alpha$ into (\ref{eq:A6}), while the conditional posterior $P(\bm{\rho}|\alpha,\mathbf{R})$ for $\bm{\rho}$, given $\alpha,$ remains the same in both cases. Thus, the marginal posteriors $P(\bm{\rho}|\mathbf{R})$ and $\hat{P}(\bm{\rho}|\mathbf{R})$ for $\bm{\rho}$ arise as mixtures of the same conditional posterior $P(\bm{\rho}|\alpha,\mathbf{R})$ with respect to two different mixing distributions, $P(\alpha|\mathbf{R})$ and $\hat{P}(\alpha|\mathbf{R})$.

It is obvious from (\ref{eq:A6}) and (\ref{eq:A7}) that $\hat{P}(\alpha|\mathbf{R})=P(\alpha|\mathbf{R})$ would hold if the ratio $Z_n(\alpha)/\hat{Z}_n(\alpha)$ would be exactly a constant in $\alpha$, and this would also entail the exact equality $\hat{P}(\bm{\rho}|\mathbf{R}) = P(\bm{\rho}|\mathbf{R})$. It was established in (\ref{eq:Zlimit}) that, in the IS scheme, $Z_n(\alpha)/\hat{Z}_n(\alpha)\rightarrow1$ as $K\rightarrow\infty$. Thus, for large enough $K$, $(Z_n(\alpha)/\hat{Z}_n(\alpha))^N\approx1$ holds as an approximation (see Proposition \ref{prop:Zlimit}). Importantly, however, (\ref{eq:A6}) shows that the approximation is only required to hold well on the effective support of $P(\alpha|\mathbf{R})$, and this support is narrow when $N$ is large. This is demonstrated clearly in Figures \ref{fig:sim1_1_alpha} and \ref{fig:sim1_2_alphaRho} (left), and in Figure 3 (left) of the Supplementary Material. On this support, because of uniform continuity in $\alpha$, also the integrand $P(\bm{\rho}|\alpha,\mathbf{R})$ in (\ref{eq:A4}) remains nearly a constant. In fact, experiments (results not shown) performed by varying $\alpha$ over a much wider range of fixed values, while keeping the same $\mathbf{R}$, gave remarkably stable results for the conditional posterior $P(\bm{\rho}|\alpha,\mathbf{R})$. This contributes to the high degree of robustness in the posterior inferences on $\bm{\rho}$, making requirements of using large values of $K$ much less stringent.  

In Figures \ref{fig:sim1_1_rho} and \ref{fig:sim1_2_alphaRho} (right), and in Figure 3 (right) of the Supplementary Material, we considered and compared the marginal posterior CDF's of the distance $d(\bm{\rho},\bm{\rho}_{\text{true}})$ under the schemes $P(\cdot|\mathbf{R})$ and $\hat{P}(\cdot|\mathbf{R})$. Using the shorthand $d^*= d(\bm{\rho},\bm{\rho}_{\text{true}})$, let
\begin{eqnarray}\label{eq:A8}
F_{d^*}(x|\alpha,\mathbf{R}) & \equiv & P( d(\bm{\rho},\bm{\rho}_{\text{true}}) \leq x|\alpha,\mathbf{R}) = \sum_{\{\bm{\rho}:d(\bm{\rho},\bm{\rho}_{\text{true}}) \leq x\}} P(\bm{\rho}|\alpha,\mathbf{R}),\\
F_{d^*}(x|\mathbf{R}) & \equiv & \sum_{\{\bm{\rho}:d(\bm{\rho},\bm{\rho}_{\text{true}}) \leq x\}} P(\bm{\rho}|\mathbf{R}) = \int F_{d^*}(x|\alpha,\mathbf{R}) P(\alpha|\mathbf{R}) d\alpha, \nonumber\\
\hat{F}_{d^*}(x|\mathbf{R}) & \equiv & \sum_{\{\bm{\rho}:d(\bm{\rho},\bm{\rho}_{\text{true}}) \leq x\}} \hat{P}(\bm{\rho}|\mathbf{R}) = \int F_{d^*}(x|\alpha,\mathbf{R}) \hat{P}(\alpha|\mathbf{R}) d\alpha.  \nonumber
\end{eqnarray}
For example, in Figure \ref{fig:sim1_1_rho} we display, for different priors, the CDF’s $F_{d^*}(x|\mathbf{R})$ on the left, and $\hat{F}_{d^*}(x|\mathbf{R})$ in the middle and on the right, corresponding to two different IS approximations of the partition function. Like the marginal posteriors $P(\bm{\rho}|\mathbf{R})$ and $\hat{P}(\bm{\rho}|\mathbf{R})$ above, $F_{d^*}(x|\mathbf{R})$ and $\hat{F}_{d^*}(x|\mathbf{R})$ can be thought of as mixtures of the same function, here $F_{d^*}(x|\alpha,\mathbf{R})$, but with respect to two different mixing distributions, $P(\alpha|\mathbf{R})$ and $\hat{P}(\alpha|\mathbf{R})$. The same arguments, which were used above in support of the robustness of the posterior inferences on $\bm{\rho}$, apply here as well. Extensive empirical evidence for their justification is provided in Figures \ref{fig:sim1_1_rho} and \ref{fig:sim1_2_alphaRho} (right), and in Figure 3 (right) of the Supplementary Material. Finally note that these arguments also strengthen considerably our earlier conclusion of the lack of sensitivity of the posterior inferences on $\bm{\rho}$ to the specification of the prior for $\alpha$. For this, we only need to consider alternative priors, say, $\pi(\alpha)$ and $\hat{\pi}(\alpha)$, in place of the mixing distributions $P(\alpha|\mathbf{R})$ and $\hat{P}(\alpha|\mathbf{R})$.

\section{Extensions to Partial Rankings and Heterogeneous Assessor Pool}\label{sec:extensions}

We now relax two assumptions of the previous Sections, namely that each assessor ranks all $n$ items and that the assessors are exchangeable, all sharing a common consensus ranking. This allows us to treat the important situation of pairwise comparisons, and of multiple classes of assessors, as incomplete data cases, within the same Bayesian Mallows framework. 

\subsection{Ranking of the Top Ranked Items}\label{sec:PartialRanks}
Often only a subset of the items is ranked: ranks can be missing at random, the assessors may only have ranked the, in-their-opinion, top-$k$ items, or can be presented with a subset of items that they have to rank. 
These situations can be handled conveniently in our Bayesian framework, by applying data augmentation techniques. We start by explaining the method in the case of the top-$k$ ranks, and then show briefly how it can be generalized to the other cases mentioned. 

Suppose that each assessor $j$ has ranked the subset of items $\mathcal{A}_{j} \subseteq \{A_{1},A_{2},\dots,A_{n}\}$, giving them top ranks from $1$ to $n_j=|\mathcal{A}_{j}|$. Let $R_{ij} = \mathbf{X}_{j}^{-1}(A_{i})$ if $A_{i} \in \mathcal{A}_{j}$, while for $A_i \in \mathcal{A}^c_{j}$, $R_{ij}$ is unknown, except for the constraint $R_{ij} > n_j$, $j=1,\dots,N,$ and follows a symmetric prior on the permutations of $(n_j+1,\ldots,n)$. We define augmented data vectors $\tilde{\mathbf{R}}_{1},\dots,\tilde{\mathbf{R}}_{N}$ by assigning ranks to these non-ranked items randomly, using an MCMC algorithm, and do this in a way which is compatible with the rest of the data. Let $\mathcal{S}_{j} = \{ \tilde{\mathbf{R}}_{j} \in \mathcal{P}_{n} : \tilde{R}_{ij} = \mathbf{X}_{j}^{-1}(A_{i} ) \text{ if } A_{i} \in \mathcal{A}_{j} \}, ~ j=1,\dots,N$, be the set of possible augmented random vectors, that is the original partially ranked items together with the allowable ``fill-ins'' of the missing ranks. Our goal is to sample from the posterior distribution
\begin{align*}
P\left(\alpha, \bm{\rho}| \mathbf{R}_{1},\dots, \mathbf{R}_{N}\right) = \sum_{\tilde{\mathbf{R}}_{1} \in \mathcal{S}_{1}} \dots \sum_{\tilde{\mathbf{R}}_{N} \in \mathcal{S}_{N}} P\left( \alpha, \bm{\rho} , \tilde{\mathbf{R}}_{1},\dots, \tilde{\mathbf{R}}_{N}|\mathbf{R}_{1},\dots, \mathbf{R}_{N}\right).
\end{align*}
Our MCMC algorithm alternates between sampling the augmented ranks given the current values of $\alpha$ and $\bm{\rho},$ and sampling $\alpha$ and $\bm{\rho}$ given the current values of the augmented ranks. For the latter, we sample from the posterior $P( \alpha, \bm{\rho} | \tilde{\mathbf{R}}_{1},\dots, \tilde{\mathbf{R}}_{N})$ as in Section \ref{sec:MHAlgorithm}. For the former, fixing $\alpha$ and $\bm{\rho}$ and the observed ranks $\mathbf{R}_{1},\dots, \mathbf{R}_{N}$, we see that $\tilde{\mathbf{R}}_{1},\dots, \tilde{\mathbf{R}}_{N}$ are conditionally independent, and moreover, that each $\tilde{\mathbf{R}}_{j}$ only depends on the corresponding ${\mathbf{R}}_{j}$. This enables us to consider the sampling of new augmented vectors $\tilde{\mathbf{R}}^\prime_{j}$ separately for each $j, j =1,\dots,N$. Specifically, given the current $\tilde{\mathbf{R}}_{j}$ (which embeds information contained in ${\mathbf{R}}_{j}$) and the current values for $\alpha$ and $\bm{\rho}$, $\tilde{\mathbf{R}}^\prime_{j}$ is sampled in $\mathcal{S}_j$ from a uniform proposal distribution, meaning that the highest ranks from $1$ to $n_j$ have been reserved for the items in $\mathcal{A}_{j}$, while compatible ranks are randomly drawn for items in $\mathcal{A}_j^c$. The proposed $\tilde{\mathbf{R}}^\prime_{j}$ is then accepted with probability
\begin{equation}\label{eq:MHratioAug}
\min\left\{1,\exp\left[-\frac{\alpha}{n}\left(d(\tilde{\mathbf{R}}^\prime_{j},\rho)-d(\tilde{\mathbf{R}}_{j},\rho)\right)\right]\right\}.
\end{equation}
The MCMC algorithm described above and used in the case of partial rankings is given in Algorithm \ref{algo:partialMH} of Appendix \ref{sec:algoSupp}. 
Our algorithm can also handle situations of generic partial ranking, where each assessor is asked to provide the mutual ranking of some subset $\mathcal{A}_j\subset\{A_1,...,A_n\}$ consisting of $n_j\leq n$  items, not necessarily the top-$n_j$. In this case, we can only say that in $\tilde{\mathbf{R}}_j=(\tilde{{R}}_{1j},...,\tilde{{R}}_{nj})$ the order between items
$A_i\in  \mathcal{A}_j$ must be preserved as in  ${\mathbf{R}}_j$, whereas the ranks of the augmented ``fill-ins'' $A_i\in  \mathcal{A}_j^c$ are left open. More exactly, the latent rank vector $\tilde{\mathbf{R}}_j$ takes values in the set $\mathcal{S}_j=\{\tilde{\mathbf{R}}_j\in\mathcal{P}_n:\text{if }R_{i_1j}<R_{i_2j},\text{ with } A_{i_1},A_{i_2}\in \mathcal{A}_j\Rightarrow \tilde{R}_{i_1j}<\tilde{R}_{i_2j}\}$. The MCMC is then easily adjusted so that the sampling of each   $\tilde{\mathbf{R}}_j$ is restricted to the corresponding   $\mathcal{S}_j$, thus respecting the mutual rank orderings in the data.

\subsubsection{Effects of Unranked Items on Consensus Ranking}\label{sec:Preselection}

In applications in which the number of items is large there are often items which none of the assessors included in their top-list. What is the exact role of such ``left-over'' items in the top-$k$ consensus ranking of all items? Can we ignore such ``left-over'' items and consider only the items explicitly ranked by at least one assessor? In the following we first show that only items explicitly ranked by the assessors appear in top positions of the consensus ranking. We then show that, when considering the MAP consensus ranking, excluding the left-over items from the ranking procedure already at the start has no effect on how the remaining ones will appear in such consensus ranking.  

For a precise statement of these results, we need some new notation. Suppose that assessor $j$ has ranked a subset $\mathcal{A}_j$ of $n_j$ items. Let $\mathcal{A} = \bigcup_{j=1,\ldots,N} \mathcal{A}_j,$ and denote $n = |\mathcal{A}|.$ Let $n^*$ be the total number of items, including left-over items which have not been explicitly ranked by any assessor. Denote by $\mathcal{A}^*=\{A_i; i=1,\ldots,n^*\}$ the collection of all items, and by $\mathcal{A}^c=\mathcal{A}^*\setminus\mathcal{A}$ the left-over items. Each rank vector $\mathbf{R}_j$ for assessor $j$ contains, in some order, the ranks from 1 to $n_j$ given to items in $\mathcal{A}_j$. In the original data the ranks of all remaining items are left unspecified, apart from the fact that implicitly, for assessor $j,$ they would have values which are at least as large as $n_j+1.$ 

The results below are formulated in terms of the two different modes of analysis, which we need to compare and which correspond to different numbers of items being included. The first alternative is to include in the analysis the complete set $\mathcal{A}^*$ of $n^*$ items, and to complement each data vector $\mathbf{R}_j$ by assigning (originally missing) ranks to all items which are not included in $\mathcal{A}_j$; their ranks will then form some permutation of the sequence $(n_j+1,\ldots,n^*)$. We call this mode of analysis \emph{full analysis}, and denote the corresponding probability measure by $P_{n^*}$. The second alternative is to include in the analysis only the items which have been explicitly ranked by at least one assessor, that is, items belonging to the set $\mathcal{A}.$ We call this second mode \emph{restricted analysis}, and denote the corresponding probability measure by $P_n$. The probability measure $P_n$ is specified as before, including the uniform prior on the consensus ranking $\bm{\rho}$ across all $n!$ permutations of $(1,2,\ldots,n),$ and the uniform prior of the unspecified ranks $R_{ij}$ of items $A_i\in\mathcal{A}_j^c$ across the permutations of $(n_j+1,\ldots,n).$ The definition of $P_{n^*}$ is similar, except that then the uniform prior distributions are assumed to hold in the complete set $\mathcal{A}^*$ of items, that is, over permutations of $(1,2,\ldots,n^*)$ and $(n_j+1,\ldots,n^*),$ respectively. In the posterior inference carried out in both modes of analysis, the augmented ranks, which were not recorded in the original data, are treated as random variables, with values being updated as part of the MCMC sampling.

\begin{myprop}\label{prop:preselection}
Consider two latent consensus rank vectors $ \bm{\rho}$ and $\bm{\rho}^\prime$ such that
\begin{itemize}
\item[(i)] in the ranking $\bm{\rho}$ all items in $\mathcal{A}$ have been included among the top-$n$-ranked, while those in $\mathcal{A}^c$ have been assigned ranks between $n+1$ and $n^*$,
\item[(ii)] $\bm{\rho}^\prime$ is obtained from $\bm{\rho}$ by a permutation, where the rank in $\bm{\rho}$ of at least one item belonging to $\mathcal{A}$ has been transposed with the rank of an item in $\mathcal{A}^c$.
\end{itemize}
Then, $P_{n^*}(\bm{\rho}|\text{data}) \geq P_{n^*}(\bm{\rho}^\prime|\text{data}),$ for the footrule, Kendall and Spearman distances in the \emph{full analysis} mode.
\end{myprop}

\noindent {\bf Remark.} The above proposition says, in essence, that any consensus lists of top-$n$ ranked items, which contains one or more items with their ranks completely missing in the data (that is, the item was not explicitly ranked by any of the assessors), can be improved \emph{locally}, in the sense of increasing the associated posterior probability with respect to $P_{n^*}$. This happens by trading such an item in the top-$n$ list against another, which had been ranked but which had not yet been selected to the list. In particular, the MAP estimate(s) for consensus ranking assign $n$ highest ranks to explicitly ranked items in the data (which corresponds to the result in \citet{MeilaBao2010} for Kendall distance). The following statement is an immediate implication of Proposition \ref{prop:preselection}, following from a marginalization with respect to $P_{n^*}$. 
\begin{mycor}\label{cor:preselectionTOP}
Consider, for $k\leq n$, collections $\{A_{i_1},A_{i_2},\ldots, A_{i_k}\}$ of $k$ items and the corresponding ranks $\{\rho_{i_1},\rho_{i_2},\ldots, \rho_{i_k}\}$. In \emph{full analysis} mode, the maximal posterior probability $P_{n^*}(\left\{\rho_{i_1},\rho_{i_2},\ldots,\right.$ $ \left.\rho_{i_k}\right\}=\{1,2,\ldots,k\}|\mathit{data})$, is attained when $\{A_{i_1},A_{i_2},\ldots, A_{i_k}\} \subset \mathcal{A}.$
\end{mycor}
Another consequence of Proposition \ref{prop:preselection} is the coincidence of the MAP estimates under the two probability measures $P_n$ and $P_{n^*}$.
\begin{mycor}\label{cor:preselectionMAP}
Denote by $\bm{\rho}^{MAP*}$ the MAP estimate for consensus ranking obtained in a \emph{full analysis,} $\bm{\rho}^{MAP*} := \argmax_{\bm{\rho} \in \mathcal{P}_{n^*}} P_{n^*}(\bm{\rho}|\text{data}),$ and by $\bm{\rho}^{MAP}$ the MAP estimate for consensus ranking obtained in a \emph{restricted analysis,} $\bm{\rho}^{MAP} := \argmax_{\bm{\rho} \in \mathcal{P}_n} P_n(\bm{\rho}|\text{data}).$ Then, $\bm{\rho}^{MAP*}\vert_{i:A_i\in\mathcal{A}}\equiv\bm{\rho}^{MAP}.$
\end{mycor}

\noindent {\bf Remark.} The above result is very useful in the context of applications, since it guarantees that the top-$n$ items in the MAP consensus ranking do not depend on which version of the analysis is performed. Recall that a full analysis cannot always be carried out in practice, due to the fact that left-over items might be unknown, or their number might be too large for any realistic computation.

\subsection{Pairwise Comparisons}\label{sec:Pairwise}

In many situations, assessors compare pairs of items rather than ranking all or a subset of items. We extend our Bayesian data augmentation scheme to handle such data. Our approach is an alternative to \citet{Lu2015}, who treated preferences by applying their Repeated Insertions Model (RIM). Our approach is simpler, it is fully integrated into our Bayesian inferential framework, and it works for any right-invariant distance. 

As an example of paired comparisons, assume assessor $j$ stated the preferences $\mathcal{B}_{j} = \{ A_{1} \prec A_{2}, A_{2} \prec A_{5},A_{4} \prec A_{5}\}$. Here $A_{r} \prec A_{s}$ means that $A_{s}$ is preferred to $A_{r}$, so that $A_{s}$ has a lower rank than $A_{r}$. Let $\mathcal{A}_{j}$ be the set of items constrained by assessor $j$, in this case $\mathcal{A}_{j} = \{A_{1},A_{2},A_{4},A_{5}\}$. Differently from Section \ref{sec:PartialRanks}, the items which have been considered by each assessor are now not necessarily fixed to a given rank. Hence, in the MCMC algorithm, we need to propose augmented ranks which obey the partial ordering constraints given by each assessor, to avoid a large number of rejections, with the difficulty that none of the items is now fixed to a given rank. Note that we can also handle the case when assessors give ties as a result of some pairwise comparisons: in such a situation, each pair of items resulting in a tie is randomized to a preference at each data augmentation step inside the MCMC, thus correctly representing the uncertainty of the preference between the two items. None of the experiments included in the paper involves ties, thus this randomization is not needed.

We assume that the pairwise orderings in $\mathcal{B}_{j}$ are mutually compatible, and define by $\text{tc}(\mathcal{B}_{j})$ the transitive closure of $\mathcal{B}_{j}$, containing all pairwise orderings of the elements in $\mathcal{A}_{j}$ induced by $\mathcal{B}_{j}$. In the example, $\text{tc}(\mathcal{B}_{j}) = \mathcal{B}_{j} \cup \{A_{1}\prec A_{5}\}$. For the case of ordered subsets of items, the transitive closure is simply the single set of pairwise preferences compatible with the ordering, for example, $\{A_{1} \prec A_{2} \prec A_{5}\}$ yields $\text{tc}(\mathcal{B}_{j}) = \{A_{1} \prec A_{2}, A_{2} \prec A_{5}, A_{1} \prec A_{5}\}$. The R packages \verb!sets! \citep{Meyer2009} and \verb!relations! \citep{Meyer2014} efficiently compute the transitive closure.

The main idea of our method for handling such data remains the same as in Section \ref{sec:PartialRanks}, and the algorithm is the same as Algorithm \ref{algo:partialMH}. However, here a ``modified'' leap-and-shift proposal distribution, rather than a uniform one, is used to sample augmented ranks which are compatible with the partial ordering constraint. Suppose that, from the latest step of the MCMC, we have a full augmented rank vector $\tilde{\mathbf{R}}_{j}$ for assessor $j$, which is compatible with $\text{tc}(\mathcal{B}_{j})$. Draw a random number $u$ uniformly from $\{1,\dots,n\}$. If $A_{u}\in \mathcal{A}_{j}$, let $l_{j} = \text{max}\{\tilde{R}_{kj} : A_k \in \mathcal{A}_{j}, k\neq u, (A_{k} \succ A_{u}  ) \in \text{tc}(\mathcal{B}_{j})\}$, with the convention that $l_{j}=0$ if the set is empty, and $r_{j} = \text{min}\{\tilde{R}_{kj}: A_k \in \mathcal{A}_{j}, k\neq u, (A_{k} \prec A_{u}  ) \in \text{tc}(\mathcal{B}_{j}) \}$, with the convention that $r_{j} =n+1$ if the set is empty. Now complete the leap step by drawing a new proposal $\tilde{R}_{uj}^{\prime}$ uniformly from the set $\{l_{j}+1,\dots,r_{j}-1\}$. Otherwise, if $A_{u} \in \mathcal{A}^c_{j}$, we complete the leap step by drawing $\tilde{R}_{uj}^{\prime}$ uniformly from $\{1,\dots,n\}$. The shift step remains unchanged. Note that this modified leap-and-shift is symmetric.

\subsection{Clustering Assessors Based on their Rankings of All Items}\label{sec:Clustering}

So far we have assumed that there exists a unique consensus ranking shared by all assessors. In many cases the assumption of homogeneity is unrealistic: the possibility of dividing assessors into more homogeneous subsets, each sharing a consensus ranking of the items, brings the model closer to reality. We then introduce a mixture of Mallows models, able to handle heterogeneity. We here assume that the data consist of complete rankings.

Let $z_1,\ldots,z_N \in \{1,\ldots,C\}$ assign each assessor to one of $C$ clusters. The assessments within each cluster $c \in \{1,\ldots,C\}$ are described by a Mallows model with parameters $\alpha_c$ and $\bm{\rho}_c$, the cluster consensus. Assuming conditional independence across the clusters, the augmented data formulation of the likelihood for the observed rankings $\mathbf{R}_1,\ldots,\mathbf{R}_N$ is given by
\begin{align*} 
P\left(\mathbf{R}_{1},\dots,\mathbf{R}_{N} | \left\{\bm{\rho}_{c},\alpha_{c}\right\}_{c=1,...,C}, z_1,\ldots,z_N \right)  =  \prod_{j=1}^N \frac{1_{\mathcal{P}_n}(\mathbf{R}_{j})}{Z_{n}(\alpha_{z_j})}\exp\left\{-\frac{\alpha_{z_j}}{n} d(\mathbf{R}_{j},\bm{\rho}_{z_j}) \right\}.
\end{align*}
For the scale parameters, we assume the prior $\pi( \alpha_1,\ldots,\alpha_C) \penalty 0 \propto \lambda^C \penalty 0 \exp(-\lambda\sum_{c=1}^C \alpha_c)$. We further assume that the cluster labels are a priori distributed according to $P(z_1,\ldots,z_N| \tau_1,\ldots,\tau_C) = \prod_{j=1}^N \tau_{z_j}$, where $\tau_c$ is the probability that an assessor belongs to the $c$-th subpopulation; $\tau_c\geq0, \ c=1,\ldots,C$ and $\sum_{c=1}^C \tau_c = 1$. Finally $\tau_1,\ldots,\tau_C$ are assigned the standard symmetric Dirichlet prior $\pi( \tau_1,\ldots,\tau_C) = \Gamma(\psi C)\Gamma(\psi)^{-C}\prod_{c=1}^C \tau_c^{\psi - 1}$, using the gamma function $\Gamma(\cdot)$.

The number of clusters $C$ is often not known, and the selection of $C$ can be based on different criteria. Here we inspect the posterior distribution of the within-cluster sum of distances of the observed ranks from the corresponding cluster consensus (see Section \ref{sec:SushiData}  for more details).
This approach is a Bayesian version of the more classical within-cluster sum-of-squares criterion for model selection, and we expect to observe an elbow in the within-cluster distance posterior distribution as a function of $C$, identifying the optimal number of clusters.

Label switching is not explicitly handled inside our MCMC, to ensure full convergence of the chain  \citep{Jasra2005, Celeux2000}. MCMC iterations are re-ordered after convergence is achieved, as in \citet{Papastamoulis2015}. The MCMC algorithm alternates between sampling $\bm{\rho}_1,\ldots,\bm{\rho}_C$ and $\alpha_1,\ldots,\alpha_C$ in a Metropolis-Hastings step, and $\tau_1,\ldots,\tau_C$ and $z_1,\ldots,z_N$ in a Gibbs sampler step. The former is straightforward, since $\left(\bm{\rho}_c,\alpha_c\right)_{c=1,\ldots,C}$ are conditionally independent given $z_1,\ldots,z_N$. 
In the latter, we exploit the fact that the Dirichlet prior for $\tau_{1},\dots,\tau_{C}$ is conjugate to the multinomial conditional prior for $z_{1},\dots,z_{N}$ given $\tau_{1},\dots,\tau_{C}$. Therefore in the Gibbs step for $\tau_{1},\ldots,\tau_C$, we sample from $\mathcal{D}(\psi+n_1,\ldots,\psi+n_C)$, where $\mathcal{D}(\cdot)$ denotes the Dirichlet distribution and $n_c=\sum_{j=1}^N 1_{c}(z_j),$ $c=1,\ldots,C$.
Finally, in the Gibbs step for $z_j$, $j=1,\dots,N$, we sample from $P(z_j = c |\tau_c,\bm{\rho}_c,\alpha_c,R_j) \propto \tau_c P(\mathbf{R}_j|\bm{\rho}_c,\alpha_c)= \tau_c Z_{n}(\alpha_c)^{-1}  \exp\{-(\alpha_c/n) d(\mathbf{R}_{j},\bm{\rho}_c) \}$. 
The pseudo-code of the clustering algorithm is sketched in Algorithm \ref{algo:clusteringMH} of Appendix \ref{sec:algoSupp}.

It is not difficult to treat situations where data are incomplete (in any way described before) and the assessors must be divided into separate clusters. Algorithms \ref{algo:clusteringMH} and \ref{algo:partialMH} are merged in an obvious way, by iterating between augmentation, clustering, and $\alpha$ and $\bm{\rho}$ updates. The MCMC algorithm for clustering based on partial rankings or pairwise preferences is sketched in Algorithm \ref{algo:clusteringPair} of Appendix \ref{sec:algoSupp}.

\subsection{Example: Preference Prediction}\label{sec:Prediction}

Consider a situation in which the assessors have expressed their preferences on a collection of items, by performing only partial rankings. Or, suppose that they have been asked to respond to some queries containing different sets of pairwise comparisons.  One may then ask how the assessors would have ranked some subset of items of interest when such ranking could not be concluded directly from the data they provided. Sometimes the interest is to predict the assessors' top preferences, accounting for the possibility that such top lists could contain items which some assessors had not seen. Problems of this type are commonly referred to as \emph{personalized ranking}, or {\em preference learning} \citep{PrefLearn}, being a step towards \emph{personalized recommendation}. There is a large and rapidly expanding literature describing a diversity of methods in this area. 

Our framework, based on the Bayesian Mallows model, and its estimation algorithms as described in the previous Sections, form a principled
approach for handling such problems. Assuming a certain degree of similarity in the individual preferences, and with different assessors providing partly complementary information, it is natural to try to borrow strength from such partial preference information from different assessors for forming a consensus. Expanding the model to include clusters allows handling heterogeneity that may be present in the assessment data \citep{Francis2010}. The Bayesian estimation procedure provides then the joint posterior distribution, expressed numerically in terms of the MCMC output consisting of sampled values of all cluster membership indicators, $z_j$, and of complete individual rankings, $\tilde{\mathbf{R}}_j$. 
For example, if assessor $j$ did not compare $A_{1}$ to $A_{2}$, we might be interested in computing $P(A_{1} \prec_{j} A_{2}| \text{data})$, the predictive probability that this assessor would have preferred item $A_{2}$ to item $A_{1}$.
This probability is then readily obtained from the MCMC output, as a marginal of the posterior $P(\tilde{\mathbf{R}}_{j}|\text{data})$.  

To illustrate how this is possible with our approach, we present a small simulated experiment, corresponding to a heterogeneous collection of assessors expressing some of their pairwise preferences, and then want to predict the full individual ranking $\tilde{\mathbf{R}}_j$ of all items, for all $j$. For this, we generated pairwise preference data from a mixture of Mallows models with footrule distance, using the procedure explained in Appendix \ref{sec:sampling}. 
We generated the data with $N=200$, $n=15$, $C=3$, $\alpha_1,...,\alpha_C=4$, $\psi_1,...,\psi_C=50$,  obtaining the true $\tilde{\mathbf{R}}_{j,\text{true}}$ for every assessor. Then, we assigned to each assessor $j$ a different number, $T_j\sim \text{TruncPoiss}(\lambda_T, T_\text{max})$, of pair comparisons, sampled from a truncated Poisson distribution with $\lambda_T=20$, denoting by $T_\text{max}=n(n-1)/2$ the total number of possible pairs from $n$ items. Each pair comparison was then ordered according to the true $\tilde{\mathbf{R}}_{j,\text{true}}$. The average number of pairs per assessor was around 20, less than 20\% of $T_\text{max}$. 

\begin{figure}[h!] 
\centering
\includegraphics[width=0.4\textwidth]{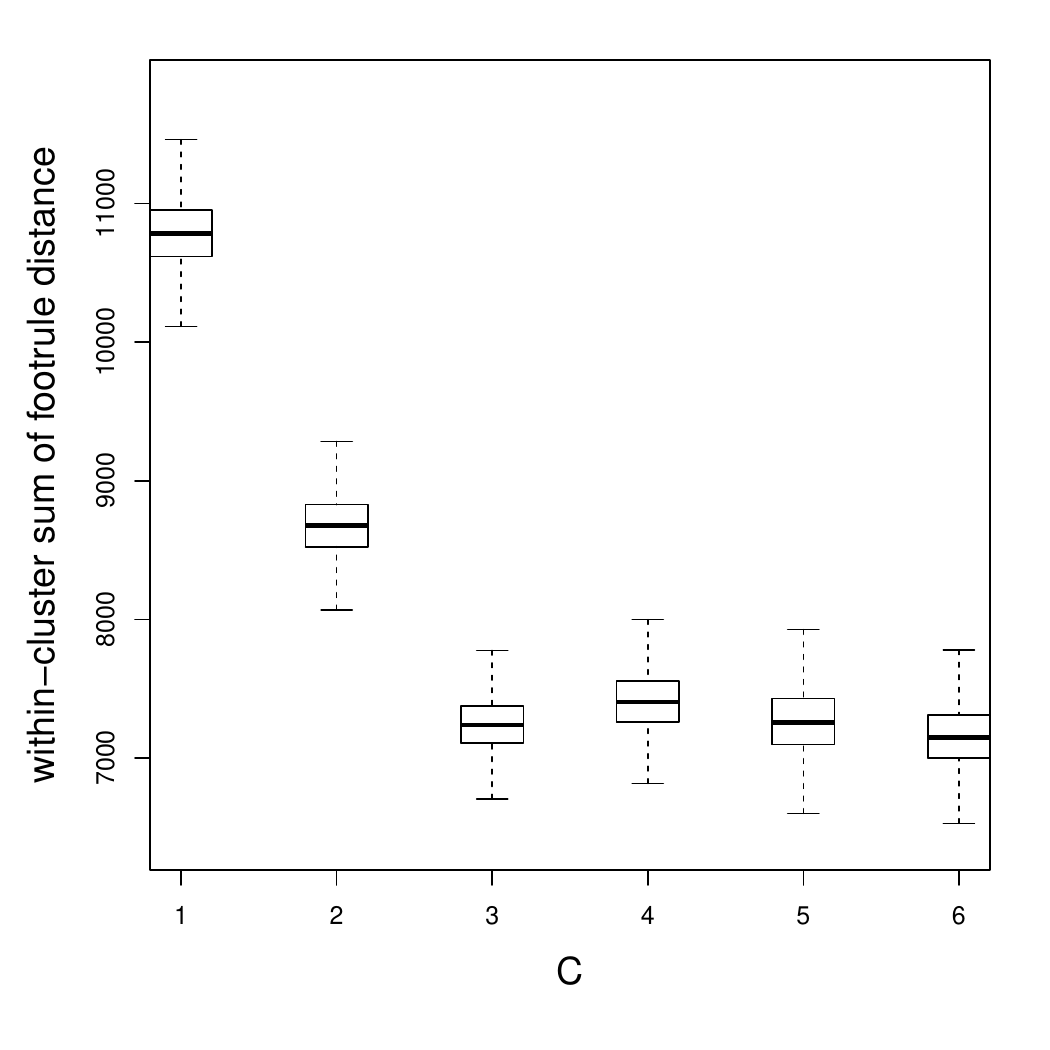}
\hspace{1cm}
\includegraphics[width=0.4\textwidth]{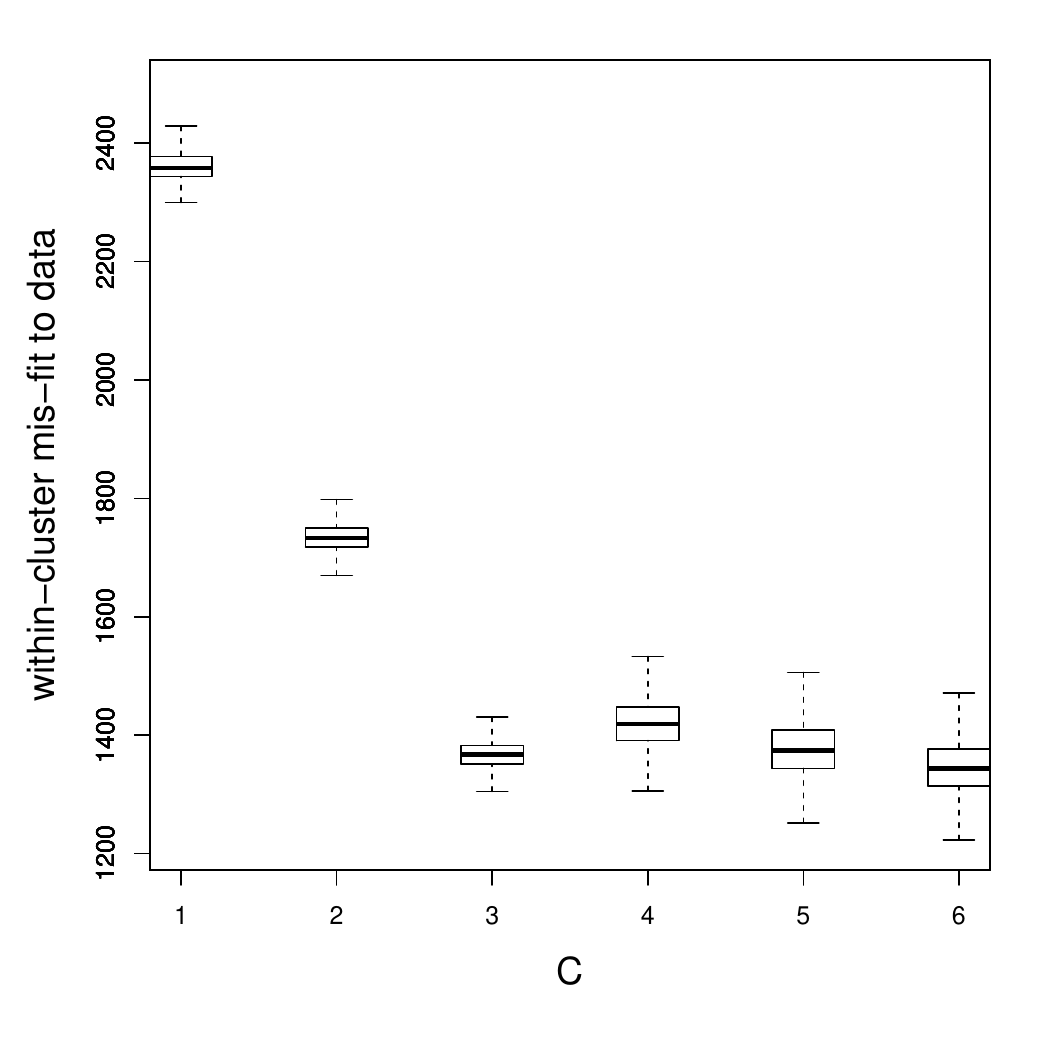}
\caption{Results of the simulation in Section \ref{sec:Prediction}. Boxplots of the posterior distribution of the within-cluster sum of footrule distances  (left), and of the within-cluster indicator of mis-fit to the data (right), for different choices of $C$.}\label{predpairs_wcsfd} \end{figure}

In the analysis, we run Algorithm \ref{algo:clusteringPair} of Appendix \ref{sec:algoSupp} on these data, using the exact partition function, for $10^5$ iterations (of which $10^4$ were for burn-in). Separate analyses were performed for $C \in \{1,\dots,6\}$. Then, in order to inspect if our  method correctly identified the true number of clusters we computed two quantities: the within-cluster sum of footrule distances, given by $\sum_{c=1}^C \sum_{j:z_j = c}d(\tilde{\mathbf{R}}_j,\bm{\rho}_{c})$, and a within-cluster indicator of mis-fit to the data, $\sum_{c=1}^C\sum_{j:z_j = c} |\{B\in\text{tc}(\mathcal{B}_{j}):B\text{ is not consistent with }\bm{\rho}_{c}\}|$, where a pair comparison $B\in\text{tc}(\mathcal{B}_{j}),B=(A_{r} \prec A_{s})$ is not consistent with $\bm{\rho}_{c}$ if $\rho_{c,s}>\rho_{c,r}$. The number of such non-consistent pairs in $\mathcal{B}_j$ gives an indication of the mis-fit of the $j$-th assessor to its cluster. Notice that, while the latter measure takes into account the data directly, the former is based on the augmented ranks $\tilde{\mathbf{R}}_j$ only. Hence, the within-cluster sum of footrule distances could be more sensitive to possible misspecifications in $\tilde{\mathbf{R}}_j$ when the data are very sparse. Notice also that the second measure is a `modified' version of the Kendall distance between the data and the cluster centers. The boxplots of the posterior distributions of these two quantities are shown in Figure \ref{predpairs_wcsfd}: the two measures are very consistent in indicating a clear elbow at $C = 3$, thus correctly identifying the value we used to generate the data.

We then studied the success rates of correctly predicting missing individual pairwise preferences. A pairwise preference between items $A_{i_1}$ and $A_{i_2}$ was considered missing for assessor $j$ if it was not among the sampled pairwise comparisons included in the data as either $A_{i_1} \prec_{j,\text{true}} A_{i_2}$ or $A_{i_2} \prec_{j,\text{true}} A_{i_1}$, nor could such ordering be concluded from the data indirectly by transitivity. Thus we computed, for all assessors $j$, the predictive probabilities $P(A_{i_1} \prec_{j} A_{i_2}| \text{data})$ for all pairs of items $\{A_{i_1}, A_{i_2}\}$ not ordered in $\text{tc}(\mathcal{B}_j)$. 
The rule for practical prediction was to always bet on the ordering with the larger predictive probability of these two probabilities, then at least 0.5. Each resulting predictive probability is a direct quantification of the uncertainty in making the bet: a value close to 0.5 expresses a high degree of uncertainty, while a value close to 1 would signal greater confidence in that the bet would turn out right. In the experiment, these bets were finally compared to the orderings of the same pairs in the simulated true rankings $\tilde{\mathbf{R}}_{j,\text{true}}.$ If they matched, this was registered as a success, and if not, then as a failure. 

\begin{figure}[t] 
\centering
\includegraphics[width=\textwidth]{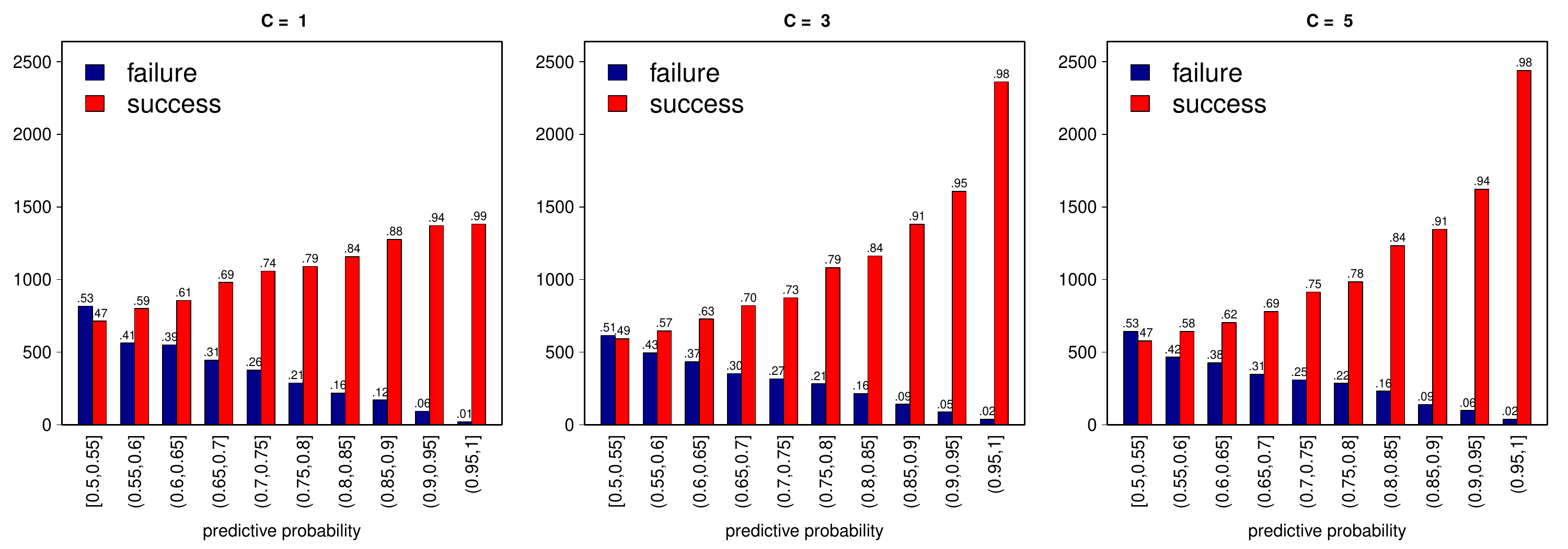}
\caption{Results of the simulation in Section \ref{sec:Prediction}. Barplots of the frequency of successes (red columns) and failures (blue columns) obtained fixing $C=1$ (left), $3$ (middle), and $5$ (right), for the data generated with $\lambda_T=20.$  For $C=1$, $75\%$ of all predictions were correct, for  $C=3$, $79.1\%$, and for $C=5$, $79\%$. }\label{predpairs2}
 \end{figure}

In Figure \ref{predpairs2} are shown the barplots of the results from this experiment, expressed in terms of the frequency of successes (red columns) and failures (blue columns), obtained by combining the outcomes from all individual assessors. For this presentation, the predictive probabilities used for betting were grouped into the respective intervals $[0.50, 0.55], (0.55, 0.60],\ldots, (0.95, 1.00]$ on the horizontal axis, so that pair preferences become more difficult to predict the more one moves to the left, along the x-axis. On top of each column the percentage of successes, or failures, of the corresponding bets is shown. For the results considered on the left, the predictions were made without assuming a cluster structure ($C = 1$) in the analysis, in the middle graph the same number ($C = 3$) of clusters was assumed in the analysis as in the data generation, and on the right, we wanted to study whether assuming an even larger number ($C = 5$) of clusters in the analysis might influence the performance of our method for predicting missing preferences. 

Two important conclusions can be made from the results of this experiment. First, from comparing the three graphs, we can see that not assuming a cluster structure ($C = 1$) in the data analysis led to an overall increased proportion of uncertain bets, in the sense of being based on predictive probabilities closer to the 0.5 end of the horizontal axis, than if either $C = 3$ or $C = 5$ was assumed. On the other hand, there is almost no difference between the graphs corresponding to $C = 3$ and $C = 5$. Thus, moderate overfitting of clusters neither improved nor deteriorated the quality of the predictions (this seems consistent with the very similar within-cluster distances in these two cases, shown in Figure \ref{predpairs_wcsfd}). A second, and more interesting, observation is that, in all three cases considered, the predictive probabilities used for betting turned out to be empirically very well calibrated (see, for example, \citet{dawid1982well} and \citet{little2011calibrated}). For example, of the bets based on predictive probabilities in the interval $(0.70, 0.75]$, 74\% were successful for $C = 1$, 73\% when $C = 3$, and 75\% when $C = 5$. By inspection, such correspondence can be seen to hold quite well on all intervals in all three graphs. That the same degree of empirical calibration holds also when an incorrect number of clusters was fitted to the data as with the correct one, signals a certain amount of robustness of this aspect towards variations in the modeling. 

We repeated the same experiment with less data, namely using $\lambda_T=10$. This gives an average number of pairs per assessor around $10\%$ of $T_\text{max}$. Results are displayed in Figure \ref{supp:pred} of the Supplementary Material, Section \ref{supp:figConvergence}. Predictive probabilities are still very well calibrated, but of course the quality of prediction is worse. Nonetheless, for $C=3$, $76.8\%$ of all predictions were correct. 

\section{Related Work}\label{sec:RelatedWork}

We briefly review the literature which uses the Mallows model, or is based on other probabilistic approaches, as these are most closely related to our method. 

The Mallows model was studied almost exhaustively in the case of Kendall distance, of which the partition function is easy to compute. Among probabilistic approaches, one of the most interesting is \cite{Meila2010}, who proposed a Dirichlet process mixture of the Generalized Mallows model of \cite{Fligner1986}
over incomplete rankings. In this paper two Gibbs sampling techniques for estimating the posterior density were studied. This framework was further extended in \citet{MeilaBao2010}, who developed an algorithm for the ML estimation of their Generalized Mallows model for infinite rankings (IGM), based on Kendall distance. They also considered Bayesian inference with the conjugate prior, showing that such inference is much harder.

In terms of focus and aim, the proposal in \citet{Lu2015} is very close to our approach: they develop a method to form clusters of assessors and perform preference learning and prediction from pairwise comparison data in the Mallows model framework. Their approach is connected to our extension to preference data (Section \ref{sec:Pairwise}), but differs most notably in the general model and algorithm. Their generalized repeated insertion model (GRIM), based on Kendall distance only, generalizes the repeated insertion method for unconditional sampling of Mallows models of \citet{doignon2004repeated}. \citet{Lu2015} perform ML estimation of the consensus ranking using  a method based on the EM algorithm, thus not providing uncertainty quantification for their estimates. Our target, on the other hand, is the full posterior distribution of the unknown consensus ranking. The fact that, for the uniform prior, the MAP estimates and the ML estimates coincide, establishes a natural link between these inferential targets. Two of our illustrations, in Sections \ref{sec:SushiData} and \ref{subsec:Movielens}, use the same data as in \citet{Lu2015}.

In the frequentist framework, the Mallows model with other distances than Kendall was studied by \citet{irurozki2014Ham} and \citet{irurozki2016sampling}, who also developed the \texttt{PerMallows} R package  \citep{irurozki2016permallows}. Moreover, mixtures of Mallows models have been used to analyze heterogeneous rank data by several authors. \citet{MurphyMartin2003} studied mixtures of Mallows with Kendall, footrule and Cayley distances, applying their method to the benchmark  American Psychological Association \citep{Diaconis1988} election data set, where only $n=5$ candidates (items) are ranked. The difficulties in the computation of the partition function for the footrule distance, which arise for larger values of $n$, were not discussed. \citet{gormley2006analysis} use mixtures of Plackett-Luce models in a maximum likelihood framework for clustering. \citet{lee2012mixtures} use mixtures of weighted distance-based models to cluster ranking data. Also \citet{BusseEtal2007} proposed a mixture approach for clustering  rank data, but focusing on the Kendall distance only.

Other probabilistic approaches, less related to the Mallows model, include the Insertion Sorting Rank (ISR) model of  \citet{Jacques2014}. It is implemented in the R package \texttt{rankcluster} \citep{Rankcluster}, and allows 
clustering of partial rankings.
 \cite{SunLebanonKidwell2012} developed a non-parametric probabilistic model on preferences, which can handle also heterogeneous assessors. 
 This work extends the non-parametric kernel density estimation approach over rankings introduced by \citet{LebanonMao2008}, enabling it then to handle ranking data of arbitrary incompleteness and tie structure. However, the approach is based on a random-censoring assumption, which could be easily violated in practice.

Among machine learning approaches, those pertaining to the area of learning to rank, or rank aggregation, are also related to ours. Their aim is to find the best consensus ranking by optimizing some objective function (for example Kemeny or Borda rankings), but they generally do not provide uncertainty quantifications of the derived point estimates. A simple comparison of our approach to two such methods is shown below, in Section \ref{subsec:comparisons}. 


\subsection{Comparisons with other methods}\label{subsec:comparisons}

The procedure we propose is Bayesian, and one of its strengths is its ability to quantify the uncertainty related to the parameter estimates and predictions. In order to compare our results with the ones obtained by other methods which provide only point estimates, we need to summarize the posterior density of the model parameters into a single point estimate, for example MAP,  mode,  mean,  cumulative probability consensus. The cumulative probability (CP) consensus ranking is the ranking arising from the following sequential scheme:  first select the item which has the maximum a posteriori marginal probability of being ranked $1^\text{st}$; then the item which has the maximum a posteriori marginal posterior probability of being ranked $1^\text{st}$ or  $2^\text{nd}$ among the remaining ones, etc. The CP consensus can be seen  as a sequential MAP. We  generated the data from the Mallows model (for details refer to Appendix \ref{sec:sampling}) with Kendall distance, since this is the unique distance handled by existing competitors based on the Mallows model.
We compare our procedure (here denoted by \texttt{BayesMallows}) with the following methods:
\begin{itemize}
\itemsep0em
\item[-] \texttt{PerMallows} \citep{irurozki2016permallows}:  MLE of the Mallows and the Generalized Mallows models, with some right-invariant distance functions, but not footrule nor Spearman. 
\item[-] \texttt{rankcluster} \citep{Rankcluster}: Inference for the Insertion Sorting Rank (ISR) model. 
\item[-] \texttt{RankAggreg} \citep{rankAggreg}: Rank aggregation via several different algorithms. Here we use the Cross-Entropy Monte Carlo algorithm.
\item[-] Borda count \citep{de1781memoire}: Easy and classic way to aggregate ranks. Basically equivalent to the average rank method, thus not a probabilistic approach. 
\end{itemize}

\begin{table}[t]
\centering
\footnotesize\begin{tabular}{|c|l|l|l|l|}
  \hline
${\alpha_\text{{T}}}$ &method& $\hat{\alpha}$ or $\hat{\pi}$ & \small{$\frac{1}{n}d(\hat{\bm{\rho}},\bm{\rho}_{\text{T}})$} &{\small$T(\hat{\bm{\rho}},\;\mathbf{R})$}
\\[0.5ex]
 \hline\hline
  \multirow{6}{*}{1}& \texttt{BayesMallows} - CP & \multirow{2}{*}{1.01  (0.22)}& 0.53 (0.26) & 19.07 (0.54) \\   \cline{2-2}\cline{4-5}
& \texttt{BayesMallows} - MAP &  & 0.57  (0.31) & 19.07   (0.56) \\ 
   \cline{2-5} &    \texttt{PerMallows}  & 1.10  (0.19) & 0.54   (0.26) & 19.12  (0.56) \\ 
     \cline{2-5} 
&  \texttt{rankcluster} & 0.60  (0.02) & 0.86  (0.34) & 19.4 (0.58) \\ 
  \cline{2-5} 
 &  \texttt{RankAggreg} &n.a.   & 0.66  (0.27) & 19.25  (0.58) \\ 
   \cline{2-5} 
 & Borda & n.a.   & 0.54   (0.27) & 19.12  (0.56) \\ 
\hline\hline
  \multirow{6}{*}{2}& \texttt{BayesMallows} - CP  &\multirow{2}{*}{2.05 (0.18)} & 0.17  (0.12) & 16.29  (0.47) \\ 
  \cline{2-2}\cline{4-5}
& \texttt{BayesMallows} - MAP  & & 0.18  (0.13) & 16.28  (0.47) \\ 
  \cline{2-5}
&  \texttt{PerMallows} & 2.07 (0.17) & 0.23  (0.13) & 16.33  (0.46) \\ 
  \cline{2-5} 
&  \texttt{rankcluster} & 0.66 (0.02) & 0.37  (0.22) & 16.52  (0.54) \\
  \cline{2-5} 
  &\texttt{RankAggreg} &n.a.  & 0.29  (0.14) & 16.41  (0.49) \\ 
    \cline{2-5} 
& Borda &n.a. & 0.23  (0.14) & 16.33  (0.46) \\ 
\hline\hline
  \multirow{6}{*}{3}& \texttt{BayesMallows} - CP & \multirow{2}{*}{3.02   (0.07)} & 0.06    (0.08) & 13.88    (0.5) \\ \cline{2-2}\cline{4-5}
& \texttt{BayesMallows} - MAP &  & 0.07    (0.09) & 13.87    (0.5) \\ 
  \cline{2-5}

  &  \texttt{PerMallows}  & 3.02   (0.21) & 0.09     (0.08) & 13.9    (0.51)\\ 
  \cline{2-5}
 &\texttt{rankcluster} &0.72   (0.01) & 0.15  (0.11) &13.96     (0.49)\\
   \cline{2-5}  
  & \texttt{RankAggreg} &n.a.   & 0.14    (0.11) & 13.94    (0.52)\\ 
    \cline{2-5} 
 & Borda & n.a.   &  0.09   (0.08) &  13.91  (0.51) \\ 
\hline\hline
  \multirow{6}{*}{4}& \texttt{BayesMallows} - CP & \multirow{2}{*}{ 3.96   (0.20) }& 0.02    (0.05) & 11.83    (0.41) \\ \cline{2-2}\cline{4-5}
& \texttt{BayesMallows} - MAP && 0.02    (0.04) & 11.83    (0.41) \\ 
  \cline{2-5}
 &  \texttt{PerMallows}  & 3.95   (0.20) & 0.03     (0.05) & 11.85    (0.4)\\ 
   \cline{2-5} 
&  \texttt{rankcluster} &0.76    (0.01) & 0.08  (0.08) &11.9     (0.44)\\  
  \cline{2-5} 
&   \texttt{RankAggreg} &n.a.   &0.06    (0.05) & 11.87    (0.42)\\ 
  \cline{2-5} 
&  Borda & n.a.   &  0.03   (0.05) &  11.85  (0.4) \\ 
   \hline
\end{tabular}
\caption{Results of the simulations of Section \ref{subsec:comparisons}.
$\hat{\alpha}$ refers to the posterior mean (row: \texttt{BayesMallows}) or to MLE (row: \texttt{PerMallows}).  $\hat{\pi}$ is the dispersion parameter of ISR. $\hat{\bm{\rho}}$ is the consensus ranking estimated by the different procedures: MAP (row: \texttt{BayesMallows} (MAP)), CP (row: \texttt{BayesMallows} (CP)), MLE (row: \texttt{PerMallows} and \texttt{rankcluster}), point estimate (row:  \texttt{RankAggreg} and Borda). Standard deviations are reported in parenthesis. Parameters setting: $N=100$, $n=10$.}\label{tab:comp}
\end{table}

The results of the comparisons are shown in Table \ref{tab:comp}.  The  \texttt{BayesMallows} estimates are obtained through Algorithm \ref{algo:basicMH} of Appendix \ref{sec:algoSupp}, with the available exact partition function corresponding to Kendall distance, and for $10^5$ iterations (after a burn-in of $10^4$ iterations). All quantities shown are averages over 50 independent repetitions of the whole simulation experiment. $\hat{\alpha}$ is the posterior mean (for \texttt{BayesMallows}) or the  MLE (for \texttt{PerMallows}), while $\hat{\pi}$ is the MLE estimate of the dispersion parameter of ISR (for \texttt{rankcluster}). $\hat{\bm{\rho}}$ is the consensus ranking estimated by the different procedures: for \texttt{BayesMallows} it is either given by the CP consensus (\texttt{BayesMallows} - CP), or by the MAP (\texttt{BayesMallows} - MAP). We compare the goodness of fit of the methods by evaluating two quantities: first, the normalized Kendall distance between the estimated consensus ranking and the true one, used to generate the data, $d(\hat{\bm{\rho}},\bm{\rho}_{\text{T}})/n$. Second, the average of Kendall distances between the data points and the estimated consensus ranking, $T(\hat{\bm{\rho}},\;\mathbf{R})=\frac{1}{N}\sum_{j=1}^Nd(\hat{\bm{\rho}},\mathbf{R}_j)$. This quantity makes sense here, being independent on the likelihood assumed by the different models.

The first remark about the results in Table \ref{tab:comp} is the clear improvement of the performance in terms of {$\frac{1}{n}d(\hat{\bm{\rho}},\bm{\rho}_{\text{T}})$}, of all the methods, for increasing $\alpha$. This obvious result is a consequence of the easier task of rank aggregation when the assessors are more concentrated around the consensus. 
Because the data were generated with the same model which \texttt{BayesMallows} and \texttt{PerMallows} used for inference, we expected that the Mallows-based methods would perform better than the rank aggregation methods we considered. The results of Table  \ref{tab:comp} confirm this claim: \texttt{BayesMallows} and \texttt{PerMallows} outperform the other rank aggregation methods, with the exception of Borda count, which gives the same results as \texttt{PerMallows}. This is not surprising, since the \texttt{PerMallows} MLE of the consensus is approximated though the Borda algorithm.
Moreover, when the summary of the Bayesian posterior is the CP consensus, the performance of \texttt{BayesMallows}, both in terms of $\frac{1}{n}d(\hat{\bm{\rho}},\bm{\rho}_{\text{T}})$ and $T(\hat{\bm{\rho}},\;\mathbf{R})$, was better than the others. This is another advantage of our approach on the competitors: being the output a full posterior distribution of the consensus, we can select any strategy to summarize it, possibly driven by the application at hand. To conclude, our approach gives slightly better results than the other existing methods, and in the worst cases the performance is still equivalent. 
In Section \ref{sec:Experiments} we will compare inferential results on real data, not necessarily  generated from the Mallows model.

\section{Experiments}\label{sec:Experiments}
The experiments considered in this Section illustrate the use of our approach in various situations corresponding to different data structures.   
\subsection{Meta-Analysis of Differential Gene Expression}\label{sec:MetaAnalysisGenes}


Studies of differential gene expression between two conditions produce lists of genes, ranked according to their level of differential expression as measured by, for example, $p$-values. There is often little overlap between gene lists found by independent studies comparing the same condition. This situation raises the question of whether a consensus top list over all available studies can be found.

We handle this situation in our Bayesian Mallows model by considering each study $j \in \{1,\dots,N\}$ to be an assessor, providing a top-$n_{j}$ list of differentially expressed genes, which are the ranked items. This problem was studied by \citet{DeConde2006}, \citet{Deng2014}, and \citet{Lin2009}, who all used the same 5 studies comparing prostate cancer patients with healthy controls \citep{Dhanasekaran2001,Luo2001,Singh2002,True2006,Welsh2001}. We consider the same 5 studies, and we aim at estimating a consensus with uncertainty. Data consist of the top-$25$ lists of genes from each study, in total $89$ genes. Here we perform a restricted analysis (see \ref{sec:Preselection}), and in this case $n_j=25$ for all $j=1,\ldots,5,$ and $n = 89$. 

\begin{figure}[h!]
\begin{floatrow}
\ffigbox{\includegraphics[width=0.5\textwidth]{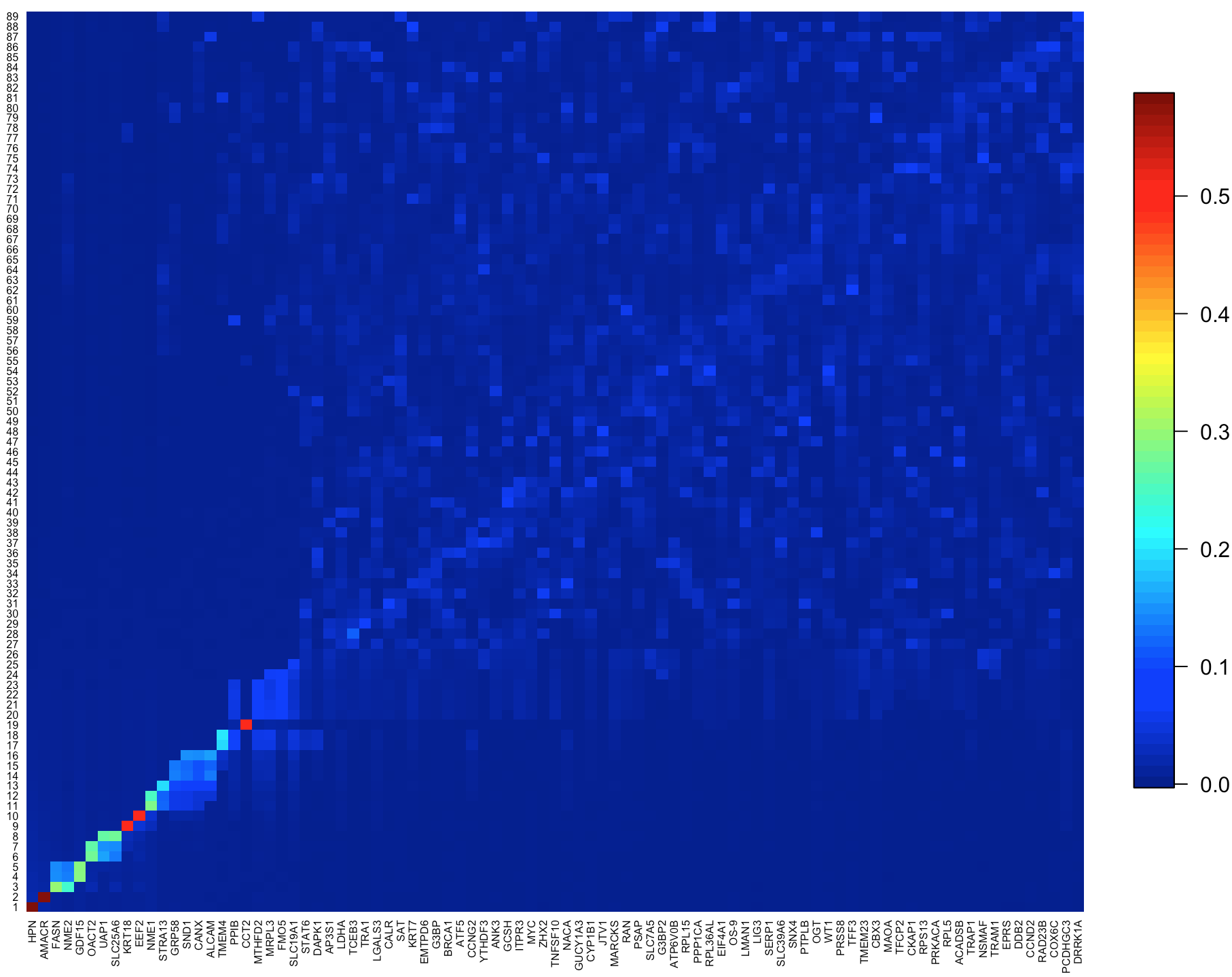}}
{\caption{Heat plot of the posterior probabilities, for $89$ genes, for being ranked as the $k-$th most preferred, for $k=1,...,89$. On the x-axis the genes are ordered according to the estimated CP  consensus.}\label{genes/trace}}
\capbtabbox{\tiny 
\begin{tabular}{|r|c|c|c|c|}
\hline
\textbf{Rank} &\textbf{MAP} & $P(\rho \leq i)$  & $P(\rho \leq 10)$ & $P(\rho \leq 25)$ \\
\hline
1 & HPN & 0.58 & 0.72 & 0.84 \\
2 & AMACR & 0.59 & 0.69 & 0.8 \\
3 & NME2 & 0.26 & 0.56 & 0.64 \\
4 & GDF15 & 0.32 & 0.67 & 0.79 \\
5 & FASN & 0.61 & 0.65 & 0.76 \\
6 & SLC25A6 & 0.19 & 0.63 & 0.71 \\
7 & OACT2 & 0.61 & 0.63 & 0.71 \\
8 & UAP1 & 0.62 & 0.64 & 0.74 \\
9 & KRT18 & 0.6 & 0.61 & 0.72 \\
10 & EEF2 & 0.64 & 0.64 & 0.75 \\
11 & GRP58 & 0.13 & 0.07 & 0.61 \\
12 & NME1 & 0.68 & 0.15 & 0.79 \\
13 & STRA13 & 0.49 & 0.06 & 0.56 \\
14 & ALCAM & 0.33 & 0.05 & 0.65 \\
15 & SND1 & 0.51 & 0.07 & 0.71 \\
16 & CANX & 0.59 & 0.07 & 0.64 \\
17 & TMEM4 & 0.34 & 0.05 & 0.58 \\
18 & DAPK1 & 0.15 & 0.04 & 0.21 \\
19 & CCT2 & 0.59 & 0.05 & 0.62 \\
20 & MRPL3 & 0.36 & 0.06 & 0.6 \\
21 & MTHFD2 & 0.43 & 0.06 & 0.58 \\
22 & PPIB & 0.51 & 0.06 & 0.57 \\
23 & SLC19A1 & 0.42 & 0.06 & 0.53 \\
24 & FMO5 & 0.58 & 0.05 & 0.59 \\
25 & TRAM1 & 0.14 & 0.04 & 0.14 \\
\hline
\end{tabular}}
{\caption{Top-$25$ genes in the MAP consensus ranking from a  total of $89$ genes. The cumulative probability of each gene in the top-$25$ positions in the MAP of being in that position, or higher, is shown in the third column of the Table, $P(\rho \leq i)$. The probabilities of being among the top-10 and top-25 are also shown for each gene.}
\label{tab:GeneList}}
\end{floatrow}
\end{figure}

Table \ref{tab:GeneList} shows the result of analyzing the five gene lists with the Mallows footrule model for partial data (Section \ref{sec:PartialRanks}). We run 20 different chains, for a total of $10^7$ iterations (computing time was $16'4''$), and discarded the first $5\cdot10^4$ iterations of each as burn-in. For the partition function, we used the IS approximation $Z_n^K(\alpha)$ with $K=10^7$, computed off-line on a grid of $\alpha$'s in $(0,40]$. After some tuning, we set $L=40,$ $\sigma_\alpha=0.95$, $\lambda=0.05$ and $\alpha_\text{jump}=1,$ and used the footrule distance. Like \citet{DeConde2006}, \citet{Deng2014}, and \citet{Lin2009}, our method ranked the genes HPN and AMACR first and second in the MAP consensus ranking. The low value of the posterior mean of $\alpha$, being 0.56 (mode 0.43, high posterior density, HPD, interval $(0.04,1.29)$), is an indicator of a generally low level of agreement between the studies. In addition, the fact that $n>N$, and having partial data, both contribute to keeping $\alpha$ small. 
However, the posterior probability for each gene to be among the top-$10$ or top-$25$ is not so low, thus demonstrating that our approach can provide a valid criterion for consensus. In the hypothetical situation in which we had included in our analysis all $n^*$ genes following a \emph{full analysis} mode, with $n^*$ being at least 7567, the largest number of genes included in in any of the five original studies \citep{DeConde2006}, this would have had the effect of making the posterior probabilities in Table \ref{tab:GeneList} smaller. On the other hand, because of Corollary \ref{cor:preselectionMAP}, the ranking order obtained from such a hypothetical analysis based on all $n^*$ genes would remain the same as in Table \ref{tab:GeneList}.

\begin{table}[h!]
\tiny
\begin{minipage}{0.48\textwidth}
\centering
\begin{tabular}{|r|c|c|}
 \hline
\textbf{rank} &\textbf{CE algorithm} & \textbf{GA algorithm} \\
\hline
1 & HPN & HPN \\ 
  2 & AMACR & AMACR \\ 
  3 & FASN & NME2 \\ 
  4 & GDF15 & 0ACT2 \\ 
  5 & NME2 & GDF15 \\ 
  6 & 0ACT2 & FASN \\ 
  7 & KRT18 & KRT18 \\ 
  8 & UAP1 & SLC25A6 \\ 
  9 & NME1 & UAP1 \\ 
  10 & EEF2 & SND1 \\ 
  11 & STRA13 & EEF2 \\ 
  12 & ALCAM & NME1 \\ 
  13 & GRP58 & STRA13 \\ 
  14 & CANX & ALCAM \\ 
  15 & SND1 & GRP58 \\ 
  16 & SLC25A6 & TMEM4 \\ 
  17 & TMEM4 & CCT2 \\ 
  18 & PPIB & FM05 \\ 
  19 & CCT2 & CANX \\ 
  20 & MRPL3 & DYRK1A \\ 
  21 & MTHFD2 & MTHFD2 \\ 
  22 & SLC19A1 & CALR \\ 
  23 & FM05 & MRPL3 \\ 
  24 & PRSS8 & TRA1 \\ 
  25 & NACA & NACA \\ 
   \hline
   \end{tabular}
\end{minipage}%
\quad
\begin{minipage}{0.48\textwidth}
\centering
\begin{tabular}{|r|c|c|c|c|}
  \hline
\textbf{rank} & \textbf{mean} & \textbf{median} & \textbf{geo.mean} & \textbf{l2norm} \\ 
  \hline
1 & HPN & HPN & HPN & HPN \\ 
  2 & AMACR & AMACR & AMACR & AMACR \\ 
  3 & GDF15 & FASN & FASN & GDF15 \\ 
  4 & FASN & KRT18 & GDF15 & NME1 \\ 
  5 & NME1 & GDF15 & NME2 & FASN \\ 
  6 & KRT18 & NME1 & SLC25A6 & KRT18 \\ 
  7 & EEF2 & EEF2 & EEF2 & EEF2 \\ 
  8 & NME2 & UAP1 & 0ACT2 & NME2 \\ 
  9 & 0ACT2 & CYP1B1 & OGT & UAP1 \\ 
  10 & SLC25A6 & ATF5 & KRT18 & 0ACT2 \\ 
  11 & UAP1 & BRCA1 & NME1 & SLC25A6 \\ 
  12 & CANX & LGALS3 & UAP1 & STRA13 \\ 
  13 & GRP58 & MYC & CYP1B1 & CANX \\ 
  14 & STRA13 & PCDHGC3 & ATF5 & GRP58 \\ 
  15 & SND1 & WT1 & CBX3 & SND1 \\ 
  16 & OGT & TFF3 & SAT & ALCAM \\ 
  17 & ALCAM & MARCKS & CANX & TMEM4 \\ 
  18 & CYP1B1 & OS-9 & BRCA1 & MTHFD2 \\ 
  19 & MTHFD2 & CCND2 & GRP58 & MRPL3 \\ 
  20 & ATF5 & DYRK1A & MTHFD2 & PPIB \\ 
  21 & CBX3 & TRAP1 & STRA13 & OGT \\ 
  22 & SAT & FM05 & LGALS3 & CYP1B1 \\ 
  23 & BRCA1 & ZHX2 & ANK3 & SLC19A1 \\ 
  24 & MRPL3 & RPL36AL & GUCY1A3 & ATF5 \\ 
  25 & LGALS3 & ITPR3 & LDHA & CBX3 \\ 
   \hline
\end{tabular}
\caption{Results given by the \texttt{RankAggreg} R package (left) and by the \texttt{TopKLists} R package (right).}
\label{tab:GeneOtherMethods}
\end{minipage}\end{table}

Next we compared the result shown in Table \ref{tab:GeneList} with other approaches: Table \ref{tab:GeneOtherMethods} (left) reports results obtained with \texttt{RankAggreg} \citep{rankAggreg}, which is specifically designed to target meta-analysis problems, while in Table \ref{tab:GeneOtherMethods} (right) different aggregation methods implemented in \texttt{TopKLists} \citep{schimek2015topklists} are considered. 
The results obtained via \texttt{RankAggreg} turned out unstable,
with the final output changing in every run, and the list shown in Table  \ref{tab:GeneOtherMethods} differs from that in \citet{rankAggreg}. Overall, apart from the genes ranked to the top$-2$ places, there is still considerable variation in the
exact rankings of the genes. Rather than considering such exact rankings, however, it may in practice be of
more interest to see to what extent the same genes are shared between different top$-k$ lists. Here the
results are more positive. For example, of the 10 genes on top of the MAP consensus list of Table \ref{tab:GeneList}, always
9 genes turned out to be in common with each of the lists of Table \ref{tab:GeneOtherMethods}, with the exception of the median (column 3 of Table \ref{tab:GeneOtherMethods}, right), where only 7 genes are shared. Column 4 of Table \ref{tab:GeneList} provides additional
support to the MAP selection of the top$-10$: all genes included in that list have posterior probability at least
0.56 for being among the top$-10$, while for those outside the list it is maximally 0.15.

In order to have a quantification of the quality of the different estimates, we compute the footrule distance for partial data \citep[p. 30]{critchlow2012metric} between $\bm{\rho}$ and $\mathbf{R}_j$, averaged over the assessors, defined as follows 
$$T_{\text{partial}}({\bm{\rho}},\;\mathbf{R})= \frac{1}{N}\sum_{j=1}^N\sum_{i=1}^n |\nu_{R_{ij}}-\nu_{\rho_i}|,$$
where $\nu_{\bm{\rho}}$, $\nu_{\mathbf{R}_j}\in\mathcal{P}_n$ are equal to $\bm{\rho}$ and $\mathbf{R}_j$ in their top$-n_j$ ranks (top$-25$ in the case of gene lists), while the rank $\frac{n+n_j+1}{2}$ is assigned to the items whose rank in $\bm{\rho}$ and $\mathbf{R}_j$ is not in their top$-n_j$. Note that $\frac{n+n_j+1}{2}$ (equal to 57.5 in this case) is the average of the ranks of the excluded items. Table \ref{tab:partial} reports the values of $T_{\text{partial}}$ for the various methods. We notice that the minimum value is achieved by the Mallows MAP consensus list. 

\begin{table}[ht]
\small\centering
\begin{tabular}{c|c|c|c|c|c|c|c|}
  \cline{2-8}
  &\textbf{MAP} & \textbf{CE} & \textbf{GA} & \textbf{mean} & \textbf{median} & \textbf{geo.mean} & \textbf{l2norm}\\
  \hline
\multicolumn{1}{|c|}{$T_{\text{partial}}({\bm{\rho}},\;\mathbf{R})$} &12.56 & 12.67 & 12.98 & 13.52 & 15.26 & 14.05 & 13.04 \\ 
   \hline
\end{tabular}\caption{Values of the average footrule distance for partial data $T_{\text{partial}}$ between the partial gene lists and the different estimated consensus rankings.}
\label{tab:partial}
\end{table}

\subsection{Beach preference data}\label{sec:Beaches}

\begin{figure}[h!]
\begin{floatrow}
\ffigbox{\includegraphics[width=0.44\textwidth]{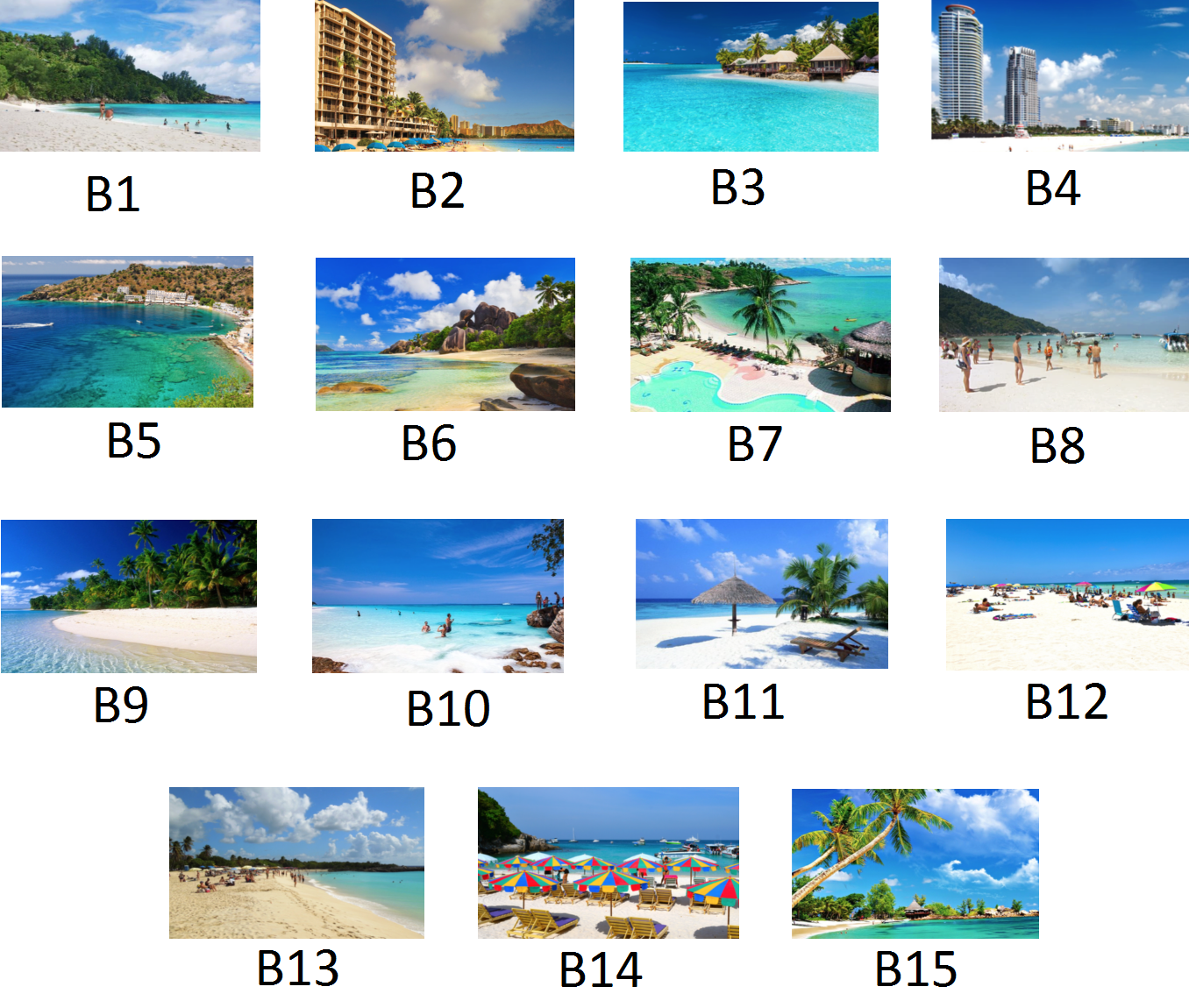}}
{\caption{The 15 images used for producing the {Beach} dataset.}\label{beaches}}
\hspace{1cm}
\capbtabbox{\footnotesize \begin{tabular}{|c|c|c|c|}\hline
$\bm{\rho}$&$\textbf{CP}$ & {${P(\rho_i \leq i)}$}  & \textbf{95\% HPDI}  \\
\hline
1 & B9 & 0.81 & (1,2) \\ 
  2 & B6 & 1 & (1,2) \\ 
  3 & B3 & 0.83 & (3,4) \\ 
  4 & B11 & 0.75 & (3,5) \\ 
  5 & B15 & 0.68 & (4,7) \\ 
  6 & B10 & 0.94 & (4,7) \\ 
  7 & B1 & 1 & (6,7) \\ 
  8 & B13 & 0.69 & (8,10) \\ 
  9 & B5 & 0.55 & (8,10) \\ 
  10 & B7 & 1 & (8,10) \\ 
  11 & B8 & 0.41 & (11,14) \\ 
  12 & B4 & 0.62 & (11,14) \\ 
  13 & B14 & 0.81 & (11,14) \\ 
  14 & B12 & 0.94 & (12,15) \\ 
  15 & B2 & 1 & (14,15) \\ 
\hline
\end{tabular}}
{\caption{Results of the pair comparisons. Beaches arranged according to the CP consensus ordering together with the corresponding 95\% highest posterior density intervals.}\label{tabCons}}
\end{floatrow}
\end{figure}

Here we consider pair comparison data (Section \ref{sec:Pairwise}) generated as follows: first we chose $n = 15$ images of tropical beaches, shown in Figure  \ref{beaches}, such that they differ in terms of presence of building and people. For example, beach B9 depicts a very isolated scenery, while beach B2 presents a large hotel seafront.   

The pairwise preference data were collected as follows. Each assessor was shown a sequence of $25$ pairs of images, and asked on every pair the question: \emph{''Which of the two beaches would you prefer to go to in your next vacation?''}. Each assessor was presented with a random set of pairs, arranged in random order. As there are 105 possible pairs, 25 pairs is less than 25\% of the total.
We collected $N=60$ answers. Seven assessors did not answer to all questions, but we kept these responses as our method is able to analyze also incomplete data. Nine assessors returned orderings which contained at least one non-transitive pattern of comparisons. In this analysis we dropped the non-transitive patterns from the data. Systematic methods for dealing with non-transitive rank data will be considered elsewhere.

We run the MCMC for $10^6$ iterations, and discarded the first $10^5$ iterations as burn-in. We set $L=2,$ $\sigma_\alpha=0.1$, $\lambda=0.1$ and $\alpha_\text{jump}=100.$ Computing time was less than $2'$.
The posterior mean of $\alpha$ was $\mathbb{E}(\alpha|\text{data})=3.38\, (2.94, 3.82)$. In Table \ref{tabCons} we report the CP consensus ranking of the beaches (column 2), the cumulative probability of each item $i$ to be in the top$-i$ positions, i.e., $P(\rho_i \leq i)$ (column 3), and the 95\% HPDI for each item (column 4), which represents the posterior uncertainty. In Table \ref{igraph} we give the consensus ranking obtained by two other methods, for comparison.

\begin{figure}[ht]
\centering\tiny
\centerline{\includegraphics[width=0.9\textwidth]{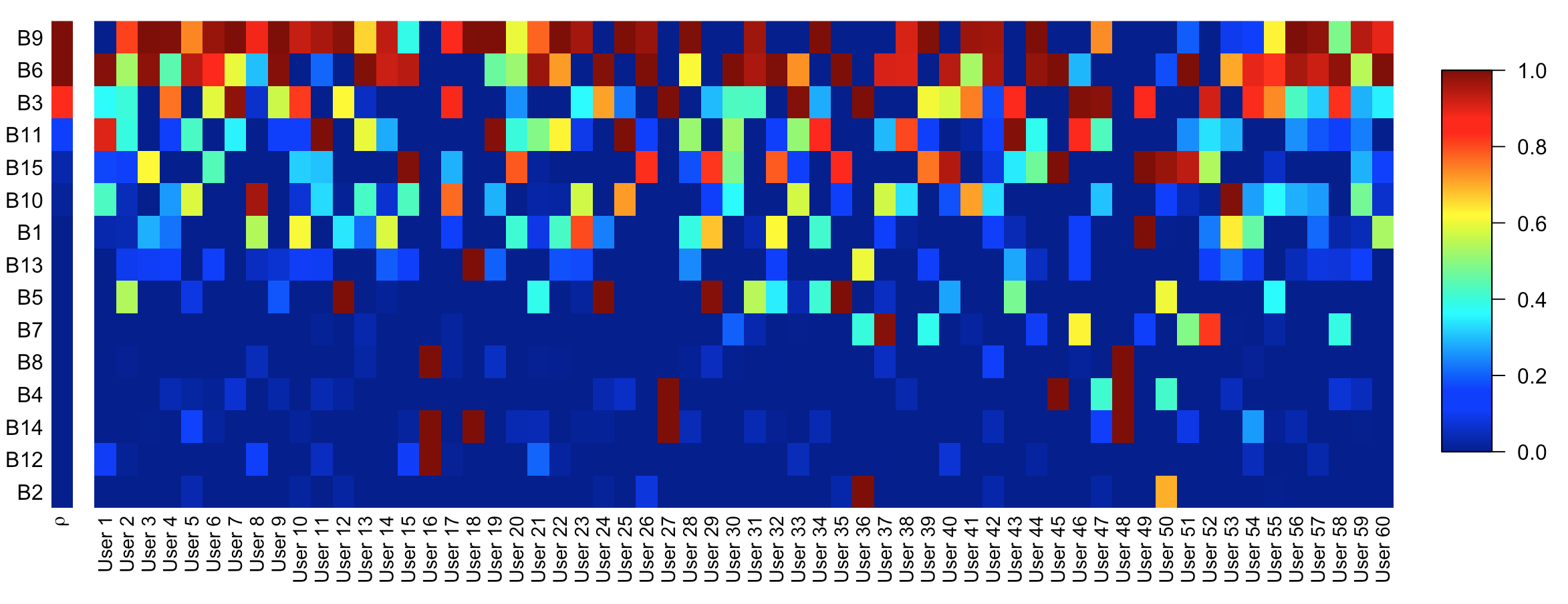}}
\caption{Posterior probability, for each beach, of being ranked among the top-3 in $\bm{\rho}$ (column 1),  and in $\mathbf{R}_j$, $j=1,...,60$ (next columns).}\label{top3}
\end{figure}

With our method we also estimate the latent full ranking of each assessor. Figure \ref{top3} was obtained as follows: in the separate column on the left, we display the posterior probability $\mathbb{P}(\rho_{\text{B}i}\leq3|\text{data})$ that a given beach B$i$, $i = 1,...,15$, is among the top$-3$ in the consensus $\bm{\rho}$. In the other columns we show, for each beach B$i$, the individual posterior probabilities $\mathbb{P}(\tilde{R}_{j,\text{B}i}\leq3|\text{data})$, of being among the top$-3$ for each assessor $j$, $j=1,...,60$. We see for example that beach B5, which was ranked only 9th in the consensus, had, for 4 assessors, posterior probability very close to 1 of being included among their top$-3$ beaches.   

\begin{table}
\footnotesize
\begin{tabular}{|c||c|c|c|c|c|c|c|c|c|c|c|c|c|c|c|}\hline
  $\bm{\rho}$&1       &2       &3      &4      &5      &6       &7       &8 
    &9      &10      &11       &12      &13      &14 &15\\\hline
     \textbf{\texttt{BT}}&B6       &B9       &B3      &B11      &B10      &B15       &B1       &B5 
    &B7      &B13      &B4       &B8      &B14      &B12 &B2\\
    \hline
  \textbf{\texttt{PR}} &B6&  B9&B10&B15&B3&B1&B11&B13&B7&B5&B8&B12&B4&B14&B2\\
\hline
\end{tabular}
\caption{Consensus ordering given by other methods: \texttt{BT} is the Bradley Terry given by the \texttt{BradleyTerry2} R package \citep{firth2012bradley}, \texttt{PR} is the popular Google PageRank output \citep{brin1998anatomy} given by the \texttt{igraph} R package \citep{csardi2013package}. Most preferred to the left.}\label{igraph}
\end{table}

\subsection{Sushi Data}\label{sec:SushiData}

\begin{figure}[h]
\centering
\includegraphics[scale=.4]{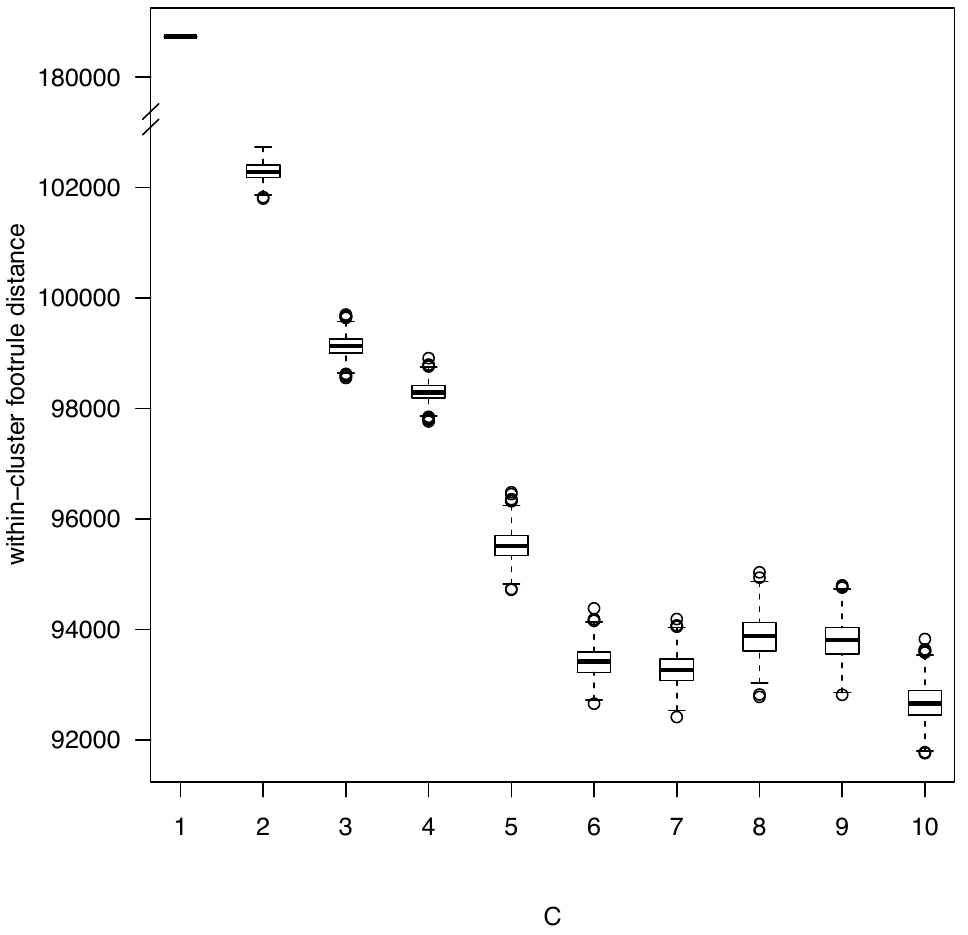}
\caption{Results of the Sushi experiment. Boxplots of the posterior distributions of the within-cluster sum of footrule distances of assessors' ranks from the corresponding cluster consensus for different choices of $C$ (note the y-axis break, for better visualization). 
}
\label{fig:sushiINDEXES}
\end{figure}

We illustrate clustering based on full rankings using the benchmark dataset of sushi preferences collected across Japan \citep{Kamishima2003}, see also \citet{Lu2015}. $N = 5000$ people were interviewed, each giving a complete ranking of $n = 10$ sushi variants. Cultural differences among Japanese regions influence food preferences, so we expect the assessors to be clustered according to different shared consensus rankings. We analyzed the sushi data using mixtures of Mallows models (Section \ref{sec:Clustering}) with the footrule distance (with  the exact partition function of the Mallows model, see Section \ref{sec:Distances}). We run the MCMC for $10^6$ iterations, and discarded the first $10^5$ iterations as burn-in. After some tuning, we set $L=1,$ $\sigma_\alpha=0.1$, $\lambda=0.1$ and $\alpha_\text{jump}=100.$ In the Dirichlet prior for $\boldsymbol{\tau}$, we set the hyper-parameter $\psi = N/C$, thus favoring high-entropy distributions. Computing time varied depending on $C$, from a minimum of $1h04'$ to a maximum of $10h45'$ for $C=10$. For each possible number of clusters $C \in \{1,\dots,10\}$, we used a thinned subset of MCMC samples to compute the posterior footrule distance between $\bm{\rho}_{c}$ and the ranking of each assessor assigned to that cluster, $\sum_{c=1}^C \sum_{j:z_j = c}d(\mathbf{R}_j,\bm{\rho}_{c})$. The posterior of this quantity, over all assessors and cluster centers, was then used for choosing the appropriate value for $C$, see Figure \ref{fig:sushiINDEXES}. We found an elbow at $C = 6$, which was then used to further inspect results. 

\begin{table}[h]
\centering
\begin{scriptsize}
\begin{tabular}{|r|l|l|l|l|l|l|}
\hline
& $c=1$ & $c=2$ & $c=3$ & $c=4$ & $c=5$ & $c=6$ \\
\hline
$\tau_c$ & 0.243 (0.23,0.26) & 0.131 (0.12,0.14) & 0.107 (0.1,0.11) & 0.117 (0.11,0.12) & 0.121 (0.11,0.13) & 0.278 (0.27,0.29) \\ 
$\alpha_c$ & 3.62 (3.52,3.75) & 2.55 (2.35,2.71) & 3.8 (3.42,4.06) & 4.02 (3.78,4.26) & 4.46 (4.25,4.68) & 1.86 (1.77,1.94) \\   \hline
1 & fatty tuna & shrimp & sea urchin & fatty tuna & fatty tuna & fatty tuna \\ 
  2 & sea urchin & sea eel & fatty tuna & salmon roe & tuna & tuna \\ 
  3 & salmon roe & egg & shrimp & tuna & tuna roll & sea eel \\ 
  4 & sea eel & squid & tuna & tuna roll & shrimp & shrimp \\ 
  5 & tuna & cucumber roll & squid & shrimp & squid & salmon roe \\ 
  6 & shrimp & tuna & tuna roll & egg & sea eel & tuna roll \\ 
  7 & squid & tuna roll & salmon roe & squid & egg & squid \\ 
  8 & tuna roll & fatty tuna & cucumber roll & cucumber roll & cucumber roll & sea urchin \\ 
  9 & egg & salmon roe & egg & sea eel & salmon roe & egg \\ 
  10 & cucumber roll & sea urchin & sea eel & sea urchin & sea urchin & cucumber roll \\ 
   \hline
\end{tabular}
\caption{Results of the Sushi experiment when setting $C=6$. Sushi items arranged according to the MAP consensus ranking found from the posterior distribution of $\bm{\rho}_c$, $c=1,\ldots,6.$ At the top of the Table, corresponding MAP estimates for $\boldsymbol{\tau}$ and $\boldsymbol{\alpha}$, with 95\% HPDIs (in parenthesis). Results are based on $10^6$ MCMC iterations.}
\label{tab:Sushi6Clusters}
\end{scriptsize}
\end{table}

\begin{figure}[h]
\centering
\includegraphics[scale=.51]{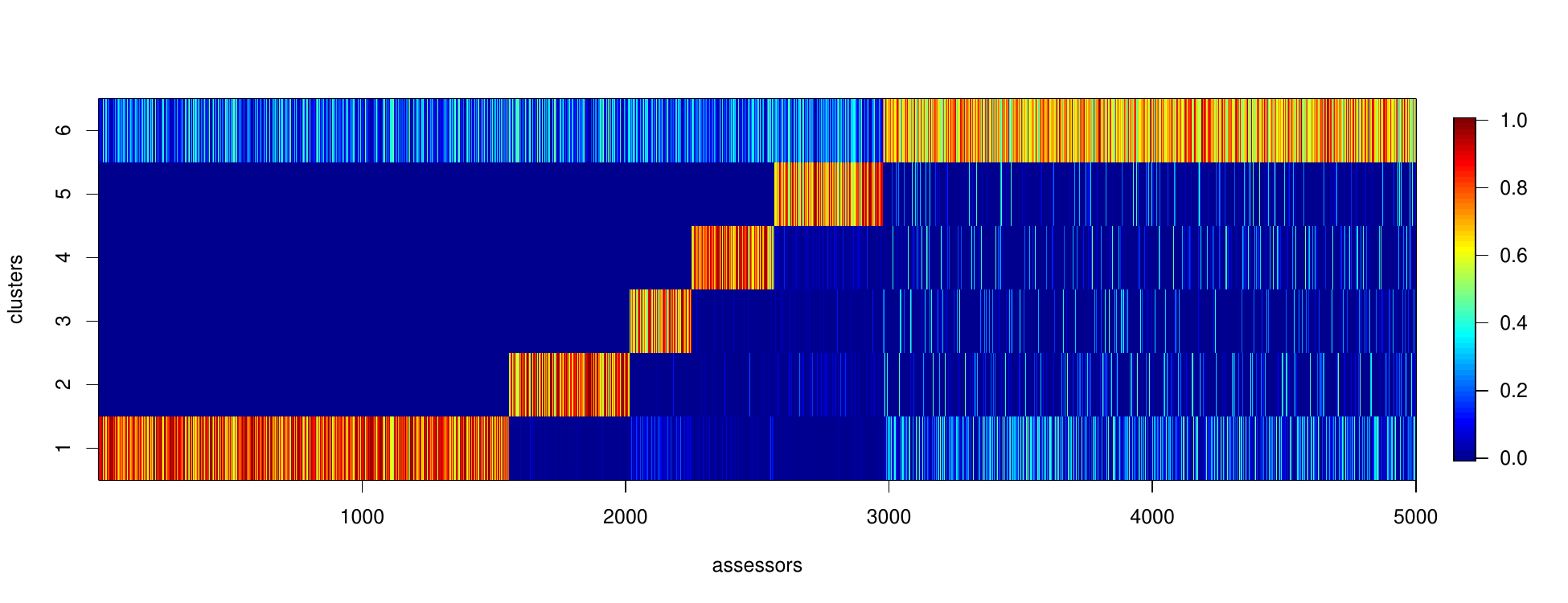}
\caption{Heatplot of posterior probabilities for all 5000 assessors (on the x-axis) of being assigned to each cluster ($c=1,\dots,6$ from bottom to top).}
\label{fig:sushiZ}
\end{figure}

Table \ref{tab:Sushi6Clusters} shows the results when the number of clusters is set to $C = 6$: for each cluster, the MAP estimates for $\boldsymbol{\tau}$ and $\boldsymbol{\alpha}$, together with their 95\% HPDIs, are shown on the top of the Table. Table \ref{tab:Sushi6Clusters} also shows the sushi items, arranged in cluster-specific lists according to the MAP consensus ordering (in this case equal to the CP consensus). Our results can be compared with the ones in \citet{Lu2015} (Table 1 in Section 5.3.2): the correspondence of the clusters could be 1-4, 2-1,3-2,4-5,5-4,6-0. Note that the dispersion parameter $\alpha$ in our Bayesian Mallows model is connected to the dispersion parameter $\phi$ in \citet{Lu2015} by the link $\alpha = -n\log(\phi)$. Hence, we can also observe that the cluster-specific $\alpha$ values reported in Table \ref{tab:Sushi6Clusters} are quite comparable to the  dispersion parameters of \citet{Lu2015}.

We investigate the stability of the clustering in Figure \ref{fig:sushiZ}, which shows the heatplot of the posterior probabilities, for all 5000 assessors (on the x-axis), of being assigned to each of the $6$ clusters in Table \ref{tab:Sushi6Clusters} (clusters $c=1,\ldots,6$ from bottom to top in Figure \ref{fig:sushiZ}): most of these individual probabilities were concentrated on some particular preferred value of $c$ among the six possibilities, indicating a reasonably stable behavior in the cluster assignments.

\subsection{Movielens Data}\label{subsec:Movielens}

The Movielens dataset\footnote{\texttt{www.grouplens.org/datasets/}.} contains movie ratings from $6040$ users. In this example, we focused on the $n = 200$ most rated movies, and on the $N = 6004$ users who rated (not equally) at least 3 movies. Each user had considered only a subset of the $n$ movies (30.2 on average). We converted the ratings given by each user from a 1-5 scale to pairwise preferences as described in \citet{Lu2015}: each movie was preferred to all movies which the user had rated strictly lower. We selected users whose rating included at least 3 movies, because two of them were needed to create at least a pairwise  comparison, and the third one was needed for prediction, as explained in the following.

\begin{figure}[h]
\centering
\includegraphics[scale=.5]{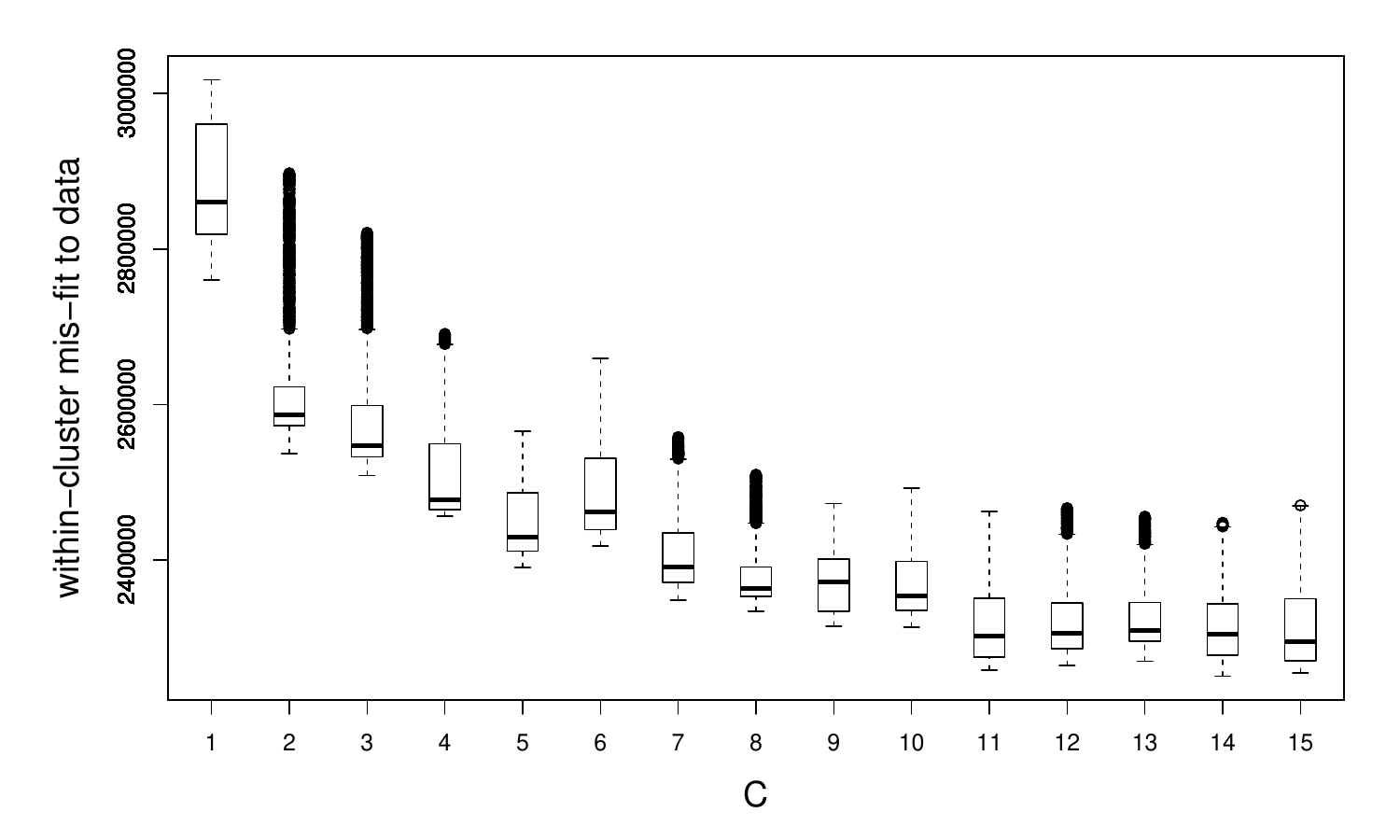}
\caption{Results of the Movielens experiment. Boxplots of the posterior distributions of the within-cluster indicator of mis-fit to the data, as introduced in Section \ref{sec:Prediction}, for different choices of $C$.}
\label{fig:movieINDEXES}
\end{figure}

Since we expected heterogeneity among users, due to age/gender/social factors/education, we applied the clustering scheme for pairwise preferences, with the footrule distance. Since $n=200$, we used the asymptotic approximation for $Z_{n}(\alpha)$ described in \cite{mukherjee2016} and in Section 2 of the Supplementary Material. We run the MCMC for $10^5$ iterations, after a burn-in of $5\cdot10^4$ iterations. We set: $L=20$, $\sigma_\alpha=0.05,$ $\alpha_\text{jump}=10$ and $\lambda=0.1$, after some tuning. Note that the label switching problem only affects inference on cluster-specific parameters, but it does not affect predictive distributions \citep{Celeux2006}. We varied the number $C$ of clusters in the set $\{1,\ldots,15\},$ and inspected the within-cluster indicator of mis-fit to the data, $\sum_{c=1}^C\sum_{j:z_j = c} |\{B\in\text{tc}(\mathcal{B}_{j}):B\text{ is not consistent with }\bm{\rho}_{c}\}|,$ introduced in Section \ref{sec:Prediction}, see Figure \ref{fig:movieINDEXES}: the posterior within-cluster indicator shows two possible elbows: $C=5,$ and $C=11$. Hence, according to these criteria, both choices seemed initially conceivable. However, it is beyond the scope of this paper to discuss ways to decide the number of clusters.

\begin{figure}[h]
\centering
\includegraphics[width=0.8\columnwidth]{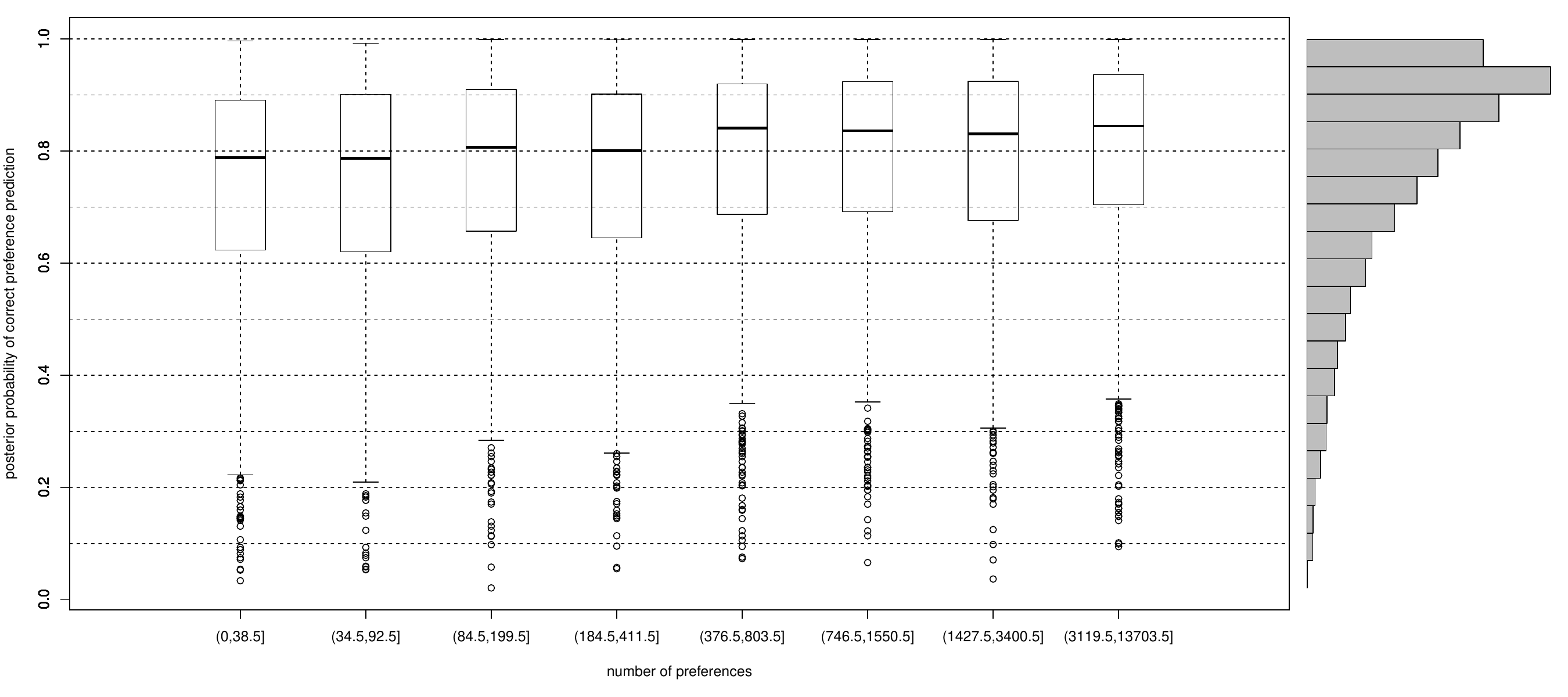}%
\caption{Results of the Movielens experiment. Boxplots of the posterior probability for correctly predicting the discarded preference conditionally on the number of preferences stated by the user, for the model with $C=5$. The histogram on the right shows the marginal posterior probability for correct preference prediction.}
\label{fig:moviesPRED}
\end{figure}

In order to select one of these two models, we examined their predictive performance. Before converting ratings to preferences, we discarded for each user $j$ one of the rated movies at random. Then, we randomly selected one of the other movies rated by the same user, and used it to create a pairwise preference involving the discarded movie. This preference was then not used for inference. After running the Bayesian Mallows model, we computed for each user the predictive probabilities $P(\tilde{\mathbf{R}}_{j}|\text{data})$, and thereby the probabilities for correctly predicting the discarded preference. The median, across all users, of these probabilities was $0.8225$ for the model with $C=5$ clusters, and $0.796$ for $C=11$ clusters. Moreover, for $C=5,$ $88~\%$ of these probabilities were higher than $0.5$. These are very positive results, and they suggest that the predictive performance of the model with $5$ clusters is slightly better than the one with $11$ clusters. It appears that the larger number of clusters in the latter model leads to a slight overfitting, and this is likely to be the main cause of the loss in the predictive success. Figure \ref{fig:moviesPRED} shows the boxplots of the posterior distribution of the probability for correct preference prediction of the left out comparison, stratified with respect to the number of preferences given by each user, for the model with $C=5$. The histogram on the right shows the same posterior probability for correctly predicting the discarded preference for all users, for the same model, regardless of how many preferences each user had expressed. 
Interestingly, in this data, the predictive power is rather stable and high, irrespectively from how many movies the users rated. In other applications, we would expect  the predictions to become better the more preferences are expressed  by a user. In this case, a figure similar to Figure \ref{fig:moviesPRED} could guide personal recommendation algorithms, which should not rely on estimated point preferences, if these are too uncertain, as happens for users who have given a few ratings only.

\begin{table}[h!]
\centering
\begin{tiny}
\begin{tabular}{|r|l|l|l|l|l|}
\hline
& $c=1$ & $c=2$ & $c=3$& $c=4$ & $c=5$ \\
\hline
$\tau_c$ &  0.325 (0.32,0.33) & 0.219 (0.21,0.23) & 0.156 (0.15,0.17) & 0.145 (0.14,0.15) & 0.155 (0.15,0.16) \\\hline
$\alpha_c$ &    2.53 (2.36,2.7) & 3.33 (3.2,3.48) & 2.58 (2.27,2.81) & 1.87 (1.67,2.02) & 2.68 (2.47,2.89) \\  \hline
1   & A Christmas Story& Citizen Kane & The Sting & Indiana Jones (I) & Shawshank Redemption \\ 
2   & Schindler's List & The Godfather & Dr. Strangelove & A Christmas Story & Indiana Jones (I) \\ 
 3  & The Godfather & Pulp Fiction & 2001: A Space Odyssey & Star Wars (IV)  & Braveheart \\ 
 4  & Casablanca & Dr. Strangelove & The Maltese Falcon & The Princess Bride & Star Wars (IV) \\ 
 5  & Star Wars (IV)  & A Clockwork Orange & Casablanca & Schindler's List & Saving Private Ryan \\ 
 6  & Shawshank Redemption & Casablanca & Taxi Driver & The Matrix & The Green Mile  \\ 
 7  & Saving Private Ryan & The Usual Suspects & Citizen Kane & Shawshank Redemption & Schindler's List \\ 
 8  &  The Sting & 2001: A Space Odyssey & Schindler's List & Indiana Jones (III) & The Sixth Sense \\ 
 9  & The Sixth Sense & American Beauty & Chinatown & The Sting & The Matrix \\ 
 10  & American Beauty & Star Wars (IV) & The Godfather & The Sixth Sense& Star Wars (V) \\ 
   \hline
\end{tabular}
\end{tiny}
\caption{Results of the Movielens experiment. Movies arranged according to the CP consensus ranking, from the posterior distribution of $\bm{\rho}_c$, $c=1,\ldots,5.$}
\label{tab:Movielens5Clusters}
\end{table}

In Table \ref{tab:Movielens5Clusters} the MAP estimates for $\boldsymbol{\tau}$ and $\boldsymbol{\alpha}$, together with their 95\% HPDIs, are shown at the top. The Table also shows a subset of the movies, arranged in cluster-specific top$-10$ lists according to the CP consensus ranking, from the posterior distribution of $\bm{\rho}_c$, $c=1,\ldots,5.$ We note that all $\alpha$ values correspond to a reasonable within-cluster variability. Moreover, the lists reported in Table \ref{tab:Movielens5Clusters} characterize the users in the same cluster as individuals sharing a reasonably well interpretable preference profile. Since in the Movielens dataset additional information on the users is available, we compared the estimated cluster assignments with the age, gender, and the occupation of the users. While occupation showed no interesting patterns, the second and fifth clusters had more males than expected, in contrast to the first and fourth clusters which included more females than average, the former above 45 and the latter below 35 of age.

\section{Discussion}\label{sec:Discussion}

In this paper, we developed a fully Bayesian hierarchical framework for the analysis of rank data. An important advantage of the Bayesian approach is that it offers coherently propagated and directly interpretable ways to quantify posterior uncertainties of estimates of any quantity of interest. Earlier Bayesian treatments of the Mallows rank model are extended in many ways: we develop an importance sampling scheme for $Z_{n}(\alpha)$ allowing the use of other distances than Kendall's, and our MCMC algorithm efficiently samples from the posterior distribution of the unknown consensus ranking and of the latent assessor-specific full rankings. We also develop various extensions of the model, motivated by applications in which data take particular forms.

The Mallows model performs very well with a large number of assessors $N$, as we show in the Sushi experiment of Section \ref{sec:SushiData}, and in the Movielens experiment of Section \ref{subsec:Movielens}. On the other hand, it may not be computationally feasible when the number of items is extremely large, for example $n \geq 10^{4}$, which is not uncommon in certain applications \citep{Volkovs2014}. For the footrule and Spearman distances, there exist asymptotic approximations for $Z_{n}(\alpha)$ as $n\to \infty$ \citep{mukherjee2016}, which we successfully used in Section \ref{subsec:Movielens}, although the MCMC algorithm converges slowly in such large spaces. Maximum likelihood estimation of $\bm{\rho}$ runs into the same problem when $n$ gets large \citep{Aledo2013,Ali2012}. \citet{Volkovs2014} developed the multinomial preference model (MPM) for cases with very large $n$, which can be efficiently computed by maximizing a concave log-likelihood function. The MPM thus seems a useful choice when $n$ is very large and real time performance is needed.

All methods presented have been implemented in C++, and run efficiently on a desktop computer, with the exception of the Movielens experiment, which needed to be run on a cluster. Obtaining a sufficiently large sample from the posterior distribution takes from a few seconds, for small problems, to several minutes, in the examples involving massive data augmentation. We are also working on distributed versions of the MCMC on parallel synchronous and asynchronous machines. 

Many of the extensions we propose for solving specific problems (for example, clustering, preference prediction, pairwise comparisons) are needed jointly in real applications, as we illustrate for example in the Movielens data. Our general framework is flexible enough to handle such extensions.

There are many situations in which rankings vary over time, as in political surveys \citep{Regenwetter1999} or book bestsellers \citep{Caron2012}. We have extended our approach to this setting \citep{asfaw2017time}. We assume to observe ranks at discrete time-points indexed by $t=0,1,\dots,T$ and let $\bm{\rho}^{(t)}$ and $\alpha^{(t)}$ denote the parameters of the Mallows model at time $t$. Interestingly, this model allows for prediction (with uncertainty quantification) of rankings in future time instances.

A natural generalization of our model is to allow for item-specific $\alpha$'s.
This is known as generalized Mallows's model, first implemented in \cite{Fligner1986}, for Kendall and Cayley distances, and further extended in \cite{MeilaBao2010}, for Kendall distance only, to the Bayesian framework. 
To our knowledge, the Mallows model with footrule and Spearman has not yet been generalized to handle  item-specific $\alpha$'s, mostly because of the obvious computational difficulties. Within our framework this appears as feasible. 


\acks{{\O}ystein S{\o}rensen and Valeria Vitelli contributed equally to this paper and are joint first authors. Marta Crispino visited OCBE at University of Oslo during this project. The authors thank Tyler Lu and Craig Boutilier for their help with the Movielens data, and Magne Thoresen for helpful discussions.}

\newpage

\setcounter{table}{0}
\setcounter{section}{0}
\setcounter{figure}{0}
\setcounter{equation}{0}

\renewcommand{\thesection}{A\arabic{section}}  
\renewcommand{\thetable}{A\arabic{table}}  
\renewcommand{\thefigure}{A\arabic{figure}} 
\renewcommand{\theequation}{A\arabic{equation}} 

\newpage
\appendix

\section{Proofs of results from Section \ref{sec:Preselection}}\label{sec:algoSupp}

\begin{proof} \textbf{of Proposition \ref{prop:preselection}.}\\
Having assumed the uniform prior across all permutations of latent consensus ranks, the desired result will hold if and only if $\sum_{j=1,\ldots,N} d(\mathbf{R}_j, \bm{\rho}) \leq \sum_{j=1,\ldots,N} d(\mathbf{R}_j, \bm{\rho}^\prime)$. This is true if $d(\mathbf{R}_j, \bm{\rho})\leq d(\mathbf{R}_j, \bm{\rho}^\prime)$ holds separately for each assessor $j,$ for $j=1,\ldots,N.$ We consider first the footrule distance $d$, and then show that the result holds also for the Kendall and Spearman distances. This proof follows Proposition 4 in \citet{MeilaBao2010}.

Suppose first, for simplicity, that all assessors have ranked the same $n$ items, that is, $\mathcal{A}_1=\mathcal{A}_2=\ldots=\mathcal{A}_N=\mathcal{A}.$ Later we allow the sets $\mathcal{A}_j$ of ranked items to be different for different assessors. Thus there are $n^* - n$ items, which nobody ranked in the original data.

We now introduce synthetic rankings for all these items as well, that is, we augment each $\mathbf{R}_j$ as recorded in the data by replacing the missing ranks of the items $A_i\in \mathcal{A}^c$ by some permutation of their possible ranks from $n+1$ to $n^*$. We then show that the desired inequality holds regardless of how these ranks $\{R_{ij}, A_i\in \mathcal{A}^c\}$ were assigned. The proof is by induction, and it is carried out in several steps.

For the first step, let $\bm{\rho}$ be a rank vector were the ranks from 1 to $n$, in any order, have been assigned to the items in $\mathcal{A}$, and the ranks $R_{ij}$ between $n+1$ and $n^*$ are given to items in $\mathcal{A}^c$. Let $\bm{\rho}^\prime$ be a rank vector obtained from $\bm{\rho}$ by a transposition of the ranks of two items, say, of $A_{i_0} \in \mathcal{A}^c$ and $A_{i_1} \in \mathcal{A}$, with $\rho_{i_0}=\rho_{i_1}^\prime \geq n+1$ and $\rho_{i_1}=\rho_{i_0}^\prime \leq n$. Fixing these two items, we want to show that $d(\mathbf{R}_j, \bm{\rho})\leq d(\mathbf{R}_j, \bm{\rho}^\prime)$. For the footrule distance we have to show that $\sum_{i=1}^n |R_{ij}- \rho_i| \leq \sum_{i=1}^n |R_{ij}- \rho_i^\prime|$. Since $\bm{\rho}$ and $\bm{\rho}^\prime$ coincide for all their coordinates $i\neq i_0,i_1$, it is enough to compare here the terms $|R_{i_0j}-\rho_{i_0}|$ and $|R_{i_1j}-\rho_{i_1}|$ on the left to the corresponding terms $|R_{i_0j}-\rho^\prime_{i_0}|$ and $|R_{i_1j}-\rho^\prime_{i_1}|$ on the right. We need to distinguish between two situations:
\begin{itemize}
\item[(i)] Suppose $R_{i_1j}\leq\rho_{i_1}.$ Then, $\rho_{i_1}^\prime-R_{i_1j}>\rho_{i_1}-R_{i_1j}$. On the other hand, $\rho_{i_0}\geq n+1$ implies that $A_{i_0}\in \mathcal{A}^c,$ and it is therefore ranked by assessor $j$ with $R_{i_0j}\geq n+1$. Therefore, $|R_{i_0j}-\rho_{i_0}^\prime|\geq|R_{i_0j}-\rho_{i_0}|.$ By combining these two results we get that $|R_{i_0j}-\rho_{i_0}| + |R_{i_1j}-\rho_{i_1}| \leq |R_{i_0j}-\rho_{i_0}^\prime| + |R_{i_1j}-\rho_{i_1}^\prime|$.
\item[(ii)] Now, suppose that $R_{i_1j}>\rho_{i_1}.$ Then, $R_{i_1j}-\rho_{i_1} \leq n - \rho_{i_1} \leq R_{i_0j}-\rho_{i_0}^\prime.$ Moreover, since $|R_{i_0j}-\rho_{i_0}| \leq |R_{i_1j}-\rho_{i_0}| = |R_{i_1j}-\rho^\prime_{i_1}|,$ we have that again $|R_{i_0j}-\rho_{i_0}| + |R_{i_1j}-\rho_{i_1}| \leq |R_{i_0j}-\rho_{i_0}^\prime| + |R_{i_1j}-\rho_{i_1}^\prime|$ holds.
\end{itemize}
The same reasoning holds also for the Kendall distance, since the Kendall distance between the two rank vectors, which are obtained from each other by a transposition of a pair of items, is the same as the footrule distance. For the Spearman distance, we only need to form squares of the distance between pairs of items, and the inequality remains valid.

For the general step of the induction, suppose that $\bm{\rho}$ has been obtained from its original version with all items in $\mathcal{A}$ ranked to the first $n$ positions, via a sequence of transpositions between items originally in $\mathcal{A}$ and items originally in $\mathcal{A}^c$. Let $\bm{\rho}^\prime$ be a rank vector where one more transposition of this type from $\bm{\rho}$ to $\bm{\rho}^\prime$ has been carried out. Then the argument of the proof can still be carried through, and the conclusion $d(\mathbf{R}_j, \bm{\rho})\leq d(\mathbf{R}_j, \bm{\rho}^\prime)$ holds. This argument needs to be complemented by considering the uniform random permutations, corresponding to the assumed prior of the ranks originally missing in the data, across their possible values from $n+1$ to $n^*$. But this is automatic, because the conclusion holds separately for all permutations of such ranks.

Finally, the argument needs to be extended to the situation in which the sets $\mathcal{A}_j$ of ranked items can be different for different assessors. In this case we are led to consider, as a by-product of the data augmentation scheme, a joint distribution of the rank vectors $\{\tilde{\mathbf{R}}_j; j=1,\ldots,N\}.$ Here, for each $j$, the $n_j$ items which were ranked first have been fixed by the data. The remaining $n-n_j$ items are assigned augmented random ranks with values between $n_j+1$ and $n$, where the probabilities, corresponding to the model $P_{n^*}$, are determined by the inference from the assumed Mallows model and the data. The conclusion remains valid regardless of the particular way in which the augmentation was done, and so it holds also when taking an expectation with respect to $P_{n^*}$. 
\end{proof}

\begin{proof} \textbf{of Corollary \ref{cor:preselectionMAP}.}\\
It follows from Proposition \ref{prop:preselection} that the $n$ top ranks in $\bm{\rho}^{MAP*}$ are all assigned to items $A_i \in \mathcal{A}.$ Therefore, using shorthand $\bm{\rho}_\mathcal{A}=(\bm{\rho}_i; A_i\in\mathcal{A})$ and $\bm{\rho}_{\mathcal{A}^c}=(\bm{\rho}_i; A_i\in\mathcal{A}^c)$ we see that $\bm{\rho}^{MAP*}$ must be of the form $\bm{\rho}^{MAP*} = (\bm{\rho}^{MAP*}_\mathcal{A}, \bm{\rho}^{MAP*}_{\mathcal{A}^c}) = (\bm{\pi}, \bm{\pi}^\prime),$ where $\bm{\pi}$ is a permutation of the set $(1,2,\ldots,n),$ and similarly $\bm{\pi}^\prime$ is some permutation of $(n+1,\ldots,n^*).$

To prove the statement, we show the following: (i) the posterior probabilities $P_{n^*}(\bm{\rho}_\mathcal{A}=\bm{\pi},\bm{\rho}_{\mathcal{A}^c} = \bm{\pi}^\prime | \text{data})$ and $P_{n^*}(\bm{\rho}_\mathcal{A}=\bm{\pi}|\bm{\rho}_{\mathcal{A}^c} = \bm{\pi}^\prime, \text{data})$ are invariant under permutations of $\bm{\pi}^\prime$, and (ii) the latter conditional probabilities $P_{n^*}(\bm{\rho}_\mathcal{A}=\bm{\pi}|\bm{\rho}_{\mathcal{A}^c} = \bm{\pi}^\prime, \text{data})$ coincide with $P_n(\bm{\rho}_\mathcal{A}=\bm{\pi}|\text{data})$. As a consequence, a list of top-$n$ items obtained from the \emph{full analysis} estimate $\bm{\rho}^{MAP*}$ qualifies also as the \emph{restricted analysis} estimate $\bm{\rho}^{MAP},$ and conversely, $\bm{\rho}^{MAP}$ can be augmented with any permutation $\bm{\pi}^\prime$ of $(n+1,\ldots,n^*)$ to jointly form $\bm{\rho}^{MAP*}$.

The first part of (i) follows by noticing that the likelihood in the \emph{full analysis}, when considering consensus rankings of the form $\bm{\rho}=(\bm{\rho}_\mathcal{A}, \bm{\rho}_{\mathcal{A}^c})=(\bm{\pi},\bm{\pi}^\prime),$ only depends on the observed data via $\bm{\pi}.$ Since the assessors act independently, each imposing a uniform prior on their unranked items, also the posterior $P_{n^*}(\bm{\rho}_\mathcal{A}=\bm{\pi},\bm{\rho}_{\mathcal{A}^c} = \bm{\pi}^\prime | \text{data})$ will depend only on $\bm{\pi}.$  The second part follows from the first, either by direct conditioning in the joint distribution, or by first computing the marginal $P_{n^*}(\bm{\rho}_{\mathcal{A}^c}=\bm{\pi}^\prime|\text{data})$ by summation, and then dividing. (ii) follows then because, for both posterior probabilities, the sample space, the prior, and the likelihood are the same.
\end{proof}

\section{Pseudo-codes of the algorithms}\label{sec:algoSupp}
We here report the pseudo-codes of the algorithms. 
The available distance functions are: Kendall, footrule, Spearman, Cayley and Hamming. For Kendall, Cayley and Hamming, there is no need to run the IS to approximate  $Z_n(\alpha)$, as it is implemented the available closed form  \citep{Fligner1986}. For footrule ($n\leq 50$) and Spearman ($n\leq 14$) the algorithm exploits the results presented in Section \ref{sec:Distances}. For footrule ($n> 50$) and Spearman ($n> 14$) the IS procedure has to be run off-line, before the MCMC.

\IncMargin{1em}
\begin{algorithm}
\begin{tiny}
\DontPrintSemicolon
\SetKwData{Left}{left}\SetKwData{This}{this}\SetKwData{Up}{up}
\SetKwFunction{Union}{Union}\SetKwFunction{FindCompress}{FindCompress}
\SetKwInOut{Input}{input}\SetKwInOut{Output}{output}
\Input{$\mathbf{R}_1,\ldots,\mathbf{R}_N$; $\lambda$, $\sigma_\alpha$, $\alpha_\text{jump},$ $L$, $d(\cdot,\cdot)$, $Z_n(\alpha)$, $M$.}
\Output{Posterior distributions of $\bm{\rho}$ and $\alpha$.}
\BlankLine
\textbf{Initialization of the MCMC:} randomly generate $\bm{\rho}_0$ and $\alpha_0$.\;\;
\For{$m\leftarrow 1$ \KwTo M}{
\textbf{M-H step: update $\bm{\rho}$:}\;
sample: $\bm{\rho^\prime} \sim \text{L\&S}(\bm{\rho}_{m-1},L)$ and $u \sim \mathcal{U}(0,1)$\;
compute: $ratio \leftarrow$ equation (\ref{eq:MHrho}) with $\bm{\rho}\leftarrow\bm{\rho}_{m-1}$ and $\alpha \leftarrow \alpha_{m-1}$\;
\lIf{u $<$ ratio}{$\bm{\rho}_m\leftarrow\bm{\rho}^\prime$}\lElse{$\bm{\rho}_m\leftarrow\bm{\rho}_{m-1}$}
\; \lIf{m$\mod \alpha_\text{jump}$ = 0} {\textbf{M-H step: update $\alpha$:}\;
sample: $\alpha^\prime \sim \log\mathcal{N}(\alpha_{m-1},\sigma^2_\alpha)$ and $u \sim \mathcal{U}(0,1)$\;
compute: $ratio \leftarrow$ equation (\ref{eq:MHalpha}) with $\bm{\rho}\leftarrow\bm{\rho}_{m}$ and $\alpha \leftarrow \alpha_{m-1}$\;
\lIf{u $<$ ratio}{$\alpha_m\leftarrow\alpha^\prime$}\lElse{$\alpha_m\leftarrow\alpha_{m-1}$}
}
}
\caption{Basic MCMC Algorithm for Complete Rankings}\label{algo:basicMH}
\end{tiny}
\end{algorithm}
\DecMargin{1em}

\clearpage


\begin{algorithm}
\begin{tiny}
\DontPrintSemicolon
\SetKwData{Left}{left}\SetKwData{This}{this}\SetKwData{Up}{up}
\SetKwFunction{Union}{Union}\SetKwFunction{FindCompress}{FindCompress}
\SetKwInOut{Input}{input}\SetKwInOut{Output}{output}
\Input{$\mathbf{R}_1,\ldots,\mathbf{R}_N$; $C$, $\psi$, $\lambda$, $\sigma_\alpha$, $\alpha_\text{jump}$, $L$, $d(\cdot,\cdot)$, $Z_n(\alpha),$ $M$.}
\Output{Posterior distributions of $\bm{\rho}_1,\ldots,\bm{\rho}_C$, $\alpha_1,\ldots,\alpha_C$, $\tau_1,\ldots,\tau_C,$ $z_1,\ldots,z_N$.}
\BlankLine
\textbf{Initialization of the MCMC:} randomly generate $\bm{\rho}_{1,0},\ldots,\bm{\rho}_{C,0},$ $\alpha_{1,0},\ldots,\alpha_{C,0},$ $\tau_{1,0},\ldots,\tau_{C,0},$ and $z_{1,0},\ldots,z_{N,0}.$\;
\;\For{$m\leftarrow 1$ \KwTo M}{
\textbf{Gibbs step: update $\tau_1,\ldots,\tau_C$}\;
compute: $n_c = \sum_{j=1}^N 1_{c}(z_{j,m-1}),$ for $c=1,\ldots,C$\;
sample: $\tau_1,\ldots,\tau_C \sim \mathcal{D}(\psi+n_1,\ldots,\psi+n_C)$\;
\;
\For{$c\leftarrow 1$ \KwTo C}{
\textbf{M-H step: update $\bm{\rho}_c$}\;
sample: $\bm{\rho}_c^\prime \sim \text{L\&S}(\bm{\rho}_{c,m-1},L)$ and  $u \sim \mathcal{U}(0,1)$\;
compute: $ratio \leftarrow$ equation (\ref{eq:MHrho}) with $\bm{\rho}\leftarrow\bm{\rho}_{c,m-1}$ and $\alpha \leftarrow \alpha_{c,m-1}$, and where the sum is over $\{j:z_{j,m-1}=c\}$\;
\lIf{u $<$ ratio}{$\bm{\rho}_{c,m}\leftarrow\bm{\rho}_c^\prime$}\lElse{$\bm{\rho}_{c,m}\leftarrow\bm{\rho}_{c,m-1}$}\;
\lIf{m$\mod \alpha_\text{jump}$ = 0} {\textbf{M-H step: update $\alpha_c$}
sample: $\alpha_c^\prime \sim \mathcal{N}(\alpha_{c,m-1},\sigma^2_\alpha)$ and $u \sim \mathcal{U}(0,1)$\;
compute: $ratio \leftarrow$ equation (\ref{eq:MHalpha}) with $\bm{\rho}\leftarrow\bm{\rho}_{c,m}$ and $\alpha\leftarrow \alpha_{c,m-1}$, and where the sum is over $\{j:z_{j,m-1}=c\}$\;
\lIf{u $<$ ratio}{$\alpha_{c,m}\leftarrow\alpha_c^\prime$}\lElse{$\alpha_{c,m}\leftarrow\alpha_{c,m-1}$}
}
}
\;
\textbf{Gibbs step: update $z_1,\ldots,z_N$}\;
\For{$j\leftarrow 1$ \KwTo N}{
\lForEach{$c\leftarrow 1$ \KwTo C}{
compute cluster assignment probabilities: $p_{cj} = \frac{\tau_{c,m}}{Z_n(\alpha_{c,m})} \exp\left[\frac{-\alpha_{c,m}}{n}d(\mathbf{R}_j,\bm{\rho}_{c,m}) \right]$
}
sample: $z_{j,m} \sim \mathcal{M}(p_{1j},\ldots,p_{Cj})$\;
}
}
\caption{MCMC Algorithm for Clustering Complete Rankings}\label{algo:clusteringMH}
\end{tiny}
\end{algorithm}

\hspace{2cm}

\IncMargin{1em}
\begin{algorithm}
\begin{tiny}
\DontPrintSemicolon
\SetKwData{Left}{left}\SetKwData{This}{this}\SetKwData{Up}{up}
\SetKwFunction{Union}{Union}\SetKwFunction{FindCompress}{FindCompress}
\SetKwInOut{Input}{input}\SetKwInOut{Output}{output}
\Input{$\{\mathcal{S}_1,\ldots,\mathcal{S}_N\}$ or $\{\text{tc}(\mathcal{B}_{1}),\ldots,\text{tc}(\mathcal{B}_{N})\}$; $\lambda$, $\sigma_\alpha$, $\alpha_\text{jump}$, $L$, $d(\cdot,\cdot)$, $Z_n(\alpha),$ $M$.}
\Output{Posterior distributions of $\bm{\rho},$ $\alpha$ and $\tilde{\mathbf{R}}_1,\ldots,\tilde{\mathbf{R}}_N$.}
\textbf{Initialization of the MCMC:} randomly generate $\bm{\rho}_0$ and $\alpha_0$.\;
\;\eIf{$\{\mathcal{S}_1,\ldots,\mathcal{S}_N\}$ among inputs}{\lForEach{$j\leftarrow 1$ \KwTo N}{randomly generate $\tilde{\mathbf{R}}^0_j$ in $\mathcal{S}_j$}}{\lForEach{$j\leftarrow 1$ \KwTo N}{randomly generate $\tilde{\mathbf{R}}^0_j$ compatible with $\text{tc}(\mathcal{B}_{j})$}}
\;\For{$m\leftarrow 1$ \KwTo M}{
\textbf{M-H step: update $\bm{\rho}$:}\;
sample: $\bm{\rho^\prime} \sim \text{L\&S}(\bm{\rho}_{m-1},L)$ and $u \sim \mathcal{U}(0,1)$\;
compute: $ratio \leftarrow$ equation (\ref{eq:MHrho}) with $\bm{\rho}\leftarrow\bm{\rho}_{m-1}$ and $\alpha \leftarrow \alpha_{m-1}$\;
\lIf{u $<$ ratio}{$\bm{\rho}_m\leftarrow\bm{\rho}^\prime$}\lElse{$\bm{\rho}_m\leftarrow\bm{\rho}_{m-1}$}
\;\lIf{m$\mod \alpha_\text{jump}$ = 0} {\textbf{M-H step: update $\alpha$:}\;
sample: $\alpha^\prime \sim \mathcal{N}(\alpha_{m-1},\sigma^2_\alpha)$ and $u \sim \mathcal{U}(0,1)$\;
compute: $ratio \leftarrow$ equation (\ref{eq:MHalpha}) with $\bm{\rho}\leftarrow\bm{\rho}_{m}$ and $\alpha \leftarrow \alpha_{m-1}$\;
\lIf{u $<$ ratio}{$\alpha_m\leftarrow\alpha^\prime$}\lElse{$\alpha_m\leftarrow\alpha_{m-1}$}
}
\;\textbf{M-H step: update $\tilde{\mathbf{R}}_1,\ldots,\tilde{\mathbf{R}}_N$:}\;
\For{$j\leftarrow 1$ \KwTo N}{
\lIf{$\{\mathcal{S}_1,\ldots,\mathcal{S}_N\}$ among inputs}{sample: $\tilde{\mathbf{R}}^\prime_j$ in $\mathcal{S}_j$ from the leap-and-shift distribution centered at $ \tilde{\mathbf{R}}^{m-1}_j$}\lElse{sample: $\tilde{\mathbf{R}}^\prime_j$ from the leap-and-shift distribution centered at $\tilde{\mathbf{R}}^{m-1}_j$ and compatible with $\text{tc}(\mathcal{B}_{j})$}
compute: $ratio \leftarrow$ equation (\ref{eq:MHratioAug}) with $\bm{\rho}\leftarrow\bm{\rho}_{m}$, $\alpha \leftarrow \alpha_{m}$ and $\tilde{\mathbf{R}}_j \leftarrow \tilde{\mathbf{R}}^{m-1}_j$\;
sample: $u \sim \mathcal{U}(0,1)$\;
\lIf{u $<$ ratio}{$\tilde{\mathbf{R}}^m_j\leftarrow\tilde{\mathbf{R}}^\prime_j$}\lElse{$\tilde{\mathbf{R}}^m_j\leftarrow\tilde{\mathbf{R}}^{m-1}_j$}
}
}
\caption{MCMC Algorithm for Partial Rankings or Pairwise Preferences}\label{algo:partialMH}
\end{tiny}
\end{algorithm}
\DecMargin{1em}

\clearpage

\IncMargin{1em}
\begin{algorithm}[h!]
\begin{tiny}
\DontPrintSemicolon
\SetKwData{Left}{left}\SetKwData{This}{this}\SetKwData{Up}{up}
\SetKwFunction{Union}{Union}\SetKwFunction{FindCompress}{FindCompress}
\SetKwInOut{Input}{input}\SetKwInOut{Output}{output}
\Input{$\{\mathcal{S}_1,\ldots,\mathcal{S}_N\}$ or $\{\text{tc}(\mathcal{B}_{1}),\ldots,\text{tc}(\mathcal{B}_{N})\}$; $C$, $\psi$, $\lambda$, $\sigma_\alpha$, $\alpha_\text{jump}$, $L$, $d(\cdot,\cdot)$, $Z_n(\alpha),$ $M$.}
\Output{Posterior distributions of $\bm{\rho}_1,\ldots,\bm{\rho}_C$, $\alpha_1,\ldots,\alpha_C$, $\tau_1,\ldots,\tau_C,$ $z_1,\ldots,z_N,$ and $\tilde{\mathbf{R}}_1,\ldots,\tilde{\mathbf{R}}_N$.}
\BlankLine
\textbf{Initialization of the MCMC:}\;
randomly generate $\bm{\rho}_{1,0},\ldots,\bm{\rho}_{C,0},$ $\alpha_{1,0},\ldots,\alpha_{C,0},$ $\tau_{1,0},\ldots,\tau_{C,0},$ and $z_{1,0},\ldots,z_{N,0}.$\;
\;\eIf{$\{\mathcal{S}_1,\ldots,\mathcal{S}_N\}$ among inputs}{\lForEach{$j\leftarrow 1$ \KwTo N}{randomly generate $\tilde{\mathbf{R}}^0_j$ in $\mathcal{S}_j$}}{\lForEach{$j\leftarrow 1$ \KwTo N}{randomly generate $\tilde{\mathbf{R}}^0_j$ compatible with $\text{tc}(\mathcal{B}_{j})$}}
\;
\For{$m\leftarrow 1$ \KwTo M}{
\textbf{Gibbs step: update $\tau_1,\ldots,\tau_C$}\;
compute: $n_c = \sum_{j=1}^N 1_{c}(z_{j,m-1}),$ for $c=1,\ldots,C$\;
sample: $\tau_1,\ldots,\tau_C \sim \mathcal{D}(\psi+n_1,\ldots,\psi+n_C)$\;
\;
\For{$c\leftarrow 1$ \KwTo C}{
\textbf{M-H step: update $\bm{\rho}_c$}\;
sample: $\bm{\rho}_c^\prime \sim \text{L\&S}(\bm{\rho}_{c,m-1},L)$  and $u \sim \mathcal{U}(0,1)$\;
compute: $ratio \leftarrow$ equation (\ref{eq:MHrho}) with $\bm{\rho}\leftarrow\bm{\rho}_{c,m-1}$ and $\alpha \leftarrow \alpha_{c,m-1}$, and where the sum is over $\{j:z_{j,m-1}=c\}$\;
\lIf{u $<$ ratio}{$\bm{\rho}_{c,m}\leftarrow\bm{\rho}_c^\prime$}\lElse{$\bm{\rho}_{c,m}\leftarrow\bm{\rho}_{c,m-1}$}
\;
\lIf{m$\mod \alpha_\text{jump}$ = 0} {\textbf{M-H step: update $\alpha_c$}\;
sample: $\alpha_c^\prime \sim \mathcal{N}(\alpha_{c,m-1},\sigma^2_\alpha)$ and $u \sim \mathcal{U}(0,1)$\;
compute: $ratio \leftarrow$ equation (\ref{eq:MHalpha}) with $\bm{\rho}\leftarrow\bm{\rho}_{c,m}$ and $\alpha \leftarrow \alpha_{c,m-1}$, and where the sum is over $\{j:z_{j,m-1}=c\}$\;
\lIf{u $<$ ratio}{$\alpha_{c,m}\leftarrow\alpha_c^\prime$}\lElse{$\alpha_{c,m}\leftarrow\alpha_{c,m-1}$}
}
}
\;
\textbf{Gibbs step: update $z_1,\ldots,z_N$}\;
\For{$j\leftarrow 1$ \KwTo N}{
\lForEach{$c\leftarrow 1$ \KwTo C}{
compute cluster assignment probabilities: $p_{cj} = \frac{\tau_{c,m}}{Z_n(\alpha_{c,m})} \exp\left[\frac{-\alpha_{c,m}}{n}d(\tilde{\mathbf{R}}^{m-1}_j,\bm{\rho}_{c,m}) \right]$
}
sample: $z_{j,m} \sim \mathcal{M}(p_{1j},\ldots,p_{Cj})$\;
}
\;
\textbf{M-H step: update $\tilde{\mathbf{R}}_1,\ldots,\tilde{\mathbf{R}}_N$:}\;
\For{$j\leftarrow 1$ \KwTo N}{
\lIf{$\{\mathcal{S}_1,\ldots,\mathcal{S}_N\}$ among inputs}{sample: $\tilde{\mathbf{R}}^\prime_j$ in $\mathcal{S}_j$ from the leap-and-shift distribution centered at $ \tilde{\mathbf{R}}^{m-1}_j$}\lElse{sample: $\tilde{\mathbf{R}}^\prime_j$ from the leap-and-shift distribution centered at $\tilde{\mathbf{R}}^{m-1}_j$ and compatible with $\text{tc}(\mathcal{B}_{j})$}
compute: $ratio \leftarrow$ equation (\ref{eq:MHratioAug}) with $\bm{\rho}\leftarrow\bm{\rho}_{z_{j,m},m}$, $\alpha \leftarrow \alpha_{z_{j,m},m}$ and $\tilde{\mathbf{R}}_j \leftarrow \tilde{\mathbf{R}}^{m-1}_j$ \;
sample: $u \sim \mathcal{U}(0,1)$\;
\lIf{u $<$ ratio}{$\tilde{\mathbf{R}}^m_j\leftarrow\tilde{\mathbf{R}}^\prime_j$}\lElse{$\tilde{\mathbf{R}}^m_j\leftarrow\tilde{\mathbf{R}}^{m-1}_j$}
}
\;
}
\caption{MCMC Algorithm for Clustering Partial Rankings or Pairwise Preferences}\label{algo:clusteringPair}
\end{tiny}
\end{algorithm}
\DecMargin{1em}

\section{Sample from Mallows model}\label{sec:sampling}
We here explain our proposed procedure to sample rankings from the Mallows model. 

To sample full rankings $\mathbf{R}_1,...,\mathbf{R}_N\sim\text{Mallows}(\bm{\rho},\alpha)$, we use the following scheme (sketched in Algorithm \ref{algo:basicMHS}).
We run a basic Metropolis-Hastings algorithm with fixed consensus  $\bm{\rho}\in\mathcal{P}_n$, $\alpha>0$ and with a given distance measure, $d(\cdot,\cdot)$, until convergence. Once convergence is achieved, we continue sampling, and store the so obtained rankings at regular intervals (large enough to achieve  independence) until we have reached the desired data dimension. 
%

\begin{algorithm}
\begin{tiny}
\DontPrintSemicolon
\SetKwData{Left}{left}\SetKwData{This}{this}\SetKwData{Up}{up}
\SetKwFunction{Union}{Union}\SetKwFunction{FindCompress}{FindCompress}
\SetKwInOut{Input}{input}\SetKwInOut{Output}{output}
\Input{$\bm{\rho}$, $\alpha$, $d$, N, L}
\Output{$\mathbf{R}_1,...,\mathbf{R}_N$}
\BlankLine
\textbf{Initialization of the MCMC:} randomly generate $\mathbf{R}_{1,0},...,\mathbf{R}_{N,0}$\;
\For{$m\leftarrow 1$ \KwTo M}{
\For{$j\leftarrow 1$ \KwTo N}{\BlankLine
sample $\mathbf{R}_j^\prime \sim \text{L\&S}(\mathbf{R}_{j,m-1},L)$\;
compute: $ratio = \frac{P_L(\mathbf{R}_j | \mathbf{R}_j^{\prime})}{P_L(\mathbf{R}_j^{\prime} | \mathbf{R}_j) }\exp\left\{-\frac{\alpha}{n} \sum_{j=1}^{N}\left[d(\mathbf{R}^{\prime}_{j},\bm{\rho}) - d(\mathbf{R}_{j},\bm{\rho}) \right]\right\} $ with $\mathbf{R}_j\leftarrow\mathbf{R}_{j,m-1}$\;
sample: $u \sim \mathcal{U}(0,1)$\;
\BlankLine
\eIf{u $<$ ratio}{$\mathbf{R}_{j,m}\leftarrow\mathbf{R}_j^\prime$}{$\mathbf{R}_{j,m}\leftarrow\mathbf{R}_{j,m-1}$}
\;
}
}
\caption{MCMC Sampler for full rankings}\label{algo:basicMHS}
\end{tiny}
\end{algorithm}

In case of heterogeneous rankings, we sample from Algorithm \ref{algo:clusterMHSCl}. As inputs, we give the number of clusters $C$, the fixed consensuses  $\bm{\rho}_\text{1},...,\bm{\rho}_\text{C}$, the fixed $\alpha_\text{1},...,\alpha_\text{C}$, the hyper-parameter $\bm{\psi}=(\psi_1,...,\psi_C)$ of the Dirichlet density over the proportion of assessors in the clusters,  and  $d(\cdot,\cdot)$.
The algorithm then returns the rankings $\mathbf{R}_1,...,\mathbf{R}_N$, sampled from a Mixture of Mallows models, as well as the the cluster assignments $z_1,...,z_N$.

\begin{algorithm}
\begin{tiny}
\DontPrintSemicolon
\SetKwData{Left}{left}\SetKwData{This}{this}\SetKwData{Up}{up}
\SetKwFunction{Union}{Union}\SetKwFunction{FindCompress}{FindCompress}
\SetKwInOut{Input}{input}\SetKwInOut{Output}{output}
\Input{$C$, $\bm{\rho}_{1:C}$, $\alpha_{1:C}$, $\bm{\psi}$, $d$, N, L}
\Output{$\mathbf{R}_1,...,\mathbf{R}_N$ and $z_1,...,z_N$}
\BlankLine
\textbf{Initialization of the MCMC:} randomly generate $\mathbf{R}_{1,0},...,\mathbf{R}_{N,0}$\;
randomly generate $\tau_1,...,\tau_C\sim\text{Dir}(\bm{\psi})$\;
randomly generate $z_1,...,z_N\sim\text{Mn}(1,\tau_1,...,\tau_C)$\;
\For{$m\leftarrow 1$ \KwTo M}{
\For{$c\leftarrow 1$ \KwTo C}{\BlankLine
compute: $N_c = \sum_{j=1}^N \mathbb{1}_{c}(z_{j}),$\;
sample $N_c$ ranks with  Algortihm \ref{algo:basicMHS}\;
 }
}
\caption{MCMC Sampler for full rankings with clusters} \label{algo:clusterMHSCl}
\end{tiny}
\end{algorithm}

For generating top-k rankings, we simply generate $\mathbf{R}_1,...,\mathbf{R}_N$ with  Algorithm \ref{algo:basicMHS}, and then keep only the top$-k$ items.
In case of clusters, we do the same as above, but starting with Algorithm \ref{algo:clusterMHSCl}.  

Finally, to sample sets of pairwise comparisons, $\mathcal{B}_1,...,\mathcal{B}_N$, we  first generate $\mathbf{R}_1,...,\mathbf{R}_N$ with  Algortihm \ref{algo:basicMHS}. We then select the number  of pairwise comparisons, $T_1,...,T_N$, that each assessor will evaluate\footnote{Here it is possible to choose the same number of comparisons $T_j=T\leq n(n-1)/2$, $\forall j=1,...,N$ but also to have a different number of pairs per assessor. In this paper, for a given mean parameter $\lambda_T$, we independently sample $T_1,...,T_N\sim \text{TruncPoiss}(\lambda_T, n(n-1)/2)$. }. Finally, given $\mathbf{R}_1,...,\mathbf{R}_N$ and $T_1,...,T_N$, we randomly sample $T_j$ pairs (for each assessor $j=1,\ldots,N$) from the collection of all possible $n(n-1)/2$ pairs, and obtain pairwise preferences by ordering all pairs according to $\mathbf{R}_j$. For generating pairwise comparisons with clusters, we follow the previous procedure, but starting with Algorithm \ref{algo:clusterMHSCl}.

\newpage
\vskip 0.2in
\bibliography{references}                                                                        

\end{document}